\documentclass{gtsThesis}[1999/05/22]
\title{SYMETRIE, GEOMETRIE,\\ TOPOLOGIE ET SPINS}
\subtitle{Spins de Heisenberg à la limite continue\\ Membranes magnétiques}
\author{\textsc{Jérôme~BENOIT}}
\thesisdate{le~29~juin~1999}
\thesisdiploma{Doctorat de l'Université de Cergy-Pontoise\\
  {\normalsize(spécialité physique)}}
\date{%
  \texttt{Version~{\sgtsversion} compilée~le~\today}\\
  \texttt{cond-mat}/9909129
  }
\thesischairman{{\Mistername} \textsc{Jean-François~SADOC}}
\firstreporter{{\Mistername}  \textsc{Karol~PENSON}}
\secondreporter{{\Mistername} \textsc{Avadh~SAXENA}}
\thesisadvisor{{\Mistername}  \textsc{Rossen~DANDOLOFF}}
\firstreader{{\Madamname}     \textsc{Radha~BALAKRISHNAN}}
\secondreader{{\Mistername}   \textsc{Jérôme~LEON}}
\thirdreader{{\Mistername}    \textsc{Tuong~TRUONG}}
\copyright{\sgtscopyleft}
\makeauthorindex
\makewordindex
\begin{document}
\maketitle
\frontmatter
\include{gtsPreamble}
\include{gtsAbstracts}
\tableofcontents
\mainmatter
\ProvidesFile{gtsIntroduction.tex}[1999/05/20 v0.9b GTS project: Introduction]
\specialchapter{Introduction}
\begin{introduction}
Le mécanisme responsable de
la supraconductivité à haute température
n'est pas encore connu :
les théories \guillemets{conventionnelles}
de la matière condensée
ne prédisent pas de températures critiques
appartenant au domaine actuellement atteint
par les expériences (de $40 K$ à $120 K$).
À basse température
l'interaction entre les électrons
(les porteurs de charge du réseau cristallin)
et les phonons
(les vibrations élémentaires du réseau cristallin)
induit la supraconductivité \guillemets{ordinaire}.
Par contraste
l'état supraconducteur à haute température
serait dû aux spins des électrons.
\bigskip

Dans ce contexte
un argument simple de topologie suggère l'existence
pour un réseau bidimensionnel antiferromagnétique
d'états caractérisés par un invariant topologique :
il s'agit en fait de recouvrir
la sphère~unité~$S^2$ (la variété des spins)
en parcourant
l'hypersphère~unité~$S^3$ (le plan-temps compactifié)
---
le nombre de recouvrement étant
l'invariant topologique de Hopf.
Une théorie de champs continue d'un tel système
est généralement décrite par le modèle $\sigma$ non-linéaire.
l'émergence lors du passage à une théorie de champs continue
de l'invariant de Hopf dans le Lagrangien
(conjecturée par Dzyaloshinski, Polyakov et Wiegmann)
est l'objet de multiples controverses :
cela modifierait la statistique du système
et révélerait l'existence d'une transmutation de Fermi-Bose
\cite{FBT}.
Pour les systèmes de basse dimension
ce mécanisme se généraliserait
en suivant la hiérarchie des algèbres~à~divisions
\cite{PPFBT}.
Avant d'envisager tout passage à une théorie des champs continue,
la construction d'un paramètre d'ordre local pertinent
est primordiale :
la première partie de ce mémoire
suggère la construction
d'un nouveau paramètre d'ordre local
qui reproduit la hiérarchie des algèbres~à~divisions
et dont le Lagrangien peut exhiber le terme de Hopf.
\bigskip

De plus les états statiques
d'un réseau bidimensionnel antiferromagnétique
sont ici caractérisé par l'indice de Pontrjagin
qui décompte les recouvrements de la variété des spins
lorsque le réseau bidimensionnel est parcouru :
à la limite continue
il s'agit d'envoyer une membrane compactifiée
(la surface magnétique)
sur la sphére~unité~$S^2$ (la variété des spins)
\textlatin{via} le modèle $\sigma$~non-linéaire
(le Hamiltonien magnétique).
Certaines membranes courbes admettent
des distributions de spins de type soliton
caractérisées par un paramètre de forme.
La deuxième partie de ce mémoire
étudie la compétition entre
l'énergie magnétique de la distribution des spins
et l'énergie élastique de la membrane
en suggérant une approche topologique.
Enfin un rétrécissement global
avec des renflements localisés aux solitons
est mis en évidence.
\end{introduction}

\ProvidesFile{gtsChapter1.tex}[1999/05/06 v0.6b GTS project: Chapter 1]
\chapter{Le réseau antiferromagnétique}
\label{chp/AFL}

\section{Le modèle antiferromagnétique de Heisenberg}
\begin{introduction}
Dans les solides
les interactions entre électrons
sont souvent importantes
et d'une extraordinaire complexité.
En physique de la matière condensée,
ce problème d'interaction électronique
se réduit dans la plupart des cas
à un problème de \wi[spin]{spins}
couplés sur \wi[reseau@réseau]{réseau} \cite{Fradkin,Auerbach,Levy}.
Ici
les spins seront identiques et
occuperont les \wi{n{\oe}uds}
de \wi[reseau@réseau ! régulier isotrope]{réseaux réguliers isotropes}
de basse dimension.
\end{introduction}
\wih{reseau@réseau ! antiferromagnétique}
\wih{modèle antiferromagnétique de Heisenberg}
\swih[modèle antiferromagnétique de Heisenberg]{Heisenberg}
\swih[spin]{moment angulaire ! intrinsèque}

\subsection{Spins sur réseau}
L'étude du magnétisme nécessite l'introduction
d'un objet physique typiquement quantique :
le \wibf[spin]{moment angulaire intrinsèque}
communément appelé \wibf{spin}.
Du point de vue de la théorie des champs,
le spin d'une particule élémentaire
correspond à la
\wi[charge ! de N{\oe}ther]{\guillemets{charge} de N{\oe}ther}
\swih[charge de N{\oe}ther]{N{\oe}ther}
associée à l'\wi{isotropie} de l'espace-temps
propre à cette particule élémentaire
\cite{BailinLove,Goldstein,Sivardiere}.
Autrement dit,
le spin d'une particule élémentaire traduit
la rotation de ladite particule élémentaire
sur elle-même,
sa rotation dans l'espace-temps étant décrite
par son \wibf[moment angulaire ! orbital]%
  {moment angulaire orbital}.
Notons au passage que
le moment angulaire orbital
possède un équivalent classique
contrairement au moment angulaire intrinsèque.
Dans le cadre présent,
à chaque n{\oe}ud $\left(\gtsVertexI\right)$
du réseau est associé un spin représenté
par un opérateur vectoriel ${\gtsSpin_{\!\gtsVertexI}}^{\alpha}$
qui engendre une \wiit{algèbre de Lie}\swih[algèbre de Lie]{Lie}
\cite{Sattinger} de dimension $\gtsSpinDim$
selon la relation de commutation
\begin{equation}
\label{Spin/commutation/crude}
\left[{\gtsSpin_{\!\gtsVertexI}}^{\alpha},
{\gtsSpin_{\!\gtsVertexJ}}^{\beta}\right]=
\ci\Planck\:
\gtsastr^{\alpha\beta\gamma}\Kronecker_{\gtsVertexI\gtsVertexJ}\;
{\gtsSpin_{\!\gtsVertexI}}_{\gamma}
\end{equation}
avec $\Planck$ la constante de Planck,
$\gtsastr^{\alpha\beta\gamma}$ le tenseur unitaire
parfaitement antisymétrique d'ordre trois
et $\Kronecker_{\gtsVertexI\gtsVertexJ}$
le symbole de Kronecker.
De fait l'\wiit[opérateur ! de Casimir]{opérateur de Casimir} $\gtsSpin^{2}$
\swih[opérateur de Casimir]{Casimir}
\cite{Sattinger}
fixe la dimension
du \wiit[degrés de liberté ! local]{degrés de liberté local}
du réseau :
\begin{equation}
\label{Spin/Casimir/crude}
\gtsSpin^{2}\equiv%
  {\gtsSpin_{\!\gtsVertexI}}^{\alpha}%
  {\gtsSpin_{\!\gtsVertexI}}_{\alpha}=%
    \gtsSpinDim\left(\gtsSpinDim+1\right)
\qquad\forall\gtsVertexI.
\end{equation}

Les réseaux envisagés sont
les \wiit[pavage compact régulier]{pavages compacts régulier}
(sans lacune ni recouvrement)
des espaces Euclidiens de basse dimension ($\gtsDimension=1,2,3,4$)
avec des \wiit[polytope régulier]{polytopes réguliers} identiques
\cite{Coxeter,Sivardiere}.
\enlargethispage*{2\baselineskip}
Dans le cadre présent,
les réseaux seront donc caractérisés
par un polytope régulier :
\begin{itemize}
  \item le réseau aura la dimension $\gtsDimension$
    du polytope régulier;
  \item les symétries qui préservent le polytope régulier
    préserveront le réseau;
  \item les translations qui reproduisent
    le réseau à partir d'un exemplaire du polytope régulier
    préserveront le réseau;
  \item la longueur des arêtes du polytope régulier correspondra
    au \wiit[reseau@réseau ! pas du]{pas}
    \swih[réseau]{pas} $\gtsLStep$ du réseau.
\end{itemize}
Observant que
les polytopes réguliers de dimension $\gtsDimension$
sont également des \wi[pavage compact régulier]{pavages réguliers}
de la sphère $S^{\gtsDimension-1}$
\cite{duVal,Coxeter},
introduisons la notion
de \wibf[surreseau@\guillemets{sur-réseau}]{\guillemets{sur-réseau}} :
nous entendrons par \guillemets{sur-réseau}
de dimension $\gtsDimension$
un réseau de même dimension
dont peut s'extraire
un réseau régulier de dimension $\gtsDimension$ en assimilant
ses \wibf[surnoeud@\guillemets{sur-n{\oe}ud}]{\guillemets{sur-n{\oe}uds}}
à des sphères $S^{\gtsDimension-1}$ identiquement
et régulièrement pavées (\figurename~\ref{fig/para/bidim}).
Il est évident qu'à
un \wi[reseau@réseau ! régulier]{réseau régulier}
peut être associé plusieurs \guillemets{sur-réseaux},
et réciproquement.

Le \wi[magnétisme]{comportement magnétique}
sur de longues distances
(longues devant le \wi[reseau@réseau ! pas du]{pas} $\gtsLStep$ du réseau)
sera appréhendé
en passant à une théorie de champs continue
\wih{limite ! continue}
selon une \witextbf{prescription due à Affleck}
\wih{prescription d'Affleck}
\swih[prescription d'Affleck]{Affleck}
\cite{SGSBA,QSCHG}.
Il s'agit de faire tendre conjointement
le \wi[reseau@réseau ! pas du]{pas} $\gtsLStep$
du réseau vers zéro
et le nombre~quantique~$\gtsSpinDim$ vers l'infini
tout en maintenant constantes
les quantités physiques mesurables expérimentalement.
Cette prescription traite implicitement
les spins à la \wiit[limite ! classique]{limite classique}.
En effet
la relation de commutation (\ref{Spin/commutation/crude}),
qui doit s'écrire pour garder un sens
\begin{equation}
\label{Spin/commutation/nested}
\left[\frac{{\gtsSpin_{\!\gtsVertexI}}^{\alpha}}{\gtsSpinDim},
\frac{{\gtsSpin_{\!\gtsVertexJ}}^{\beta}}{\gtsSpinDim}\right]=
\ci\frac{\Planck}{\gtsSpinDim}\:
\gtsastr^{\alpha\beta\gamma}\Kronecker_{\gtsVertexI\gtsVertexJ}\;
\frac{{\gtsSpin_{\!\gtsVertexI}}_{\gamma}}{\gtsSpinDim},
\end{equation}
s'évanouit lorsque
le nombre~quantique~$\gtsSpinDim$ s'approche de l'infini.
Nous supposerons
les résultats physiques obtenus applicables
pour des valeurs physiques
du nombre~quantique~$\gtsSpinDim$
($\gtsSpinDim=\frac{1}{2}$ ou $\gtsSpinDim=1$)
et des valeurs du pas $\gtsLStep$
caractéristiques des solides cristallins
(de l'ordre de 4 \AA).
Il est évident que la construction
d'un \wiit[paramètre d'ordre ! local]{paramètre d'ordre local}
pertinent est un prélude nécessaire et complexe
qui doit à la fois clarifier les phénomènes physiques observés
et valider la justesse du passage à une théorie de champs continue.

\subsection{Interactions d'échange : Hamiltonien de Heisenberg}
\wih{echange@échange électronique}
Le \wi[magnétisme]{magnétisme électronique} provient
de l'\wiit[interaction ! Coulombienne]{interaction Coulombienne}
entre électrons qui force les \wi[spin]{spins}
dans des états ordonnés :
le \wiit[Pauli~(principe de)]{principe de Pauli}
impose des états complètement antisymétriques
lorsque deux fermions sont échangées.
Généralement
le Hamiltonien initiale
d'un solide cristallin
qui est extraordinairement complexe
peut être réduit sous une forme simple
paramétrisée seulement en termes de spins.
Le plus souvent cet Hamiltonien
effectif est de \wiit[Hamiltonien ! Heisenberg@de Heisenberg]{Heisenberg}
\swih[Hamiltonien de Heisenberg]{Heisenberg}
\cite{Levy}:
\begin{equation}
\label{Hm/Heiseinberg/generic}
\gtsHm_{H}=
-\sum\limits_{\gtsVertexI,\gtsVertexJ}%
  \gtsJ_{\gtsVertexI\gtsVertexJ}^{\alpha\beta}\,%
  {\gtsSpin_{\!\gtsVertexI}}_{\alpha}
  {\gtsSpin_{\!\gtsVertexJ}}_{\beta},
\end{equation}
où les indices $\gtsVertexI$ et $\gtsVertexJ$ étiquettent
les porteurs de spins.
Les tenseurs génériques
$\gtsJ_{\gtsVertexI\gtsVertexJ}^{\alpha\beta}$
décrivent les \wiit[echange@échange électronique]{interactions d'échange}.
\swih[échange électronique]{interaction ! d'échange}
\enlargethispage*{6\baselineskip}
Pour un \wiit[couplage ! antiferromagnétique isotrope]%
  {couplage antiferromagnétique isotrope}
entre premiers voisins
sur un réseau régulier de spins
nous lisons
\begin{equation}
\label{Hm/Heiseinberg/afnn}
\gtsHm_{H}=
\gtsJ\sum\limits_{\left<\gtsVertexI,\gtsVertexJ\right>}
  {\gtsSpin_{\!\gtsVertexI}}^{\alpha}
  {\gtsSpin_{\!\gtsVertexJ}}_{\alpha},
\end{equation}
avec $\gtsJ$ une constante
de \wiit[couplage ! spin-spin]{couplage spin-spin} positive.

À haute température
(température supérieure à la \wiit{température de Curie})
\swih[température de Curie]{Curie}
il n'y a pas
d'\wiit[ordonnancement ! magnétique]{ordonnancement magnétique} :
le système magnétique est
dans un \wibf[etat@état ! paramagnétique]{état paramagnétique}.
Le système de \wi[spin]{spins} est alors localement
isotrope et invariant par rotation :
un système paramagnétique possède
les symétries de son réseau de spins.
À basse température différents ordres peuvent émerger.
À très basse température
(température proche du \wiit[zero@zéro absolu]{zéro absolu})
les systèmes de spins réguliers et isotropes acquièrent
une orientation particulière dans l'espace :
il y a un \wibf[ordonnancement ! magnétique]{ordonnancement magnétique}.
Cette direction spontanée
\wiit[brisure de symétrie]{brise} l'invariance locale
par rotation du système.
En observant qu'un système physique
ne peut pas changer graduellement de symétrie
(une symétrie existe ou n'existe pas)
Landau a mis en évidence
qu'une \wiit{transition de phase du second ordre}
doit séparer les états de différentes symétries
\cite{Levy,Sivardiere}.
Généralement
quand une symétrie est \wiit[brisure de symétrie]{brisée}
l'\wibf[etat@état ! non-symétrique]{état non-symétrique}
est caractérisé
par un \wibf[paramètre d'ordre ! local]{paramètre d'ordre local} :
selon Landau tout paramètre local
dont la valeur moyenne
est nulle dans l'état symétrique
et non-nulle dans l'état non-symétrique
peut constituer un
\wiit[paramètre d'ordre ! local]{paramètre d'ordre local}
\cite{Sivardiere}.

\begin{figure}[bh]
  \begin{center}
    \begin{minipage}[c]{.48\linewidth}
      \includegraphics[width=\linewidth]{sgtsTriangularLattice.mps}
      \subcaption{Le réseau triangulaire paramagnétique}
    \end{minipage} \hfill
    \begin{minipage}[c]{.48\linewidth}
      \includegraphics[width=\linewidth]{sgtsSquareLattice.mps}
      \subcaption{Le réseau carré paramagnétique}
    \end{minipage}
  \caption[Les réseaux triangulaire et carré paramagnétiques]{%
    Les \witextit{réseaux triangulaire} et \witextit{carré paramagnétiques} :
    \wih{reseau@réseau ! triangulaire paramagnétique}
    \wih{reseau@réseau ! carré paramagnétique}
    l'invariance locale par rotation des deux systèmes magnétiques
    est schématisée en magnifiant les n{\oe}uds porteurs de spin.
    Pour chaque réseau le
    \wiit[surpavage@\guillemets{sur-pavage}]{\guillemets{sur-pavage}}
    représente un choix possible de
    \wiit[surreseau@\guillemets{sur-réseau}]{\guillemets{sur-réseau}} :
    les \wiit[plaquette de Néel]{plaquettes} grisées correspondant
    alors aux \wiit[surnoeud@\guillemets{sur-n{\oe}ud}]{\guillemets{sur-n{\oe}uds}}.
    }
  \label{fig/para/bidim}
  \end{center}
\end{figure}

\subsection{\'Etat antiferromagnétique : sous-réseaux de Néel}
\begin{figure}[p]
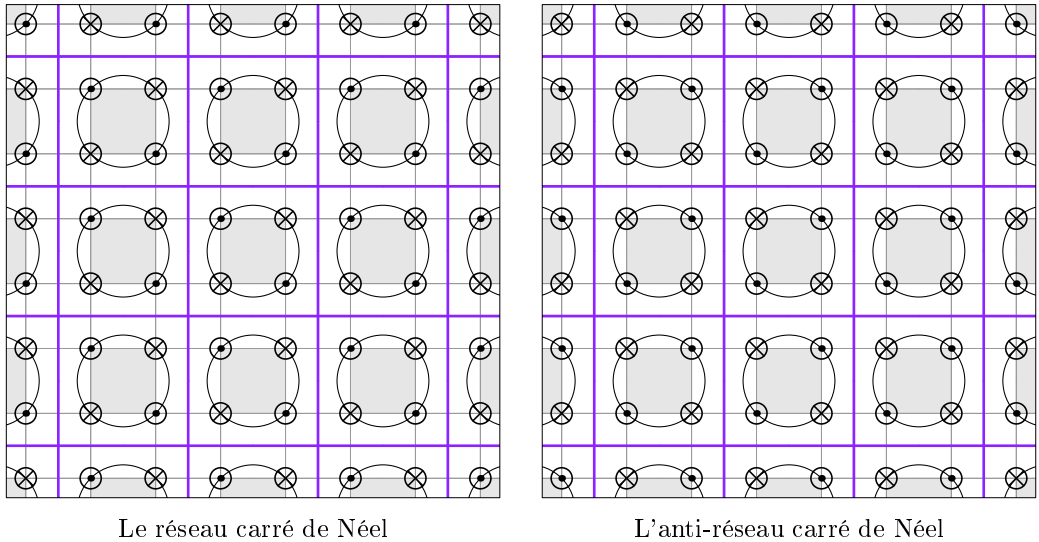

  \begin{center}
    \begin{minipage}[c]{.48\linewidth}
      \includegraphics[width=\linewidth]{sgtsSquareLatticeNeel.mps}
      \subcaption{Le réseau carré de Néel}
    \end{minipage} \hfill
    \begin{minipage}[c]{.48\linewidth}
      \includegraphics[width=\linewidth]{sgtsSquareLatticeANeel.mps}
      \subcaption{L'anti-réseau carré de Néel}
    \end{minipage}
  \caption[Le réseau carré ferromagnétique]{%
    Le \wiit[reseau@réseau ! carré ferromagnétique]%
      {réseau carré ferromagnétique} :
    l'arrangement alterné des spins est \wiit{bicoloré}.
    L'antisymétrie bicolorée est usuellement représentée
    par deux familles de vecteurs de direction identique
    mais de sens opposé :
    les \wi[sousreseau@sous-réseau de Néel]{sous-réseaux de Néel}
    ($\gtsSpinUp$ et $\gtsSpinDown$).
    }
  \label{fig/af/square}
  \end{center}
\end{figure}

Dans un solide cristallin antiferromagnétique
deux spins adjacents font de leur mieux pour s'aligner
selon des orientations opposées ($\gtsSpinUpDown$).
Il se crée ainsi un \wibf{motif alterné}
à travers le réseau formé par les spins.
Sur un réseau régulier de spins
cet arrangement alterné se traduit
par l'émergence d'une antisymétrie locale
\cite{Sivardiere} :
les n{\oe}uds de chaque polytope
d'un \guillemets{sur-réseau} du réseau régulier initial
sont colorés.
Cette \wibf{brisure de symétrie} locale caractérise
l'\wibf[etat@état ! antiferromagnétique]{état antiferromagnétique}.
Pour illustration :
la \wiit[chaîne ! paramagnétique]{chaîne paramagnétique}
de pas $\gtsLStep$
(\figurename~\ref{fig/para/chain})
est préservée par la translation de pas $\gtsLStep$,
alors que
la \wiit[chaîne ! antiferromagnétique]{chaîne antiferromagnétique}
de pas $\gtsLStep$ (\figurename~\ref{fig/af/chain})
est alternée par l'antitranslation de pas $\gtsLStep$
et préservée par la translation de pas $2\gtsLStep$
(\textlatin{i.e.} le pas des \guillemets{sur-chaînes}).
L'état classique correspondant
à l'énergie la plus basse du
\wiit[Hamiltonien ! Heisenberg@de Heisenberg antiferromagnétique]%
  {Hamiltonien de Heisenberg antiferromagnétique}
(\ref{Hm/Heiseinberg/afnn}),
communément appelé l'\wiit[etat@état ! Neel@de Néel]{état de Néel},
\swih[état de Néel]{Neel@Néel}
constitue une représentation immédiate
du réseau de spins coloré.
Les spins du \wiit[reseau@réseau ! Neel@de Néel]{réseau de Néel}
\swih[réseau de Néel]{Neel@Néel}
ayant la même orientation forment
les \wiit[sousreseau@sous-réseau de Néel]{sous-réseaux de Néel} :
\swih[sous-réseau de Néel]{Neel@Néel}
les réseaux cubiques\,%
\footnote{\label{fnt/rl/cubic/Neel}%
Dans le cadre présent :
la chaîne ($\gtsDimension\!\!=\!\!1$),
le réseau carré ($\gtsDimension\!\!=\!\!2$),
le réseau cubique ($\gtsDimension\!\!=\!\!3$)
et le réseau $4$-cubique ($\gtsDimension\!\!=\!\!4$).
} %
de Néel sont \wiit{bipartites}
(\figuresname~\ref{fig/af/chain} et~\ref{fig/af/square}),
le réseau triangulaire de Néel est \wiit{tripartite}
(\figurename~\ref{fig/af/triangular}).
Enfin les \guillemets{sur-n{\oe}uds}
du réseau de Néel sont usuellement appelés
les \wiit[plaquette de Néel]{plaquettes de Néel} :
\swih[plaquette de Néel]{Neel@Néel}
chaque plaquette de Néel correspond
en général à une description locale des sous-réseaux de Néel,
d'où une certaine ambiguïté
qui sera levée en utilisant systématiquement
la notion de \wibf[surnoeud@\guillemets{sur-n{\oe}ud}]%
  {\guillemets{sur-n{\oe}ud}}
dont le caractère local est sans équivoque
(\figuresname~\ref{fig/para/bidim} et~\ref{fig/para/chain}).
\nopagebreak
\begin{figure}[bh]
  \begin{center}
    \begin{minipage}[c]{\linewidth}
      \includegraphics[width=.98\linewidth]{sgtsChain.mps}
    \end{minipage}
  \caption[La chaîne paramagnétique]{%
    La \wiit[chaîne ! paramagnétique]{chaîne paramagnétique}.
    }
  \label{fig/para/chain}
  \end{center}
  \vspace{1.2\medskipamount}
  \begin{center}
    \begin{minipage}[c]{.98\linewidth}
      \includegraphics[width=\linewidth]{sgtsChainNeel.mps}
      \subcaption{La chaîne de Néel}
      \bigskip
    \end{minipage} \hfill
    \begin{minipage}[c]{\linewidth}
      \includegraphics[width=.98\linewidth]{sgtsChainANeel.mps}
      \subcaption{L'anti-chaîne de Néel}
    \end{minipage}
  \caption[La chaîne ferromagnétique]{%
    La \wiit[chaîne ! ferromagnétique]{chaîne ferromagnétique}.
    }
  \label{fig/af/chain}
  \end{center}
\end{figure}
\clearpage

\begin{figure}[p]
  \begin{center}
    \begin{minipage}[c]{.48\linewidth}
      \includegraphics[width=\linewidth]{sgtsTriangularLatticeNeel.mps}
      \subcaption{Le réseau triangulaire de Néel}
    \end{minipage} \hfill
    \begin{minipage}[c]{.48\linewidth}
      \includegraphics[width=\linewidth]{sgtsTriangularLatticeANeel.mps}
      \subcaption{L'anti-réseau triangulaire de Néel}
    \end{minipage}
  \caption[Le réseau triangulaire ferromagnétique]{%
    Le \wiit[reseau@réseau ! carré ferromagnétique]%
      {réseau carré ferromagnétique} :
    l'arrangement alterné des spins est \wiit{tricoloré}.
    L'antisymétrie tricolorée est usuellement représentée
    par trois familles de vecteurs dont les directions
    prises deux à deux forment un angle de $\frac{2}{3}\pi$ :
    les \wi[sousreseau@sous-réseau de Néel]{sous-réseaux de Néel}
    ($\gtsSpinArrow$, $\gtsSpinJArrow$ et $\gtsSpinJJArrow$).
    }
  \label{fig/af/triangular}
  \end{center}
\end{figure}

\clearpage

Par construction
deux antiopérations successives laissent
le réseau de spins invariant :
nous caractériserons
les \wi[surreseau@\guillemets{sur-réseau}]{\guillemets{sur-réseaux}}
par leur \wibf{hélicité}
\cite{Sivardiere}.
Conséquences immédiates :
toute antiopération autre que l'antitranslation
n'affecte que les \guillemets{sur-n{\oe}uds}
tout en changeant l'\wiit{hélicité} du \guillemets{sur-réseau},
et deux antitranslations successives
ne constituent rien d'autre
qu'une translation du \guillemets{sur-réseau}.
En d'autres termes,
puisque l'antisymétrie caractérise
l'état antiferromagnétique,
les \guillemets{sur-réseaux}
sont transparents
à l'antiferromagnétisme du système
(à l'hélicité du \guillemets{sur-réseau} près)
alors que
l'antiferromagnétisme est concentré
en leurs \guillemets{sur-n{\oe}uds}.
\textlatin{A posteriori}
le \wibf[surreseau@\guillemets{sur-réseau}]{\guillemets{sur-réseau}}
apparaît donc
comme le \wibf[reseau@réseau ! etat@de l'état antiferromagnétique]%
  {réseau de l'état antiferromagnétique} :
un \wiit[paramètre d'ordre ! local]{paramètre d'ordre local}
pertinent de
l'\wiit[etat@état ! antiferromagnétique]{état antiferromagnétique}
doit être associé à un \guillemets{sur-n{\oe}ud}
\cite{DombreReadA,DombreReadT,RGSHM}.

\subsection{Antiferromagnétisme frustré et assouvi}
Deux grands types de frustrations existent
pour les systèmes antiferromagnétiques
\cite{UPTAFA,Levy}:
\begin{itemize}
  \item il existe une compétition entre plusieurs ordres magnétiques
    de différentes natures dont un au moins est antiferromagnétique :
    \textlatin{e.g.} un ordre antiferromagnétique entre premiers voisins
    s'opposant à
    une \wiit[echange@échange électronique]{interaction d'échange}
    ferromagnétique entre seconds voisins;
  \item le réseau interdit la satisfaction
    de toutes les interactions d'échange : \textlatin{e.g.}
    sur le \wiit[reseau@réseau ! triangulaire]{réseau triangulaire}
    il est impossible d'aligner antiparallèlement
    trois spins deux à deux adjacents.
\end{itemize}
Seule la frustration due au réseau
est pertinente dans le contexte présent.
Cette \wiit[frustration ! géométrique]{frustration géométrique}
est caractérisé par un nombre
de \wiit[sousreseau@sous-réseau de Néel]{sous-réseaux de Néel}
supérieur à deux.
Le \wi[surreseau@\guillemets{sur-réseau}]{\guillemets{sur-réseau}}
est donc transparent à la frustration géométrique
\cite{DombreReadT,RGSHM}.

\section{Le \guillemets{sur-réseau} antiferromagnétique}
\wih{surreseau@\guillemets{sur-réseau}}
\begin{introduction}
Dans la partie précédente
le réseau des porteurs de spin apparaît
comme le \wi[reseau@réseau ! etat@de l'état paramagnétique]%
  {réseau de l'état paramagnétique},
le \guillemets{sur-réseau}
comme le \wi[reseau@réseau ! etat@de l'état antiferromagnétique]%
  {réseau de l'état antiferromagnétique}
--- les \guillemets{sur-n{\oe}uds} renfermant
la non-symétrie.

J'esquisse une mise en évidence
pour les \wi[surreseau@\guillemets{sur-réseau}]{\guillemets{sur-réseaux}}
antiferromagnétiques réguliers
de basse dimension d'une \wiit{hiérarchie algébrique}
analogue à celle du
\wi[sigmanonlin@$\sigma$~non-linéaire~(modèle)]%
 {modèle $\sigma$~non-linéaire}
\cite{PPFBT,TPEMSDASM,GurseyTze}:
les \wi[surnoeud@\guillemets{sur-n{\oe}ud}]{\guillemets{sur-n{\oe}uds}}
se comportent ici comme des
\wiit[surparticule@\guillemets{sur-particule}]{\guillemets{sur-particules}}
à part-entière,
les \guillemets{sur-réseaux} antiferromagnétiques
comme des réseaux de \guillemets{sur-particules}.
\end{introduction}

\subsection{Les degrés de liberté antiferromagnétiques locaux}
\label{chp/AFL/prescription}
Je construis
les \wibf[degrés de liberté ! antiferromagnétiques locaux]%
  {degrés de liberté antiferromagnétiques locaux}
associés à un \guillemets{sur-n{\oe}ud}
comme étant les combinaisons linéaires
des \wiit[degrés de liberté ! local]{degrés de liberté locaux}
(\textlatin{i.e.} les \wi[spin]{spins})
associés au \wi{polytope régulier} du \guillemets{sur-n{\oe}ud}
qui réduisent
les \wiit[représentation ! irréductible]{représentations irréductibles}
du groupe des transformations laissant
le polytope régulier du \guillemets{sur-n{\oe}ud} invariant.
Il s'avère alors avantageux de se représenter
le polytope régulier du \guillemets{sur-n{\oe}ud}
comme un \wi[pavage compact régulier]{pavage régulier}
de la sphère $S^{\gtsDimension-1}$
avec $\gtsDimension$ la dimension du réseau.
Or chaque groupe des transformations laissant invariant
un pavage régulier de la sphère~$S^{\gtsDimension-1}$
admet une \wiit[représentation ! algébrique]{représentation}
dans l'\wibf{algèbre réelle à~divisions}\,%
\footnote{\label{fnt/rdivalg}%
\foreignlanguage{english}{\textit{Real Division Algebra}}
}
$\mathbb{A}_{\gtsDimension}$
de dimension~$2^{\gtsDimension-1}$
\cite{Ebbinghaus,GurseyTze,Dixon,duVal,RCPCoxeter}:
les combinaisons linéaire seront construites
dans cette algèbre~$\mathbb{A}_{\gtsDimension}$.
Je choisi \textlatin{a priori} comme
\wibf[paramètre d'ordre ! antiferromagnétique local]%
  {paramètre d'ordre antiferromagnétique local}
la \wiit{combinaison linéaire algébrique}
normée à la limite continue
qui est préservée par l'antiopération
caractérisant l'état antiferromagnétique à l'\wi{hélicité} prés.
Lors du passage à une théorie de champs continue
\wih{limite ! continue}
je suppose que les combinaisons linéaires algébriques restantes
se comportent comme si les degrés de liberté (\textlatin{i.e.} les spins)
du polytope régulier étaient effectivement décrits
par un champs continu.
Enfin j'impose une représentation locale :
les \wiit[degrés de liberté ! antiferromagnétiques locaux]%
  {degrés de liberté antiferromagnétiques locaux}
seront définis à une rotation locale prés.
La rotation locale s'appliquant
aux degrés de liberté du \guillemets{sur-n{\oe}ud}
et non à sa sphère~$S^{\gtsDimension-1}$
doit s'interpréter comme
un \wibf[jauge locale]{champ de jauge locale}
\cite{Naber,BailinLove,Nakahara}.
Cette dernière hypothèse est la plus originale :
la représentation implicitement choisie est généralement globale
\cite{DombreReadA,DombreReadT,RGSHM}.

Les interactions d'échange
n'engendrent pas ici
un \wiit[ordonnancement ! antiferromagnétique]{ordonnancement antiferromagnétique}
\witextit{global}
(c'est-à-dire \guillemets{à la Néel})
mais forcent les spins
à s'organiser en \guillemets{sur-n{\oe}uds} :
les spins appartenant à un même \guillemets{sur-n{\oe}ud}
se lient entre eux tout en se désolidarisant
des spins appartenant aux \guillemets{sur-n{\oe}uds} voisins.
Dans ce contexte
les \guillemets{sur-n{\oe}uds} acquièrent
une légitimité physique immédiate :
ils se comportent comme des
\wibf[surparticule@\guillemets{sur-particule}]{\guillemets{sur-particules}}.
Dans ce nouvel état antiferromagnétique
le comportement magnétique sur de longues distances
\wih{magnétisme}
(longues devant le rayon d'une \guillemets{sur-particule})
provient des interactions entre les \guillemets{sur-particules}
décrites ici comme une désolidarisation entre spins
liés à deux \guillemets{sur-particules} adjacentes.
En outre,
un polytope régulier de dimension $\gtsDimension$
n'étant rien d'autre
qu'un réseau régulier sur la sphère~$S^{\gtsDimension-1}$,
une \guillemets{sur-particule} n'est rien d'autre
qu'un réseau antiferromagnétique régulier
de la sphère~$S^{\gtsDimension-1}$ :
l'utilisation de la théorie des groupes
pour effectuer un changement de degrés de liberté
apparaît donc légitime.
Une représentation étant définie à une rotation prés,
l'antisymétrie antiferromagnétique
de chaque réseau sphérique
(\textlatin{i.e.} de chaque \guillemets{sur-particule})
est définie à une rotation près :
les réseaux sphériques sont désolidarisés
en imposant un \wi[jauge locale]{champs de jauge locale}.
Les rotations invoquées sont celles
qui laissent la sphère~$S^{\gtsDimension-1}$ invariante.
La théorie des groupes
permet ainsi de séparer
l'\wibf{antisymétrie imposée} par
les \wi[echange@échange électronique]{interactions d'échange}
et la \wibf{symétrie héritée} du \wi{polytope régulier},
le réseau maintenant une certaine cohésion
traduite par l'existence
d'un \wi[jauge locale]{champs de jauge locale}.
Aussi les \wi[degrés de liberté ! antiferromagnétiques locaux]%
  {degrés de liberté antiferromagnétiques locaux}
héritent-ils de la \wiit{hiérarchie algébrique}
des groupes des transformations
préservant les polytopes réguliers,
c'est-à-dire
la hiérarchie algébrique des
\wiit[algèbre réelle à~divisions]%
  {algèbres réelles à~divisions}~$\mathbb{A}_{\gtsDimension}$.
Nous nous attendons donc à rencontrer une physique exotique.

\subsection{Cas unidimensionnel : la chaîne de Haldane}
\wih{chaîne ! de Haldane}
\swih[chaîne de Haldane]{Haldane}
La \wiit[chaîne ! régulière]{chaîne régulière}
\swih[chaîne]{reseau@réseau}
\swih[réseau]{chaîne}
est pavée avec le $1$-cube $\gtsCubicPolytope^{0}$
\wi[représentation ! algébrique]{représenté}
dans l'\wiit{algèbre réelle à~divisions}~$\mathbb{R}$
par le doublet
\begin{subequations}
\label{rp/chain/alg}
\begin{align}
\gtsCubicPolytope^{0}
\label{afl/chain/rp/alg/crude}
&=\left\{+1,-1\right\},\\
\label{afl/chain/rp/alg/nested}
&=S^{0}.
\end{align}
\end{subequations}
\begin{figure}[!h]
  \begin{center}
  \includegraphics[width=.4\linewidth]{sgtsAlg1Cube.mps}
  \caption[Représentation réelle du $1$-cube]{%
    \wiit[représentation ! réelle]{Représentation réelle}
    du $1$-cube $\gtsCubicPolytope^0$.
    }
  \label{fig/alg/chain}
  \end{center}
\end{figure}

\pagebreak
\noindent
Les transformations préservant le $1$-cube
se réduisent à l'identité $1$ et à la rotation/antisymétrie $-1$.
Le choix d'une
\wi[surchaine@\guillemets{sur-chaîne}]{\guillemets{sur-chaîne}}
étant fait,
pour chaque \guillemets{sur-n{\oe}ud}~$\left(\gtsHVertexI\right)$
de la \guillemets{sur-chaîne} nous étiquetterons
par $\left(\gtsHVertexI_{0}\right)$ et $\left(\gtsHVertexI_{1}\right)$
les deux n{\oe}uds représentés respectivement
par les réels $+1$ et $-1$.
\begin{subequations}
\label{afl/chain/AFDF}
Respectant la prescription introduite
dans la {\subsectionname}~\ref{chp/AFL/prescription},
nous introduisons
le \wiit[paramètre d'ordre ! antiferromagnétique local]%
  {paramètre d'ordre antiferromagnétique local}
${\gtsAFOP_{\gtsHVertexI}}^{\alpha}$ tel que
\begin{align}
\label{afl/chain/AFDF/AFOP}
2\,\gtsSpinNorm{\gtsAFOP_{\gtsHVertexI}}^{\alpha}&=%
  {\gtsSpin_{\!\gtsHVertexI_{0}}}^{\alpha}
  -{\gtsSpin_{\!\gtsHVertexI_{1}}}^{\alpha},
\intertext{et la \wiit[représentation ! triviale]%
  {représentation triviale}
${\gtsAFTR_{\gtsHVertexI}}^{\alpha}$ telle que}
\label{afl/chain/AFDF/AFTR}
2\,\gtsLStep{\gtsAFTR_{\gtsHVertexI}}^{\alpha}&=%
  {\gtsSpin_{\!\gtsHVertexI_{0}}}^{\alpha}
  +{\gtsSpin_{\!\gtsHVertexI_{1}}}^{\alpha}.
\end{align}
\end{subequations}
\begin{subequations}
\label{afl/chain/rel}
À la \wi[limite ! classique]{limite classique}
l'\wi[opérateur ! de Casimir]{opérateur de Casimir} $\gtsSpin^{2}$
(\ref{Spin/Casimir/crude})
impose entre
les nouveaux
\wiit[degrés de liberté ! antiferromagnétiques locaux]%
  {degrés de liberté antiferromagnétiques locaux}
les relations exactes suivantes :
\begin{gather}
\label{afl/chain/rel/unit}
\gtsAFOP_{\gtsHVertexI}^2=%
  1-{\scriptstyle\frac{\gtsLStep^2}{\gtsSpinNorm^2}}\;
  \gtsAFTR_{\gtsHVertexI}^2,\\
\label{afl/chain/rel/chirality}
{\gtsAFOP_{\gtsHVertexI}}^{\alpha}{\gtsAFTR_{\gtsHVertexI}}_{\alpha}=0,
\end{gather}
\end{subequations}
avec
\begin{equation}
\label{afl/nu}
\gtsSpinNorm=\sqrt{\gtsSpinDim\left(\gtsSpinDim+1\right)}.
\end{equation}
Les deux \wiit[contrainte]{contraintes} (\ref{afl/chain/rel})
montrent clairement que
le \wiit[degrés de liberté ! nombre total de]%
 {nombre total de degrés de liberté} (égal à quatre)
est conservé
lors du changement de degrés de liberté (\ref{afl/chain/AFDF}).
En outre lorsque le pas $\gtsLStep$ du réseau
et le nombre~quantique~$\gtsSpinDim$
tendent conjointement vers zéro et l'infini respectivement
(\wi{prescription d'Affleck} \cite{SGSBA,QSCHG})
le paramètre d'ordre antiferromagnétique local
${\gtsAFOP_{\gtsHVertexI}}^{\alpha}$
est effectivement normé.
Observons enfin que
pour la théorie des champs continus
\wih{limite ! continue}
ainsi construite le paramètre d'ordre local
vit sur la sphère~unité~$S^2$.

\'Etant donnés les degrés de liberté antiferromagnétiques locaux
${\gtsAFOP_{\gtsHVertexI}}^{\alpha}$
et ${\gtsAFTR_{\gtsHVertexI}}^{\alpha}$
décrivant une \guillemets{sur-particule}
$\left(\gtsHVertexI\right)$
d'une \wi[chaîne ! antiferromagnétique]{chaîne antiferromagnétique},
les \wibf[spin ! interne]{spins internes} de ladite
\wi[surparticule@\guillemets{sur-particule}]{\guillemets{sur-particule}}
${\gtsSpin_{\!\gtsHVertexI_{0}}}^{\alpha}$
et ${\gtsSpin_{\!\gtsHVertexI_{1}}}^{\alpha}$
s'obtiennent aisément en inversant (\ref{afl/chain/AFDF});
nous lisons
\begin{subequations}
\label{afl/chain/AFDF/inv}
\begin{align}
\label{afl/chain/AFDF/inv/zero}
{\gtsSpin_{\!\gtsHVertexI_{0}}}^{\alpha}&=%
  \hphantom{-}\gtsSpinNorm\,{\gtsAFOP_{\gtsHVertexI}}^{\alpha}%
  +\gtsLStep\,{\gtsAFTR_{\gtsHVertexI}}^{\alpha},\\
\label{afl/chain/AFDF/inv/one}
{\gtsSpin_{\!\gtsHVertexI_{1}}}^{\alpha}&=
  -\gtsSpinNorm\,{\gtsAFOP_{\gtsHVertexI}}^{\alpha}%
  +\gtsLStep\,{\gtsAFTR_{\gtsHVertexI}}^{\alpha}.
\end{align}
\end{subequations}

Pour conclure,
la prescription envisagée reproduit parfaitement
la \wiit[chaîne ! de Haldane]{décomposition de Haldane}
\cite{HaldanePRL,Levy}.

\subsection{Cas d'un système frustré : le réseau triangulaire}
Le \wiit[reseau@réseau ! triangulaire]{réseau triangulaire} est pavé
avec le triangle équilatéral $\gtsTriPolytope^{1}$
représenté dans l'\wiit{algèbre réelle à~divisions} $\mathbb{C}$
par le triplet
\begin{equation}
\label{afl/tri/rp/alg}
\gtsTriPolytope^{1}=\left\{1,\cw,\cw^2\right\},
\end{equation}
avec $\cw$ la racine troisième de l'unité :
\begin{equation*}
\cw\equiv\gtse^{\ci\,\frac{2\pi}{3}}.
\end{equation*}
\begin{figure}[!h]
  \begin{center}
  \includegraphics[width=.4\linewidth]{sgtsAlgTri.mps}
  \caption[Représentation complexe du triangle équilatéral]{%
    \wiit[représentation ! complexe]{Représentation complexe}
    du triangle équilatéral $\gtsTriPolytope^1$.
    }
  \label{fig/alg/tri}
  \end{center}
\end{figure}

\noindent
Le triangle équilatéral $\gtsTriPolytope^{1}$ est préservé
par toutes combinaisons de l'opération de conjugaison
et de la rotation $\cw$.
Un \guillemets{sur-réseau} étant choisi,
les trois n{\oe}uds d'un \guillemets{sur-n{\oe}ud}
quelconque~$\left(\gtsHVertexI\right)$
représentés respectivement
par les trois complexes $1$, $\cw$ et $\cw^2$
seront étiquetés respectivement par
$\left(\gtsHVertexI_{0}\right)$,
$\left(\gtsHVertexI_{1}\right)$ et
$\left(\gtsHVertexI_{2}\right)$.
\begin{subequations}
\label{afl/tri/AFDF}
L'antiopération antiferromagnétique
n'étant rien d'autre que la rotation~$\cw$,
nous construisons
le \wiit[paramètre d'ordre ! antiferromagnétique local]%
  {paramètre d'ordre antiferromagnétique local}
${\gtsAFOP_{\gtsHVertexI}}^{\alpha}$ tel que
\begin{align}
\label{afl/tri/AFDF/AFOP}
{\scriptstyle\frac{3}{\sqrt{2}}}\,%
\gtsSpinNorm{\gtsAFOP_{\gtsHVertexI}}^{\alpha}&=%
  \gtse^{-\ci\,\gtsGP_{\gtsHVertexI}}%
  \left[%
    {\gtsSpin_{\!\gtsHVertexI_{0}}}^{\alpha}
    +\cw\,{\gtsSpin_{\!\gtsHVertexI_{1}}}^{\alpha}
    +\cw^2\,{\gtsSpin_{\!\gtsHVertexI_{2}}}^{\alpha}
  \right],
\intertext{et la \wiit[représentation ! triviale]%
  {représentation triviale}
${\gtsAFTR_{\gtsHVertexI}}^{\alpha}$ telle que}
\label{afl/tri/AFDF/AFTR}
3\,\gtsLStep^2{\gtsAFTR_{\gtsHVertexI}}^{\alpha}&=%
  {\gtsSpin_{\!\gtsHVertexI_{0}}}^{\alpha}
  +{\gtsSpin_{\!\gtsHVertexI_{1}}}^{\alpha}
  +{\gtsSpin_{\!\gtsHVertexI_{2}}}^{\alpha}.
\end{align}
\end{subequations}
Soulignons que
la \wiit[représentation ! complexe]{représentation complexe}
(\ref{afl/tri/AFDF/AFOP})
est définie à un \wi[jauge locale]{champs de jauge locale}
$\gtsGP_{\gtsHVertexI}$ prés.
\begin{subequations}
\label{afl/tri/rel}
À la limite classique
les \wiit[degrés de liberté ! antiferromagnétiques locaux]%
  {degrés de liberté antiferromagnétiques}
satisfont aux \wiit[contrainte]{contraintes} suivantes :
\begin{gather}
\label{afl/tri/rel/unit}
{\gtsAFOP_{\gtsHVertexI}}^{\alpha}{^{\dag}\gtsAFOP_{\gtsHVertexI}}_{\alpha}=%
  1-{\scriptstyle\frac{\gtsLStep^4}{\gtsSpinNorm^2}}\;
  \gtsAFTR_{\gtsHVertexI}^2,\\
\label{afl/tri/rel/chirality}
{\gtsAFOP_{\gtsHVertexI}}^{\alpha}{\gtsAFOP_{\gtsHVertexI}}_{\alpha}=%
  -{\scriptstyle 2\sqrt{2}\:\frac{\gtsLStep^2}{\gtsSpinNorm}}\,%
  \gtse^{-\ci\,\gtsGP_{\gtsHVertexI}}\;%
  {\gtsAFTR_{\gtsHVertexI}}^{\alpha}{^{\dag}\gtsAFOP_{\gtsHVertexI}}_{\alpha}.
\end{gather}
\end{subequations}
Le \wiit[degrés de liberté ! nombre total de]%
  {nombre total de degrés de liberté} (égal à six)
est donc préservé lors du changement
des degrés de libertés locaux
et le \wi[paramètre d'ordre ! antiferromagnétique local]%
  {paramètre d'ordre antiferromagnétique local}
${\gtsAFOP_{\gtsHVertexI}}^{\alpha}$
est effectivement normé à la \wi[limite ! continue]{limite continue}.

Il s'avère alors judicieux d'introduire
les combinaisons linéaires locales
\begin{subequations}
\label{afl/tri/XY}
\begin{align}
\label{afl/tri/XY/X/def}
{{\gtsAFOP_{X}}_{\gtsHVertexI}}^{\alpha}&\equiv%
  {\scriptstyle\frac{2}{3\gtsSpinNorm}}\,
  \left[%
  {\gtsSpin_{\!\gtsHVertexI_{0}}}^{\alpha}%
  -{\scriptstyle\frac{1}{2}}\left({\gtsSpin_{\!\gtsHVertexI_{1}}}^{\alpha}%
    +{\gtsSpin_{\!\gtsHVertexI_{2}}}^{\alpha}\right)%
  \right],\\
\label{afl/tri/XY/Y/def}
{{\gtsAFOP_{Y}}_{\gtsHVertexI}}^{\alpha}&\equiv%
  {\scriptstyle\frac{2}{3\gtsSpinNorm}}\,
  {\scriptstyle\frac{\sqrt{3}}{2}}%
  \left({\gtsSpin_{\!\gtsHVertexI_{1}}}^{\alpha}%
    -{\gtsSpin_{\!\gtsHVertexI_{2}}}^{\alpha}\right);
\end{align}
\end{subequations}
à la \wi[limite ! classique]{limite classique} nous avons
\begin{subequations}
\label{afl/tri/XY/rel}
\begin{gather}
\label{afl/tri/XY/rel/X}
{{\gtsAFOP_{X}}_{\gtsHVertexI}}^2=1%
  -{\scriptstyle 2\frac{\gtsLStep^2}{\gtsSpinNorm}}\,%
    {{\gtsAFOP_{X}}_{\gtsHVertexI}}^{\alpha}%
    {\gtsAFTR_{\gtsHVertexI}}_{\alpha}%
  -{\scriptstyle \frac{\gtsLStep^4}{\gtsSpinNorm^2}}\,%
    {\gtsAFTR_{\gtsHVertexI}}^{2},\\
\label{afl/tri/XY/rel/Y}
{{\gtsAFOP_{Y}}_{\gtsHVertexI}}^2=1%
  +{\scriptstyle 2\frac{\gtsLStep^2}{\gtsSpinNorm}}\,%
    {{\gtsAFOP_{X}}_{\gtsHVertexI}}^{\alpha}%
    {\gtsAFTR_{\gtsHVertexI}}_{\alpha}%
  -{\scriptstyle \frac{\gtsLStep^4}{\gtsSpinNorm^2}}\,%
    {\gtsAFTR_{\gtsHVertexI}}^{2},\\
\label{afl/tri/XY/rel/XY}
{{\gtsAFOP_{X}}_{\gtsHVertexI}}^{\alpha}%
  {{\gtsAFOP_{Y}}_{\gtsHVertexI}}_{\alpha}=
  {\scriptstyle 2\frac{\gtsLStep^2}{\gtsSpinNorm}}\,%
    {{\gtsAFOP_{Y}}_{\gtsHVertexI}}^{\alpha}%
    {\gtsAFTR_{\gtsHVertexI}}_{\alpha}.%
\end{gather}
\end{subequations}
Comme à la limite continue le couple
$\left({{\gtsAFOP_{X}}_{\gtsHVertexI}}^{\alpha},%
  {{\gtsAFOP_{Y}}_{\gtsHVertexI}}^{\alpha}\right)$
n'est rien d'autre
qu'un \wiit[diedre@dièdre orthonormé]{dièdre orthonormé},
introduisons un nouvel opérateur
${{\gtsAFOP_{Z}}_{\gtsHVertexI}}^{\alpha}$
tel que le triplet
$\left({{\gtsAFOP_{X}}_{\gtsHVertexI}}^{\alpha},%
  {{\gtsAFOP_{Y}}_{\gtsHVertexI}}^{\alpha},%
  {{\gtsAFOP_{Z}}_{\gtsHVertexI}}^{\alpha}\right)$
forme un \wiit[triedre@trièdre orthonormé]{trièdre orthonormé}
à la limite continue :
\begin{equation}
\label{afl/tri/XY/Z/def}
{{\gtsAFOP_{Z}}_{\gtsHVertexI}}_{\alpha}\equiv%
{\scriptstyle \frac{1}{2}}\,%
\gtsastr_{\alpha\beta\gamma}%
\left[%
{{\gtsAFOP_{X}}_{\gtsHVertexI}}^{\beta}{{\gtsAFOP_{Y}}_{\gtsHVertexI}}^{\gamma}%
-{{\gtsAFOP_{Y}}_{\gtsHVertexI}}^{\beta}{{\gtsAFOP_{X}}_{\gtsHVertexI}}^{\gamma}%
\right];
\end{equation}
en injectant dans (\ref{afl/tri/XY/Z/def})
les expressions (\ref{afl/tri/XY}) nous obtenons
\begin{equation}
\label{afl/tri/XY/Z/crude}
{{\gtsAFOP_{Z}}_{\gtsHVertexI}}_{\alpha}=%
{\scriptstyle \frac{2\sqrt{3}}{3\gtsSpinNorm^2}}\:%
{\scriptstyle \frac{1}{3}}\,\gtsastr_{\alpha\beta\gamma}%
\left[
{\gtsSpin_{\!\gtsHVertexI_{0}}}^{\beta}{\gtsSpin_{\!\gtsHVertexI_{1}}}^{\gamma}%
+{\gtsSpin_{\!\gtsHVertexI_{1}}}^{\beta}{\gtsSpin_{\!\gtsHVertexI_{2}}}^{\gamma}%
+{\gtsSpin_{\!\gtsHVertexI_{2}}}^{\beta}{\gtsSpin_{\!\gtsHVertexI_{0}}}^{\gamma}%
\right].
\end{equation}
La \wiit[représentation ! complexe]{représentation complexe}
(\ref{afl/tri/AFDF/AFOP}) s'écrit immédiatement
\begin{equation}
\label{afl/tri/AFDF/AFOP/nested}
{\gtsAFOP_{\gtsHVertexI}}^{\alpha}=%
  {\scriptstyle\frac{1}{\sqrt{2}}}\,\gtse^{-\ci\,\gtsGP_{\gtsHVertexI}}%
  \left[{{\gtsAFOP_{X}}_{\gtsHVertexI}}^{\alpha}%
    +\ci\,{{\gtsAFOP_{Y}}_{\gtsHVertexI}}^{\alpha}%
  \right];
\end{equation}
la \wiit[représentation ! pseudo-scalaire]{représentation pseudo-scalaire}
(\ref{afl/tri/XY/Z/def}) se réécrit alors
\begin{equation}
\label{afl/tri/XY/Z/nested}
{{\gtsAFOP_{Z}}_{\gtsHVertexI}}_{\alpha}=%
{\scriptstyle \frac{1}{2}}\,\ci\:%
\gtsastr_{\alpha\beta\gamma}%
\left[%
{^{\dag}\gtsAFOP_{\gtsHVertexI}}^{\beta}{\gtsAFOP_{\gtsHVertexI}}^{\gamma}
-{\gtsAFOP_{\gtsHVertexI}}^{\beta}{^{\dag}\gtsAFOP_{\gtsHVertexI}}^{\gamma}
\right].
\end{equation}
Conséquence immédiate :
pour la théorie des champs ainsi construite
le paramètre d'ordre local
vit dans $\mathrm{SO}(3)$,
le groupe des rotations
dans l'espace Euclidien $\mathbb{R}^3$.

Les \wiit[spin ! interne]{spins internes}
d'une \wi[surparticule@\guillemets{sur-particule}]%
{\guillemets{sur-particule}}
$\left(\gtsHVertexI\right)$
d'un \wiit[reseau@réseau ! triangulaire antiferromagnétique]%
  {réseau triangulaire antiferromagnétique}
sont donnés par les relations suivantes :
\begin{subequations}
\label{afl/tri/AFDF/inv}
\begin{align}
\label{afl/tri/AFDF/inv/zero}
{\gtsSpin_{\!\gtsHVertexI_{0}}}^{\alpha}&=%
{\scriptstyle\frac{1}{\sqrt{2}}}\,\gtsSpinNorm\:%
\left[%
\hphantom{\cw}\,\gtse^{+\ci\,\gtsGP_{\gtsHVertexI}}%
  {\gtsAFOP_{\gtsHVertexI}}^{\alpha}%
+\hphantom{\cw}\,\gtse^{-\ci\,\gtsGP_{\gtsHVertexI}}%
  {^{\dag}\gtsAFOP_{\gtsHVertexI}}^{\alpha}%
\right]%
+\gtsLStep^2\,{\gtsAFTR_{\gtsHVertexI}}^{\alpha},\\
\label{afl/tri/AFDF/inv/one}
{\gtsSpin_{\!\gtsHVertexI_{1}}}^{\alpha}&=
{\scriptstyle\frac{1}{\sqrt{2}}}\,\gtsSpinNorm\:%
\left[%
\overline{\cw}\,\gtse^{+\ci\,\gtsGP_{\gtsHVertexI}}%
  {\gtsAFOP_{\gtsHVertexI}}^{\alpha}%
+\cw\,\gtse^{-\ci\,\gtsGP_{\gtsHVertexI}}%
  {^{\dag}\gtsAFOP_{\gtsHVertexI}}^{\alpha}%
\right]%
+\gtsLStep^2\,{\gtsAFTR_{\gtsHVertexI}}^{\alpha},\\
\label{afl/tri/AFDF/inv/tow}
{\gtsSpin_{\!\gtsHVertexI_{2}}}^{\alpha}&=
{\scriptstyle\frac{1}{\sqrt{2}}}\,\gtsSpinNorm\:%
\left[%
\cw\,\gtse^{+\ci\,\gtsGP_{\gtsHVertexI}}%
  {\gtsAFOP_{\gtsHVertexI}}^{\alpha}%
+\overline{\cw}\,\gtse^{-\ci\,\gtsGP_{\gtsHVertexI}}%
  {^{\dag}\gtsAFOP_{\gtsHVertexI}}^{\alpha}%
\right]%
+\gtsLStep^2\,{\gtsAFTR_{\gtsHVertexI}}^{\alpha}.
\end{align}
\end{subequations}
En outre à chaque
\wi[surparticule@\guillemets{sur-particule}]{\guillemets{sur-particule}}
$\left(\gtsHVertexI\right)$ nous associons la \wibf{chiralité}
$\gtsAFChirality_{\gtsHVertexI}$ définie par
\cite{Sivardiere,UPTAFA}
\begin{equation}
\label{afl/tri/chirality/def}
\gtsAFChirality_{\gtsHVertexI}\equiv%
{\scriptstyle \frac{3\sqrt{3}}{2\gtsSpinNorm^3}}\:%
\gtsastr_{\alpha\beta\gamma}%
{\gtsSpin_{\!\gtsHVertexI_{0}}}^{\alpha}%
{\gtsSpin_{\!\gtsHVertexI_{1}}}^{\beta}%
{\gtsSpin_{\!\gtsHVertexI_{2}}}^{\gamma}.
\end{equation}
Un calcul rapide montre que la \wiit{chiralité}
$\gtsAFChirality_{\gtsHVertexI}$ est indépendante
du \wi[jauge locale]{champs de jauge locale} $\gtsGP_{\gtsHVertexI}$;
nous avons
\begin{equation}
\label{afl/tri/chirality/nested}
\gtsAFChirality_{\gtsHVertexI}=%
{\scriptstyle \frac{\gtsLStep^2}{\gtsSpinNorm}}\,%
{{\gtsAFOP_{Z}}_{\gtsHVertexI}}^{\alpha}%
{\gtsAFTR_{\gtsHVertexI}}_{\alpha}.
\end{equation}
Cette relation (\ref{afl/tri/chirality/nested})
nous invite à introduire la \wibf[chiralité ! nue]{chiralité nue}
\begin{equation}
\label{afl/tri/chirality/nude}
\gtsAFNChirality_{\gtsHVertexI}\equiv%
{{\gtsAFOP_{Z}}_{\gtsHVertexI}}^{\alpha}%
{\gtsAFTR_{\gtsHVertexI}}_{\alpha}
\end{equation}
qui décrirait l'état antiferromagnétique
de la \guillemets{sur-particule} :
une \guillemets{sur-particule}
idéalement antiferromagnétique
(c'est-à-dire \guillemets{à la Néel})
aurait une \wibf[chiralité ! nue nulle]{chiralité nue nulle}
indépendamment du pas $\gtsLStep$ du réseau
et du nombre~quantique~$\gtsSpinDim$.
La définition d'une \wiit[chiralité ! nue]{chiralité nue}
$\gtsAFNChirality_{\gtsHVertexI}$
pour une \guillemets{sur-particule}
$\left(\gtsHVertexI\right)$
appartenant à un réseau antiferromagnétique assouvi
est immédiate.
Selon la relation (\ref{afl/chain/rel/chirality}),
les \guillemets{sur-particules} de
la \wiit[chaîne ! antiferromagnétique]{chaîne antiferromagnétique}
ont une \wiit[chiralité ! nue nulle]{chiralité nue nulle}.

\subsection{Cas d'un système assouvi : le réseau carré}
Le \wiit[reseau@réseau ! carré]{réseau carré}
est pavé avec le $2$-cube $\gtsCubicPolytope^1$
représenté dans l'\wiit{algèbre réelle à~divisions}~$\mathbb{C}$
par le quadruplet
\begin{equation}
\label{afl/square/rp/alg}
\gtsCubicPolytope^{1}=\left\{+1,+\ci,-1,-\ci\right\}.
\end{equation}
\begin{figure}[!h]
  \begin{center}
  \includegraphics[width=.4\linewidth]{sgtsAlgSquare.mps}
  \caption[Représentation complexe du carré]{%
    \wiit[représentation ! complexe]{Représentation complexe}
    du carré $\gtsCubicPolytope^1$.
    }
  \label{fig/alg/square}
  \end{center}
\end{figure}

\noindent
Le carré $\gtsCubicPolytope^1$ est préservé
par toutes combinaisons de l'opération de conjugaison
et de la rotation $\ci$.
Un \guillemets{sur-réseau} étant choisi,
les quatre n{\oe}uds d'un \guillemets{sur-n{\oe}ud}
quelconque $\left(\gtsHVertexI\right)$
représenté respectivement
par les quatre complexes $+1$,~$+\ci$,~$-1$~et~$-\ci$
seront étiquetés respectivement par
$\left(\gtsHVertexI_{0}\right)$,
$\left(\gtsHVertexI_{1}\right)$,
$\left(\gtsHVertexI_{2}\right)$ et
$\left(\gtsHVertexI_{3}\right)$.
\begin{subequations}
\label{afl/square/AFDF}
L'antiopération antiferromagnétique étant la rotation $-1$,
nous construisons
le \wiit[paramètre d'ordre ! antiferromagnétique local]%
  {paramètre d'ordre antiferromagnétique local}
${\gtsAFOP_{\gtsHVertexI}}^{\alpha}$ tel que
\begin{align}
\label{afl/square/AFDF/AFOP}
4\,\gtsSpinNorm{\gtsAFOP_{\gtsHVertexI}}^{\alpha}&=%
\left[%
{\gtsSpin_{\!\gtsHVertexI_{0}}}^{\alpha}%
+{\gtsSpin_{\!\gtsHVertexI_{2}}}^{\alpha}%
\right]%
-\left[%
{\gtsSpin_{\!\gtsHVertexI_{1}}}^{\alpha}%
+{\gtsSpin_{\!\gtsHVertexI_{3}}}^{\alpha}%
\right],
\intertext{la \wiit[représentation ! complexe]%
  {représentation complexe}
${\gtsAFAlg_{\gtsHVertexI}}^{\alpha}$ telle que}
\label{afl/square/AFDF/AFAlg}
{\scriptstyle\frac{4}{\sqrt{2}}}\,%
\gtsLStep\gtsSpinNorm{\gtsAFAlg_{\gtsHVertexI}}^{\alpha}&=%
  \gtse^{-\ci\,\gtsGP_{\gtsHVertexI}}%
  \left[%
    {\gtsSpin_{\!\gtsHVertexI_{0}}}^{\alpha}%
    +\ci\,{\gtsSpin_{\!\gtsHVertexI_{1}}}^{\alpha}%
    -\,{\gtsSpin_{\!\gtsHVertexI_{2}}}^{\alpha}%
    -\ci\,{\gtsSpin_{\!\gtsHVertexI_{3}}}^{\alpha}%
  \right],
\intertext{et la \wiit[représentation ! triviale]%
  {représentation triviale}
${\gtsAFTR_{\gtsHVertexI}}^{\alpha}$ telle que}
\label{afl/square/AFDF/AFTR}
4\,\gtsLStep^2{\gtsAFTR_{\gtsHVertexI}}^{\alpha}&=%
  {\gtsSpin_{\!\gtsHVertexI_{0}}}^{\alpha}
  +{\gtsSpin_{\!\gtsHVertexI_{1}}}^{\alpha}
  +{\gtsSpin_{\!\gtsHVertexI_{2}}}^{\alpha}
  +{\gtsSpin_{\!\gtsHVertexI_{3}}}^{\alpha}.
\end{align}
\end{subequations}
\begin{subequations}
\label{afl/square/rel}
À la \wi[limite ! classique]{limite classique}
les \wiit[contrainte]{contraintes} sont :
\begin{gather}
\label{afl/square/rel/unit}
{\gtsAFOP_{\gtsHVertexI}}^2=%
1-{\scriptstyle\gtsLStep^2}\;%
{\gtsAFAlg_{\gtsHVertexI}}^{\alpha}{^{\dag}\gtsAFAlg_{\gtsHVertexI}}_{\alpha}
-{\scriptstyle\frac{\gtsLStep^4}{\gtsSpinNorm^2}}\;%
\gtsAFTR_{\gtsHVertexI}^2,\\
\label{afl/square/rel/ortho}
{\gtsAFOP_{\gtsHVertexI}}^{\alpha}{\gtsAFAlg_{\gtsHVertexI}}_{\alpha}=%
-{\scriptstyle\frac{\gtsLStep^2}{\gtsSpinNorm}}\,%
{\gtsAFAlg_{\gtsHVertexI}}^{\alpha}{\gtsAFTR_{\gtsHVertexI}}_{\alpha},\\
\label{afl/square/rel/chirality}
{\gtsAFOP_{\gtsHVertexI}}^{\alpha}{\gtsAFTR_{\gtsHVertexI}}_{\alpha}=%
{\scriptstyle\frac{\gtsSpinNorm}{4}}\,%
\left[%
{\gtsAFAlg_{\gtsHVertexI}}^{\alpha}%
  {\gtsAFAlg_{\gtsHVertexI}}_{\alpha}%
+{^{\dag}\gtsAFAlg_{\gtsHVertexI}}^{\alpha}%
  {^{\dag}\gtsAFAlg_{\gtsHVertexI}}_{\alpha}%
\right].
\end{gather}
\end{subequations}
Le \wiit[degrés de liberté ! nombre total de]%
  {nombre total de degrés de liberté} (égal à huit)
est donc préservé
et le paramètre d'ordre antiferromagnétique local
${\gtsAFOP_{\gtsHVertexI}}^{\alpha}$
est effectivement normé à la limite continue.

Selon la relation (\ref{afl/square/rel/chirality})
la \wiit[chiralité ! nue]{chiralité nue} $\gtsAFNChirality_{\gtsHVertexI}$
d'une \guillemets{sur-particule} $\left(\gtsHVertexI\right)$
du réseau carré est donnée par
\begin{equation}
\label{afl/square/chirality/nude}
\gtsAFNChirality_{\gtsHVertexI}=%
{\scriptstyle\frac{\gtsSpinNorm}{4}}\,%
\left[%
{\gtsAFAlg_{\gtsHVertexI}}^{\alpha}%
  {\gtsAFAlg_{\gtsHVertexI}}_{\alpha}%
+{^{\dag}\gtsAFAlg_{\gtsHVertexI}}^{\alpha}%
  {^{\dag}\gtsAFAlg_{\gtsHVertexI}}_{\alpha}%
\right].
\end{equation}
Aussi si la \guillemets{sur-particule} $\left(\gtsHVertexI\right)$
est idéalement antiferromagnétique
(\wiit[chiralité ! nue nulle]{chiralité nue nulle}),
alors ${\gtsAFAlg_{\gtsHVertexI}}^{\alpha}$
est un dièdre orthogonal et le couple
$\left({\gtsAFAlg_{\gtsHVertexI}}^{\alpha},%
  {\gtsAFOP_{\gtsHVertexI}}^{\alpha}\right)$
un trièdre orthogonal à la limite continue.
Par conséquent
pour la théorie des champs ainsi construite
le paramètre d'ordre antiferromagnétique local
pertinent vit dans $\mathrm{SO}(3)$,
comme pour le réseau triangulaire.

Les \wiit[spin ! interne]{spins internes}
d'une \guillemets{sur-particule} $\left(\gtsHVertexI\right)$
d'un \wi[reseau@réseau ! carré antiferromagnétique]%
  {réseau carré antiferromagnétique}
sont donnés par les relations suivantes :
\begin{subequations}
\label{afl/square/AFDF/inv}
\begin{align}
\label{afl/square/AFDF/inv/zero}
{\gtsSpin_{\!\gtsHVertexI_{0}}}^{\alpha}&=%
\hphantom{-}\gtsSpinNorm\,{\gtsAFOP_{\gtsHVertexI}}^{\alpha}%
+{\scriptstyle\frac{1}{\sqrt{2}}}\,\gtsLStep\gtsSpinNorm\:%
\left[
\hphantom{\ci}\,\gtse^{+\ci\,\gtsGP_{\gtsHVertexI}}%
  {\gtsAFAlg_{\gtsHVertexI}}^{\alpha}%
+\hphantom{\ci}\,\gtse^{-\ci\,\gtsGP_{\gtsHVertexI}}%
  {^{\dag}\gtsAFAlg_{\gtsHVertexI}}^{\alpha}%
\right]
+\gtsLStep^2\,{\gtsAFTR_{\gtsHVertexI}}^{\alpha},\\
\label{afl/square/AFDF/inv/one}
{\gtsSpin_{\!\gtsHVertexI_{1}}}^{\alpha}&=%
-\gtsSpinNorm\,{\gtsAFOP_{\gtsHVertexI}}^{\alpha}%
-{\scriptstyle\frac{1}{\sqrt{2}}}\,\gtsLStep\gtsSpinNorm\:%
\left[
\ci\,\gtse^{+\ci\,\gtsGP_{\gtsHVertexI}}%
  {\gtsAFAlg_{\gtsHVertexI}}^{\alpha}%
-\ci\,\gtse^{-\ci\,\gtsGP_{\gtsHVertexI}}%
  {^{\dag}\gtsAFAlg_{\gtsHVertexI}}^{\alpha}%
\right]
+\gtsLStep^2\,{\gtsAFTR_{\gtsHVertexI}}^{\alpha},\\
\label{afl/square/AFDF/inv/tow}
{\gtsSpin_{\!\gtsHVertexI_{2}}}^{\alpha}&=%
\hphantom{-}\gtsSpinNorm\,{\gtsAFOP_{\gtsHVertexI}}^{\alpha}%
-{\scriptstyle\frac{1}{\sqrt{2}}}\,\gtsLStep\gtsSpinNorm\:%
\left[
\hphantom{\ci}\,\gtse^{+\ci\,\gtsGP_{\gtsHVertexI}}%
  {\gtsAFAlg_{\gtsHVertexI}}^{\alpha}%
+\hphantom{\ci}\,\gtse^{-\ci\,\gtsGP_{\gtsHVertexI}}%
  {^{\dag}\gtsAFAlg_{\gtsHVertexI}}^{\alpha}%
\right]
+\gtsLStep^2\,{\gtsAFTR_{\gtsHVertexI}}^{\alpha},\\
\label{afl/square/AFDF/inv/three}
{\gtsSpin_{\!\gtsHVertexI_{3}}}^{\alpha}&=%
-\gtsSpinNorm\,{\gtsAFOP_{\gtsHVertexI}}^{\alpha}%
+{\scriptstyle\frac{1}{\sqrt{2}}}\,\gtsLStep\gtsSpinNorm\:%
\left[
\ci\,\gtse^{+\ci\,\gtsGP_{\gtsHVertexI}}%
  {\gtsAFAlg_{\gtsHVertexI}}^{\alpha}%
-\ci\,\gtse^{-\ci\,\gtsGP_{\gtsHVertexI}}%
  {^{\dag}\gtsAFAlg_{\gtsHVertexI}}^{\alpha}%
\right]
+\gtsLStep^2\,{\gtsAFTR_{\gtsHVertexI}}^{\alpha}.
\end{align}
\end{subequations}


\ProvidesFile{gtsChapter2.tex}[1999/06/27 v0.9b GTS project: Chapter 2]
\chapter{Membranes magnétiques}
\label{chp/HSES}

\section{Le modèle  $\sigma$~non-linéaire}
\begin{introduction}
Le \wi[sigmanonlin@$\sigma$~non-linéaire~(modèle)]%
  {modèle $\sigma$~non-linéaire},
ou la \swi[$\sigma$~non-linéaire]{forme harmonique}
pour les mathématiciens,
est un outil incontournable en physique théorique
et un domaine de recherche en mathématique
\cite{Urakawa,TPEMSDASM,Fradkin}.
En physique de la matière condensée,
le modèle $\sigma$~non-linéaire décrit
des matériaux antiferromagnétiques
isotropes
\cite{Fradkin,BelavinPolyakov,Trimper,Chakravarty,Haldane}.
\end{introduction}

\subsection{Préliminaires}
Dans le cas présent,
le \wibf[sigmanonlin@$\sigma$~non-linéaire~(modèle)]%
  {modèle $\sigma$~non-linéaire}
correspondra à la limite continue du modèle de Heisenberg
décrivant un système de spins classiques
distribués sur un réseau régulier bidimensionnel
et interagissant entre plus proche voisin \textlatin{a priori}.
Le \wibf[paramètre d'ordre ! local]{paramètre d'ordre local}
\cite{ToulouseKleman}
d'un tel système peut-être représenté par
un vecteur unitaire $\gtsOP$
généralement interprété dans la littérature
comme une \wiit{magnétisation locale} \cite{Nakahara} :
dans le {\chaptername} précédent,
je montre que la nature
du \wibf[paramètre d'ordre ! local]{paramètre d'ordre} $\gtsOP$
est plus complexe.

\vfill
\begin{figure}[!hb]
  \begin{center}
  \includegraphics[width=.6\linewidth]{sgtsSpheric.mps}
  \caption[Le paramètre d'ordre $\gtsOP$ : %
      coordonnées sphériques]{%
    Le paramètre d'ordre $\gtsOP$ :
    \wibf[coordonnées ! sphériques]{coordonnées sphériques}.
    }
  \label{fig/OP/sph}
  \end{center}
\end{figure}
\clearpage
\begin{figure}[!ht]
  \begin{center}
  \includegraphics[width=.6\linewidth]{sgtsStereographic.mps}
  \caption[Le paramètre d'ordre $\gtsOP$ : %
      coordonnées stéréographiques]{%
    Le paramètre d'ordre $\gtsOP$ :
    \wibf[coordonnées ! stéréographiques]{coordonnées stéréographiques}.
    }
  \label{fig/OP/str}
  \end{center}
\end{figure}

Le \wiit[Hamiltonien ! magnétique]{Hamiltonien magnétique} $\gtsHm_{mag}$
d'une \wibf[membrane ! magnétique]{membrane magnétique} $\gtsSurface$
sera donc décrite par la fonctionnelle \cite{VSDTS,TSGF,HSETS}
\begin{equation}
\label{Hm/mg}
\gtsHm_{mag}=
\gtsJ\!{\iint\limits_\gtsSurface}\!\sqrt{g}\,{\gtsd\Omega}\:%
g^{ij}h_{\alpha\beta}\partial_{i}n^{\alpha}\partial_{j}n^{\beta},
\end{equation}
où l'intégrale est prise sur toute la membrane $\gtsSurface$.
La constante $\gtsJ$ désigne
le \wiit[couplage ! spin-spin]{couplage spin-spin}
isotrope entre plus proches voisins,
c'est-à-dire le terme
d'\wiit[echange@échange électronique]{échange électronique} \cite{Levy}
entre deux sites voisins du réseau initial.
Les tenseurs métriques $g_{ij}$ et $h_{\alpha\beta}$
décrivent respectivement
la membrane magnétique $\gtsSurface$
et
la \wiit[variété ! du paramètre d'ordre]%
  {variété du paramètre d'ordre} $\gtsOP$,
autrement dit la sphère~unité~$S^2$.

Le paramètre d'ordre $\gtsOP$,
qui doit satisfaire la contrainte $\gtsOP^2=1$,
admet deux descriptions naturelles :
une \wiit[description ! géométrique]{description géométrique}
et une \wiit[description ! algébrique]{description algébrique}.
La \wibf[description ! algébrique]{description algébrique}
est implicitement utilisée dans le Hamiltonien (\ref{Hm/mg}),
toutefois un système de coordonnées adéquate reste à préciser.
Le \wiit[paramètre d'ordre ! local]{paramètre d'ordre} $\gtsOP$
est communément représenté par
ses \wiit[coordonnées ! sphériques]%
  {coordonnées sphériques}~$(\gtsMTheta,\gtsMPhi)$
(\figurename~\ref{fig/OP/sph}):
\begin{equation}
\label{OP/n/sph}
\gtsOP=
\left[
\sin\gtsMTheta\ \cos\gtsMPhi,
\sin\gtsMTheta\ \sin\gtsMPhi,
\cos\gtsMTheta
\right].
\end{equation}
Le paramètre d'ordre $\gtsOP$
est ainsi représenté par un point sur la sphère~unité~$S^2$;
par conséquent la métrique $h$
de la \wiit[variété ! cible]{variété cible} s'écrit
\begin{equation}
\label{OP/metric}
h=\gtsd\gtsMTheta\!\otimes\!\gtsd\gtsMTheta
+\sin^2 \gtsMTheta\ \gtsd\gtsMPhi\!\otimes\!\gtsd\gtsMPhi.
\end{equation}
Le \wiit[Hamiltonien ! magnétique]{Hamiltonien magnétique} (\ref{Hm/mg})
devient alors
\begin{equation}
\label{Hm/mg/nested}
\gtsHm_{mag}=
\gtsJ\!{\iint\limits_\gtsSurface}\!\sqrt{g}\,{\gtsd\Omega}\:g^{ij}
\left[\vphantom{\frac{1}{2}}
\partial_{i}\gtsMTheta\,\partial_{j}\gtsMTheta
+\sin^2 \gtsMTheta\ \partial_{i}\gtsMPhi\,\partial_{j}\gtsMPhi
\right].
\end{equation}
Sous cette forme,
qui incorpore la contrainte du paramètre d'ordre,
le caractère non-linéaire de la fonctionnelle (\ref{Hm/mg})
apparaît ostensiblement.

La \wibf[description ! algébrique]{description algébrique} consiste
à projeter le paramètre d'ordre $\gtsOP$
sur la sphère complexe $\mathbb{C}\cup\{\infty\}$
\textlatin{via} une \wiit{projection stéréographique}
\cite{Cohn,Nakahara}.
Rappelons au passage que
la \wiit[sphère ! complexe]{sphère complexe}
(ou \wiit[sphère ! de Riemann]{de Riemann})
\swih[sphère de Riemann]{Riemann}
s'obtient par \wibf{compactification}
du plan complexe $\mathbb{C}$ :
l'ensemble des points situés à l'infini est identifié
à un point unique, l'infini.
Compte tenu des conventions précédemment employées,
il est commode de projeter stéréographiquement
la sphère~unité~$S^2$
par rapport au pôle sud.
Les \wiit[coordonnées ! stéréographiques]%
  {coordonnées stéréographiques}~$(\gtsMWx,\gtsMWy)$
associées au paramètre d'ordre $\gtsOP$
correspondent alors
aux coordonnées cartésiennes du point intersection
du plan équatorial
et de la demi-droite issue du pôle sud
passant par
le point de coordonnées sphériques~$(\gtsMTheta,\gtsMPhi)$,
représentant du paramètre d'ordre $\gtsOP$
sur la sphère~unité~$S^2$ (\figurename~\ref{fig/OP/str}).
Notons que
le pôle nord $(\gtsMTheta=0)$ se projette sur l'origine,
les points de l'équateur $(\gtsMTheta=\pi/2)$ sur eux-même,
et le pôle sud $(\gtsMTheta=\pi)$ sur le point infini.
\begin{subequations}
\label{OP/RS/ster}
Des notions élémentaires de géométrie analytique
conduisent aux relations de passages suivantes :
\begin{align}
\label{OP/RS/ster/U}
\gtsMWx&=\tan{\scriptstyle{\frac{1}{2}}}\gtsMTheta\;\cos\gtsMPhi,\\
\label{OP/RS/ster/V}
\gtsMWy&=\tan{\scriptstyle{\frac{1}{2}}}\gtsMTheta\;\sin\gtsMPhi.
\end{align}
\end{subequations}
\begin{subequations}
Au \wiit[paramètre d'ordre ! local]{paramètre d'ordre} $\gtsOP$
est alors associé la variable complexe
\begin{equation}
\label{OP/RS/Cx/def}
\gtsMW\equiv\gtsMWx+\ci\gtsMWy
\end{equation}
qui vit effectivement
sur la sphère complexe $\mathbb{C}\cup\{\infty\}$;
nous lisons
\begin{equation}
\label{OP/RS/Cx}
\gtsMW=
  \tan{\scriptstyle{\frac{1}{2}}}\gtsMTheta\;\gtse^{\ci\gtsMPhi}.
\end{equation}
\end{subequations}
Ici,
le \wi[Hamiltonien ! magnétique]{Hamiltonien magnétique} (\ref{Hm/mg})
s'écrit immédiatement
\begin{equation}
\label{Hm/mg/RS}
\gtsHm_{mag}=
4\gtsJ\!{\iint\limits_\gtsSurface}\!\sqrt{g}\,{\gtsd\Omega}\:g^{ij}
\frac{\partial_{i}\gtsMW\,\partial_{j}\overline{\gtsMW}}
  {\left[1+\gtsMW\overline{\gtsMW}\right]^2}
\end{equation}
en incorporant dans (\ref{Hm/mg/nested})
l'égalité utile
\begin{equation}
\label{OP/RS/Cx/derivative}
\sin \gtsMTheta\ \partial_{i}\gtsMW
=\gtsMW
\left[
\partial_{i}\gtsMTheta
  +\ci\sin \gtsMTheta\ \partial_{i}\gtsMPhi
\right].
\end{equation}

\subsection[Double obstruction topologique :
    classification de Toulouse-Kléman]{%
  Double obstruction topologique :\\ classification de Toulouse-Kléman}
\swih[obstruction topologique]{classification de Toulouse-Kléman}
\swih[classification de Toulouse-Kléman]{Toulouse-Kléman}
Considérons maintenant la membrane $\gtsSurface$
comme rien d'autre qu'une variété orientable de dimension $2$
fermée ou ouverte
qui, si nécessaire, peut-être \wibf[compactification]{compactifiée}
\wih{membrane ! compactifiée}
en identifiant chacun de ses bords, si cela a un sens,
à un point unique;
nous noterons $\gtsCSurface$
la membrane $\gtsSurface$ ainsi compactifiée.
En d'autres termes,
la membrane est momentanément envisagée
comme un \wibf{objet topologique} :
\swih[objet topologique]{topologie}
seules nous intéressent ici
les propriétés qui ne changent pas
lors de \wiit[deformation@déformation ! continue]{déformations continues}.

Chaque membrane compactifiée $\gtsCSurface$
est ainsi \wibf[classification topologique]{classée} suivant
son \wibf{genre topologique} $\gtsgenus\in\mathbb{N}$ :
deux membranes compactifiées $\gtsCSurface$
\wih{membrane ! compactifiée}
de même genre topologique $\gtsgenus$ sont
déformables continûment
\wih{deformation@déformation ! continue}
de l'une à l'autre sans introduire aucune déchirure ni trou
\cite{Nakahara}.
Pour illustration :
une sphère est de genre topologique $0$,
un tore de genre topologique $1$,
deux tores soudés\wih{sommation connexe} de genre topologique $2$,
\textlatin{etc}.
Autrement dit,
l'\wi[invariant ! topologique]{invariant topologique} $\gtsgenus$
\wi[obstruction topologique]{verrouille}
\swih[obstruction topologique]{verrouillage topologique}
le nombre de trous de la membrane;
nous noterons $T_{\gtsgenus}$
la classe d'équivalence de genre topologique $\gtsgenus$.
La \wiit[topologie ! combinatoire]{topologie combinatoire}
fournit une relation simple entre
la \wiit{caractéristique d'Euler}
\swih[caractéristique d'Euler]{Euler}
et le \wiit{genre topologique} \cite{Nakahara} :
\begin{equation}
\label{Mbr/genus/Euler}
\gtsEuler\left(T_{\gtsgenus}\right)=2\left(1-\gtsgenus\right);
\end{equation}
tandis que le
\wiit[Gauss-Bonnet~(théorème de)]{théorème de Gauss-Bonnet} affirme
\cite{Struik,DubrovinI,Nakahara} :
\begin{equation}
\label{Mbr/GaussBonnet}
\gtsEuler\left(\gtsCSurface\right)=
\frac{1}{2\pi}
 {\iint\limits_\gtsCSurface}\!\sqrt{g}\,{\gtsd\Omega}\:\gtsGC
\end{equation}
avec $\gtsGC$ la
\wiit[courbure ! locale de Gauss]{courbure locale de Gauss}
(\ref{Mbr/cur/Gaussian}).

Une approche similaire existe
pour les \wibf[configuration ! spins@de spins]{configurations de spins}.
En effet,
les configurations continues
du paramètre d'ordre $\gtsOP$
ne sont autres que
les fonctions continues qui envoient
la \wi[membrane ! compactifiée]{membrane compactifiée}
$\gtsCSurface$ (la \wi[variété ! support]{variété support})
sur la sphère~unité~$S^2$ (la \wi[variété ! cible]{variété cible}).
Or l'ensemble de telles fonctions forme
le \wibf{groupe de cohomotopie}
$\pi^2\left(T_{\gtsgenus}\right)$,
avec $\gtsgenus$ le genre topologique de $\gtsCSurface$ :
les fonctions continues sont
\wibf[classification topologique]{classées} ici selon
leur \wibf{classe d'homotopie}
\cite{BCG,STHu,Nakahara}.
Conséquence immédiate :
si le \wiit{groupe de cohomotopie} $\pi^2\left(T_{\gtsgenus}\right)$
ne se réduit pas au groupe trivial,
nous nous attendons à rencontrer
une physique exotique.
En effet,
certains concepts généralement très féconds en physique
sont ici infructueux :
les perturbations \guillemets{continues} ou \guillemets{plastiques}
ne permettent pas de sauter
d'une \wiit[classification topologique]{classe de configuration}
à une autre.
Aussi,
la \wiit[classification topologique]%
  {classification des configurations de spins}
suivant leur \wiit{classe d'homotopie}
s'avère fort légitime, voire essentielle
\cite{BelavinPolyakov,Bogomolnyi,ToulouseKleman}.
Les configurations de spins
homotopes à la configuration de spins
dont la fonction continue associée $\gtsOP$
envoie la membrane compactifiée $\gtsCSurface$
en un point unique de la sphère~unité~$S^2$
(\textlatin{e.g.} au pôle nord)
correspondent
aux \wiit[configuration ! triviale]{configurations triviales}.
En fait,
l'homologie et la cohomologie nous apprennent que
les \wiit[groupe de cohomotopie]{groupes de cohomotopie}
$\pi^2\left(T_{\gtsgenus}\right)$ sont isomorphes
au groupe des entiers relatifs $\mathbb{Z}$.
Nous avons effectivement
\begin{align}
\label{Mbr/cohomtopy/sphere}
&\pi^2\left(S^2\right)=
\pi_2\left(S^2\right)
\cong\mathbb{Z},
\intertext{et,
en invoquant la \wiit[Künneth~(formule de)]{formule de Künneth},}
\label{Mbr/cohomtopy/torus}
&\pi^2\left(T^2\right)=
\pi^2\left({S^1}\times{S^1}\right)=
\pi^1\left({S^1}\right)\otimes\pi^1\left({S^1}\right)
\cong\mathbb{Z};
\intertext{donc, par \wiit{sommation connexe},}
&\pi^2\left(T_{\gtsgenus}\right)
\cong\mathbb{Z}
\qquad
\forall \gtsgenus\in\mathbb{N}.
\end{align}
L'\wiit[invariant ! topologique]{invariant topologique}
qui \wiit[obstruction topologique]{verrouille}
les configurations de spins
est l'\wibf{indice de Pontrjagin} $\gtsPontrjagin$
\swih[indice de Pontrjagin]{Pontrjagin}
qui s'écrit
\begin{equation}
\label{Mbr/inv/Pontrjagin}
\gtsPontrjagin=
\frac{1}{4\pi}
  {\iint\limits_\gtsCSurface}
    \sin\gtsMTheta\;\gtsd\gtsMTheta\,\gtsd\gtsMPhi
\end{equation}
dans le contexte présent
\cite{Felsager,Nakahara}.
Contrairement aux topologues,
les physiciens parle de
\swih[indice de Pontrjagin]{charge ! topologique}
\witextbf{charge topologique};
nous noterons $\pi^2_{\gtsgenus\gtsPontrjagin}$
la \wiit[classification topologique]{classe d'équivalence}
de charge topologique $\gtsPontrjagin$
qui envoie la classe d'équivalence $T_{\gtsgenus}$
sur la sphère~unité~$S^2$.
Cet \wiit[invariant ! topologique]{invariant topologique}
correspond au nombre de fois que la sphère~unité~$S^2$
(la \wiit[variété ! cible]{variété cible})
est recouverte entièrement quand
la membrane compactifiée $\gtsCSurface$
(la \wiit[variété ! support]{variété support})
est parcourue
\cite{Felsager}.
En adoptant le point de vue originelle,
l'\wiit{indice de Pontrjagin} décompte les retournements
du \wiit[paramètre d'ordre ! locale]{paramètre d'ordre} $\gtsOP$.
Par conséquent,
les \wiit[configuration ! triviale]{configurations triviales}
de spins seront caractérisées par une charge topologique nulle;
par contraste,
les \wi[configuration ! topologique]{configurations topologiques}
auront une charge topologique non-nulle.

Pour résumer,
la \wiit[membrane ! magnétique]{membrane magnétique}
$\gtsSurface$ est caractérisée
par deux \wiit[invariant ! topologique]{invariants topologiques} :
\begin{itemize}
  \item son \wiit{genre topologique} $\gtsgenus$
    qui \wiit[obstruction topologique]{verrouille}
    sa \wiit[configuration ! spatiale]{configuration spatiale};
  \item sa \wiit[indice de Pontrjagin]{charge topologique} $\gtsPontrjagin$
    qui \wiit[obstruction topologique]{verrouille}
    sa \wiit[configuration ! magnétique]{configuration magnétique}.
\end{itemize}
Il est évident
qu'une modélisation pertinente de la membrane magnétique
doit à la fois satisfaire et clarifier
cette double \wiit{obstruction topologique}
dont le mécanisme peut-être schématisé comme suit :
\begin{equation}
\label{diag/Mbr/inv}
\begin{CD}
\gtsSurface
  @>\text{compactification}>\text{genre topologique $\gtsgenus$}>
\gtsCSurface \in T_{\gtsgenus}
  @>\pi^2\left(T_{\gtsgenus}\right)\cong\mathbb{Z}>%
    \text{charge topologique $\gtsPontrjagin$}>
\gtsOP\in\pi^2_{\gtsgenus\gtsPontrjagin}.
\end{CD}
\end{equation}

\subsection[\'Energie magnétique : décomposition de Bogomol'nyi]%
  {\'Energie magnétique :\\ décomposition de Bogomol'nyi}
\wih{decomposition@décomposition de Bogomol'nyi}
\swih[décomposition de Bogomol'nyi]{Bogomol'nyi}
Comme des considérations de symétrie
permettent de modéliser
de nombreux systèmes physiques
à ordre symétrique
et d'élaborer des théories physiques
\cite{Sivardiere},
des considérations de topologie
doivent permettre de décrire des systèmes physiques
à \swiit[classification topologique]{ordre topologique}
\swih[obstruction topologique]{ordre topologique}
\cite{ToulouseKleman,Mermin,Sivardiere}.
Notons en revanche que,
si l'origine des \wiit[invariant ! symetrie@de symétrie]{invariants de symétrie}
est acquise,
l'émergence d'\wiit[invariant ! topologique]{invariants topologiques}
en physique demeure une énigme \cite{Ryder}.
Il est donc remarquable que nous puissions retrouver
le \wiit[obstruction topologique]{verrouillage} topologique
induit par la distribution de spins à partir du
\wiit[Hamiltonien ! magnétique]{Hamiltonien magnétique} (\ref{Hm/mg})
en appliquant une méthode suggérée indépendamment
par Belavin et Polyakov \cite{BelavinPolyakov} d'une part,
et par Bogomol'nyi \cite{Bogomolnyi} d'autre part.

Pour commencer,
introduisons le tenseur
$T_{i\hphantom{\alpha}}^{\hphantom{i}\alpha}$
tel que
\begin{equation}
\label{BPB/T/def}
T_{i\hphantom{\alpha}}^{\hphantom{i}\alpha}
\equiv
\frac{1}{\sqrt{2}}
\left[
\partial_{i}n^{\alpha}-
\gtsSign\,
E_{i\hphantom{r}}^{\hphantom{i}r}
F_{\hphantom{\alpha}\kappa}^{\alpha\hphantom{\kappa}}
\partial_{r}n^{\kappa}
\right]
\qquad
\gtsSign=\pm{1}.
\end{equation}
Les tenseurs $E_{ij}$ et $F_{\alpha\beta}$ sont
les tenseurs unitaires parfaitement antisymétriques d'ordre deux
associés respectivement
à la membrane magnétique $\gtsSurface$ et
à la variété du paramètre d'ordre $\gtsOP$;
nous lisons
\begin{equation}
\label{BPB/uast/def}
E_{ij}=\sqrt{g}\:\epsilon_{ij},
\qquad
F_{\alpha\beta}=\sqrt{h}\:\epsilon_{\alpha\beta},
\end{equation}
avec $\epsilon_{ab}$
le pseudotenseur unitaire parfaitement antisymétrique.
En choisissant pour la variété cible
la métrique sphérique (\ref{OP/metric})
et en réduisant localement la métrique $g$
à une métrique orthogonale,
il saute aux yeux
que le scalaire
$g^{ij}h_{\alpha\beta}\,%
  T_{i\hphantom{\alpha}}^{\hphantom{i}\alpha}%
  T_{j\hphantom{\beta}}^{\hphantom{j}\beta}$
est positif.
Or, puisque
$T_{i\hphantom{\alpha}}^{\hphantom{i}\alpha}$
est un tenseur,
cette propriété reste vraie
pour tout autre choix de système de coordonnées.
Nous en déduisons que
\begin{equation}
\label{BPB/trace/positive}
g^{ij}h_{\alpha\beta}\,
T_{i\hphantom{\alpha}}^{\hphantom{i}\alpha}
T_{j\hphantom{\beta}}^{\hphantom{j}\beta}
\geqslant 0.
\end{equation}
En outre,
le calcul explicite de cette trace donne
\begin{subequations}
\label{BPB/trace}
\begin{equation}
\label{BPB/trace/expanded}
g^{ij}h_{\alpha\beta}\,
T_{i\hphantom{\alpha}}^{\hphantom{i}\alpha}
T_{j\hphantom{\beta}}^{\hphantom{j}\beta}
=
g^{ij}h_{\alpha\beta}\,
\partial_{i}n^{\alpha}\partial_{j}n^{\beta}
-2\,\gtsSign\frac{\sqrt{h}}{\sqrt{g}}\,\gtsJacobian,
\end{equation}
où $\gtsJacobian$ est le Jacobien de la transformation locale
$\left(\gtsx^i\right)\to\left(n^\alpha\right)$ :
\begin{equation}
\label{BPB/trace/Jacobian}
\gtsJacobian=
\frac{1}{2}\,\epsilon_{\kappa\lambda}\epsilon^{rs}%
  \partial_{r}n^{\kappa}\partial_{s}n^{\lambda}.
\end{equation}
\end{subequations}
Des relations (\ref{BPB/trace/positive}) et (\ref{BPB/trace/expanded})
nous tirons la précieuse inégalité
\begin{equation}
\label{BPB/thepoint}
g^{ij}h_{\alpha\beta}\,
\partial_{i}n^{\alpha}\partial_{j}n^{\beta}
\geqslant
2\,\gtsSign\frac{\sqrt{h}}{\sqrt{g}}\,\gtsJacobian.
\end{equation}
L'intégration membre à membre de l'inégalité (\ref{BPB/thepoint})
sur toute la membrane compactifiée $\gtsCSurface$
en adoptant la métrique sphérique (\ref{OP/metric})
pour la variété cible
nous conduit aisément
au point remarquable de la méthode :
\begin{subequations}
\label{BPB/int/geq}
\begin{align}
\label{BPB/init/geq/crude}
{\iint\limits_\gtsCSurface}\!\sqrt{g}\,{\gtsd\Omega}\:
g^{ij}h_{\alpha\beta}\,
\partial_{i}n^{\alpha}\partial_{j}n^{\beta}
&\geqslant 2\,\gtsSign
{\iint\limits_\gtsCSurface}\!\sqrt{h}\,{\gtsd\Omega}\:\gtsJacobian,\\
\label{BPB/init/geq/nested}
&\geqslant 2\,
\left|
{\iint\limits_\gtsCSurface}
\sin\gtsMTheta\;\gtsd\gtsMTheta\,\gtsd\gtsMPhi
\right|,\\
\label{BPB/init/geq/thepoint}
&\geqslant 8\pi\left|\gtsPontrjagin\right|,
\end{align}
\end{subequations}
avec $\gtsPontrjagin$
l'\wiit{indice de Pontrjagin} (\ref{Mbr/inv/Pontrjagin}).
Ayant encore en mémoire l'expression du
\wiit[Hamiltonien ! magnétique]{Hamiltonien magnétique} (\ref{Hm/mg}),
nous écrivons sans attendre
\begin{equation}
\label{BPB/Hm/mg}
\gtsHm_{mag}
\geqslant
8\pi\gtsJ\left|\gtsPontrjagin\right|.
\end{equation}
Plus prosaïquement,
l'énergie magnétique
des \wiit[configuration ! spins@de spins]{configurations de spins}
appartenant à une même
\wiit[classification topologique]{classe d'équivalence}
$\pi^2_{\gtsgenus\gtsPontrjagin}$ est minorée
par l'\wibf[energie@énergie ! topologique]{énergie topologique}
\begin{equation}
\label{BPB/Em/mg}
\gtsTE_{\gtsPontrjagin}
\equiv
8\pi\gtsJ\left|\gtsPontrjagin\right|.
\end{equation}
Qui plus est,
par construction,
cette \wiit[energie@énergie ! topologique]{énergie topologique}
$\gtsTE_{\gtsPontrjagin}$ n'est atteinte
que par les configurations de spins pour lesquelles le tenseur
$T_{i\hphantom{\alpha}}^{\hphantom{i}\alpha}$
se confond avec le tenseur nul :
c'est-à-dire,
compte tenue de la définition (\ref{BPB/T/def})
du tenseur $T_{i\hphantom{\alpha}}^{\hphantom{i}\alpha}$,
pour lesquelles le paramètre d'ordre $\gtsOP$
vérifie les équations
\begin{equation}
\label{BPB/self-dual/n}
\partial_{i}n^{\alpha}=
\gtsSign\,
E_{i\hphantom{r}}^{\hphantom{i}r}
F_{\hphantom{\alpha}\kappa}^{\alpha\hphantom{\kappa}}
\partial_{r}n^{\kappa}.
\end{equation}
Si la métrique de la \wiit[variété ! cible]{variété cible} est
la métrique sphérique (\ref{OP/metric}),
\wih{description ! géométrique}
les \wibf[equations@équations ! auto-duales]{équations auto-duales}
(\ref{BPB/self-dual/n})
se lisent
\begin{align}
\label{BPB/self-dual/sph}
\partial_{i}\gtsMTheta
&=\gtsSign\,\sqrt{g}\sin \gtsMTheta\ %
  \epsilon_{ir}\partial^{r}\gtsMPhi
\;;\\
\intertext{tandis que leur versions algébriques
\wih{description ! algébrique}
s'écrivent}
\label{BPB/self-dual/RS}
\partial_{i}\gtsMW
&=-\ci\gtsSign\,\sqrt{g}\:\epsilon_{ir}\partial^{r}\gtsMW.
\end{align}
Les \wibf[configuration ! topologique auto-duale]%
  {configurations topologiques auto-duales}
seront donc les \wiit[configuration ! topologique]%
  {configurations topologiques}
qui minimisent
l'\wiit[energie@énergie ! magnétique]{énergie magnétique} (\ref{Hm/mg}).
Enfin,
l'\wiit[energie@énergie ! topologique]{énergie topologique}
$\gtsTE_{\gtsPontrjagin}$ s'interprète immédiatement comme
l'énergie minimale nécessaire
pour retourner $\left|\gtsPontrjagin\right|$ fois
le \wiit[paramètre d'ordre ! local]{paramètre d'ordre} $\gtsOP$,
et le facteur $8\pi\gtsJ$
comme le \textlatin{quantum} d'énergie minimale de retournement.
Avant de conclure,
notons que l'injection de l'égalité (\ref{BPB/trace})
dans la fonctionnelle originelle (\ref{Hm/mg}) donne
une écriture très pertinente du
\wiit[Hamiltonien ! magnétique]{Hamiltonien magnétique}
\cite{Felsager} :
\begin{equation}
\label{Hm/mg/BPB}
\gtsHm_{mag}=
\gtsJ\!{\iint\limits_\gtsCSurface}\!\sqrt{g}\,{\gtsd\Omega}\:
g^{ij}h_{\alpha\beta}\,
T_{i\hphantom{\alpha}}^{\hphantom{i}\alpha}
T_{j\hphantom{\beta}}^{\hphantom{j}\beta}
+\;
8\pi\gtsJ\left|\gtsPontrjagin\right|.
\end{equation}

Pour résumer,
la \wibf[decomposition@décomposition de Bogomol'nyi]{décomposition de Bogomol'nyi}
montre que
le \wiit[Hamiltonien ! magnétique]{Hamiltonien magnétique} (\ref{Hm/mg})
\wibf[levée topologique]{lève}
la \wiit{classification topologique} des configurations magnétiques :
\begin{itemize}
  \item l'énergie magnétique des configurations de spins
    appartenant à la même \wiit{classe d'homotopie}
    est minorée par l'énergie minimale
    nécessaire pour retourner
    le paramètre d'ordre $\gtsOP$;
  \item les configurations topologiques
    qui minimisent l'énergie magnétique vérifient
    les \wiit[equations@équations ! auto-duales]{équations auto-duales}
    (équations différentielles du premier ordre).
\end{itemize}
Un mécanisme comparable existe pour la classification topologique
des configurations spatiales.

\subsection[\'Energie de courbure : Hamiltonien de Helfrich-Willmore]{%
  \'Energie de courbure :\\ Hamiltonien de Helfrich-Willmore}
\wih{energie@énergie ! courbure@de courbure}
\swih[Hamiltonien ! Helfrich@de Helfrich-Willmore]{Willmore}
\swih[Hamiltonien ! Helfrich@de Helfrich-Willmore]{Helfrich}
Pour commencer,
rappelons qu'une surface bidimensionnelle est parfaitement
décrite localement par ses deux
\wibf[courbure ! principale]{courbures principales}
$\gtsPC$ et $\gtspC$.
Leur produit
\begin{equation}
\label{Mbr/cur/Gaussian}
\gtsGC\equiv\gtsPC\gtspC
\end{equation}
est appelé
la \wibf[courbure ! locale de Gauss]{courbure locale de Gauss} de la surface,
leur moyenne arithmétique
\begin{equation}
\label{Mbr/cur/MeanCurvature}
\gtsMC\equiv
\frac{1}{2}\left[\gtsPC+\gtspC\right]
\end{equation}
la \wibf[courbure ! moyenne locale]{courbure moyenne locale} de la surface.
\pagebreak

\begin{figure}[!ht]
  \begin{center}
  \includegraphics[width=.6\linewidth]{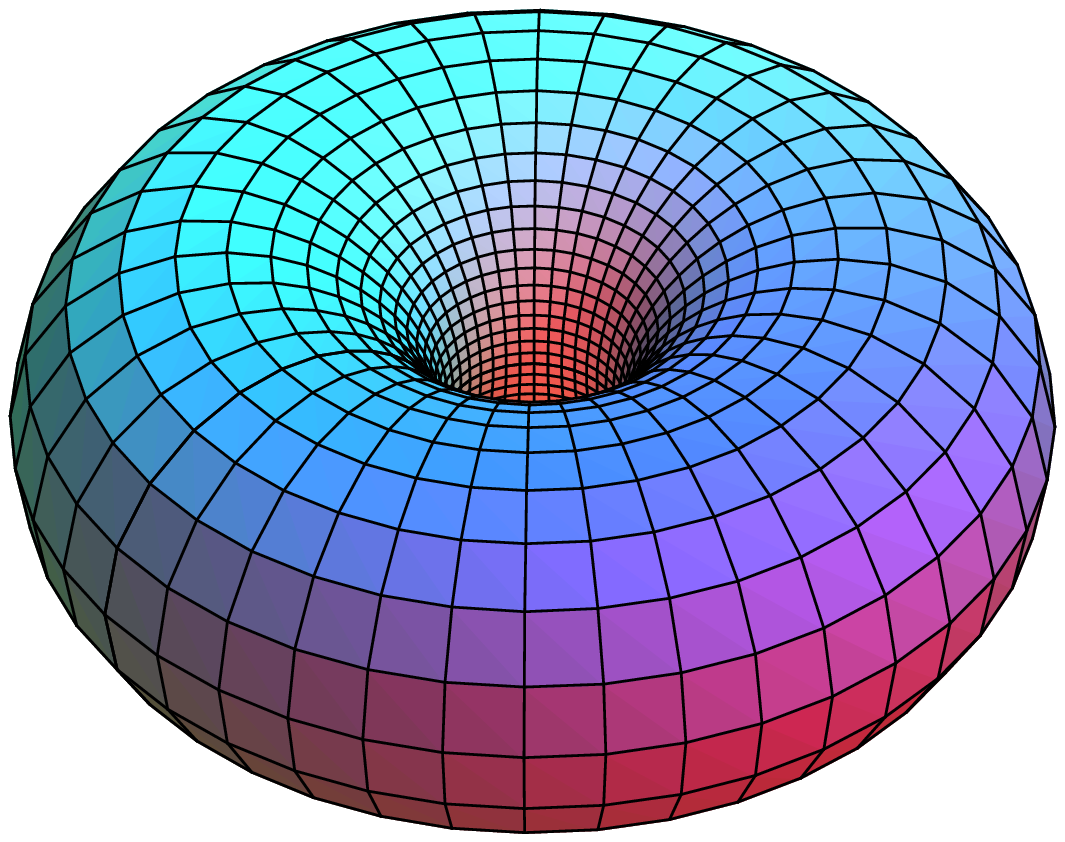}
  \caption[Tore de Clifford]{%
    Le \wibf{tore de Clifford}
    \swih[tore de Clifford]{Clifford}
    est le tore à symétrie axiale
    dont le rapport du \wiit[rayon ! axiale]{rayon axiale} $r$
    sur le \wiit[rayon ! revolution@de révolution]{rayon de révolution} $R$
    est égal à $1/\sqrt{2}$.
    Selon une
    \wibf[conjecture de Clifford]{conjecture due à Willmore}
    \swih[conjecture de Clifford]{Clifford}\cite{Willmore},
    la \wiit{fonctionnelle de Willmore} est minimisée
    soit par le \wiit{tore de Clifford}
    soit par un de ses transformés conformes.
    }
  \label{fig/torus/Clifford}
  \end{center}
\end{figure}
Le Hamiltonien gouvernant la dispersion
\wih{Hamiltonien ! elastique@élastique naïf}
des \wiit[courbure ! principale]{courbures principales}
au sens de la théorie des probabilités
\cite{IWTS,WillmoreRG}
\begin{equation}
\label{Hm/el/nude/crude}
{\gtsHm}_{el}=%
{\textstyle{\frac{1}{2}}}\gtsKn\!%
{\iint\limits_\gtsCSurface}\!\sqrt{g}\,{\gtsd\Omega}\:%
\left[\frac{\gtsPC-\gtspC}{2}\right]^2
\end{equation}
apparaît comme un candidat légitime pour décrire
les \witextit{déformations continues} et \witextit{localisées}
\wih{deformation@déformation ! continue}\wih{deformation@déformation ! locale}
subies par une
\wiit[membrane ! souple compactifiée]{membrane souple compactifiée}
$\gtsCSurface$ spontanément
à \wiit[courbure ! constante]{courbure constante}.
La constante $\gtsKn$ sera interprétée comme
la \wibf[rigidité ! courbure@de courbure nue]{rigidité de courbure nue}.
La relation immédiate
\begin{subequations}
\label{Hm/el/nude/density}
\begin{align}
\label{Hm/el/nude/density/exp}
\left[\frac{\gtsPC-\gtspC}{2}\right]^2%
&=\left[\frac{\gtsPC+\gtspC}{2}\right]^2-\gtsPC\gtspC,\\
\label{Hm/el/nude/density/nested}
&=\gtsMC^2-\gtsGC,
\end{align}
\end{subequations}
permet d'écrire
le \wibf[Hamiltonien ! elastique@élastique naïf]{Hamiltonien élastique naïf}
(\ref{Hm/el/nude/crude}) sous la forme plus pertinente :
\begin{equation}
\label{Hm/el/nude/nested}
{\gtsHm}_{el}=%
{\textstyle{\frac{1}{2}}}\gtsKn\!%
{\iint\limits_\gtsCSurface}\!\sqrt{g}\,{\gtsd\Omega}\:%
\gtsMC^2
-
{\textstyle{\frac{1}{2}}}\gtsKn\!%
{\iint\limits_\gtsCSurface}\!\sqrt{g}\,{\gtsd\Omega}\:%
\gtsGC.
\end{equation}
\clearpage\noindent
D'après le \wiit[Gauss-Bonnet~(théorème de)]{théorème de Gauss-Bonnet}
(\ref{Mbr/GaussBonnet}),
l'intégrale de droite ne dépend que du \wiit{genre topologique} $\gtsgenus$
de la \wiit[membrane ! souple compactifiée]{membrane souple compactifiée}
$\gtsCSurface$.
L'intégrale de gauche
n'est autre que
la \wibf{fonctionnelle de Willmore}
\swih[fonctionnelle de Willmore]{Willmore}
\cite{Willmore,WillmoreRG} :
pour des surfaces
à \wibf[genre topologique ! sphérique]{topologie sphérique} ($\gtsgenus=0$),
elle est minimale (et nulle) pour la \wi[membrane ! plane]{membrane plane};
pour des surfaces
à \wibf[genre topologique ! torique]{topologie torique} ($\gtsgenus=1$),
elle est minorée par $4\pi$
et serait minimale
(\wiit[conjecture de Clifford]{conjecture due à Willmore} \cite{Willmore})
soit pour le \wiit{tore de Clifford} (\figurename~\ref{fig/torus/Clifford})
soit pour un de ses transformés conformes.
Ainsi le \wiit[Hamiltonien ! elastique@élastique naïf]%
  {Hamiltonien élastique naïf} (\ref{Hm/el/nude/crude})
ne convient-il plus pour décrire
des \witextit{déformations continues} et \witextit{localisées}
\wih{deformation@déformation ! continue}\wih{deformation@déformation ! locale}
endurées par une membrane souple compactifiée $\gtsCSurface$
de topologie non sphérique ($\gtsgenus\neq 0$).
En introduisant
une \wiit[courbure ! moyenne spontanée]{courbure moyenne spontanée}
non nulle $\gtsSMC$,
Helfrich \cite{HelfrichZN1,TRTV} obtient
un \wiit[Hamiltonien ! elastique@élastique]{Hamiltonien élastique}
plus adéquate :
\begin{equation}
\label{Hm/el/heuristic/crude}
{\gtsHm}_{el}=%
{\textstyle{\frac{1}{2}}}\gtsKc\!%
{\iint\limits_\gtsCSurface}\!\sqrt{g}\,{\gtsd\Omega}\:%
\left[\gtsMC-\gtsSMC\right]^2
-
{\textstyle{\frac{1}{2}}}\gtsKg\!%
{\iint\limits_\gtsCSurface}\!\sqrt{g}\,{\gtsd\Omega}\:%
\gtsGC,
\end{equation}
où la \wiit[rigidité ! courbure@de courbure nue]{rigidité de courbure nue} $\gtsKn$
a été renormalisée.
La constante $\gtsKc$ désigne
la \wibf[rigidité ! courbure@de courbure]{rigidité de courbure},
$\gtsKg$ la \wibf[rigidité ! topologique]{rigidité topologique}.
Dorénavant,
par \wibf[Hamiltonien ! elastique@élastique]{Hamiltonien élastique}
${\gtsHm}_{el}$ nous entendrons
le Hamiltonien heuristique (\ref{Hm/el/heuristic/crude}).
En vertu du
\wiit[Gauss-Bonnet~(théorème de)]{théorème de Gauss-Bonnet}
(\ref{Mbr/GaussBonnet})
et de la relation (\ref{Mbr/genus/Euler}),
le \wibf[Hamiltonien ! Helfrich@de Helfrich-Willmore]{Hamiltonien de Helfrich-Willmore}
(\ref{Hm/el/heuristic/crude}) se lit
\begin{equation}
\label{Hm/el/heuristic/nested}
{\gtsHm}_{el}=%
{\textstyle{\frac{1}{2}}}\gtsKc\!%
{\iint\limits_\gtsCSurface}\!\sqrt{g}\,{\gtsd\Omega}\:%
\left[\gtsMC-\gtsSMC\right]^2
+
2\pi\gtsKg\left(\gtsgenus-1\right).
\end{equation}
Cette écriture du \wiit[Hamiltonien ! elastique@élastique]{Hamiltonien élastique}
 ${\gtsHm}_{el}$ doit être comparée avec l'écriture (\ref{Hm/mg/BPB})
du \wiit[Hamiltonien ! magnétique]{Hamiltonien magnétique} ${\gtsHm}_{mag}$,
et réciproquement.
En regard de quoi,
la \wibf[courbure ! moyenne spontanée]{courbure spontanée} $\gtsSMC$
qui doit décrire
les propriétés physiques de la membrane souple $\gtsSurface$
correspondra dans la suite du {\chaptername}
à la courbure moyenne d'une surface choisie
\textlatin{a priori} arbitrairement
comme étant effectivement la \wibf[forme ! spontanée]{forme spontanée}
de la membrane compactifiée $\gtsCSurface$.
Dans ce cas restrictif,
le \wiit[Hamiltonien ! magnétique]{Hamiltonien magnétique}
(\ref{Hm/el/heuristic/nested}) se contente de décrire
des \wibf[deformation@déformation ! de faible amplitude]%
  {déformations de faible amplitude},
et les \witextit{courbures renormalisées} $\gtsKc$ et $\gtsKg$
\wih{rigidité ! courbure@de courbure}\wih{rigidité ! topologique}
dépendent clairement de ce choix arbitraire.

Pour résumer,
nous pouvons construire un
\wiit[Hamiltonien ! elastique@élastique]{Hamiltonien élastique}
qui \wibf[levée topologique]{lève} la \wiit{classification topologique}
des \wi[configuration ! spatiale]{configurations spatiales} :
\begin{itemize}
  \item l'\wiit[energie@énergie ! elastique@élastique]{énergie élastique}
    est minorée par l'énergie minimale
    requise pour créer un trou topologique;
  \item les \wiit[forme ! topologique]{formes topologiques}
    qui minimisent l'énergie élastique
    correspondent aux \wiit[forme ! spontanée]{formes spontanées}
    de la \wiit[membrane ! compactifiée]{membrane compactifiée} $\gtsCSurface$.
\end{itemize}

\clearpage
\subsection{Frustration géométrique}
\wih{frustration ! géométrique}
Dans les parties antérieures nous montrons
qu'une membrane magnétique $\gtsSurface$
se caractérise par
une double \wiit{classification topologique},
chacune étant levée par un Hamiltonien :
\begin{itemize}
  \item le Hamiltonien élastique ${\gtsHm}_{el}$
    engendre des \wiit[forme ! homéomorphe]{formes homéomorphes}
    à des \wiit[forme ! spontanée]{formes spontanées}
    pour lesquelles
    l'\wiit[energie@énergie ! courbure@de courbure saturée]%
      {énergie de courbure}
    ${\gtsEg}_{el}$ est \witextit{saturée};
  \item le \wiit[Hamiltonien ! magnétique]{Hamiltonien magnétique}
    ${\gtsHm}_{mag}$ génére des
    \wiit[configuration ! homotope]{configurations de spins homotopes}
    à des \wiit[configuration ! topologique auto-duale]%
      {configurations topologiques auto-duales}
    pour lesquelles
    l'\wiit[energie@énergie ! magnétique saturée]{énergie magnétique}
    ${\gtsEg}_{mag}$ est \witextit{saturée}.
\end{itemize}
Ainsi à la double \wi{obstruction topologique} (\ref{diag/Mbr/inv})
correspond-t-il la double \wi{levée topologique} suivante :
\begin{equation}
\label{diag/Mbr/Hm}
\begin{CD}
\gtsSurface
  @>{\gtsHm}_{el}>\text{forme spontanée}>
\gtsCSurface \in T_{\gtsgenus}
  @>{\gtsHm}_{mag}>\text{configuration auto-duale}>
\gtsOP\in\pi^2_{\gtsgenus\gtsPontrjagin}.
\end{CD}
\end{equation}

Généralement,
le \wiit[Hamiltonien ! total]{Hamiltonien total} $\gtsHm$
d'une membrane magnétique $\gtsSurface$
est défini comme la somme de trois Hamiltoniens :
\begin{equation}
\label{Hm/ttl/gen}
\gtsHm\equiv{\gtsHm}_{mag}+{\gtsHm}_{el}+{\gtsHm}_{m-el},
\end{equation}
où ${\gtsHm}_{mag}$, ${\gtsHm}_{el}$ et ${\gtsHm}_{m-el}$
représentent, respectivement,
le \wiit[Hamiltonien ! magnétique]{Hamiltonien magnétique},
\wiit[Hamiltonien ! elastique@élastique]{élastique} et
\wibf[Hamiltonien ! magnéto-élastique]{magnéto-élastique}.
Or,
pour un système quasi-uni\-di\-men\-sion\-nel,
l'addition d'un terme de
\wibf[couplage ! spin-élasticité]{couplage spin-élasticité}
revient à renormaliser la constante $\gtsJ$
de \wiit[couplage ! spin-spin]{couplage spin-spin}
\cite{CHC}.
Nous définirons donc
le \wibf[Hamiltonien ! total]{Hamiltonien total} $\gtsHm$
de la membrane magnétique compactifiée $\gtsCSurface$
comme la somme
du \wiit[Hamiltonien ! magnétique]{Hamiltonien magnétique} (\ref{Hm/mg})
et
du \wiit[Hamiltonien ! elastique@élastique]{Hamiltonien élastique}
(\ref{Hm/el/heuristic/crude});
des expressions (\ref{Hm/mg/BPB}) et (\ref{Hm/el/heuristic/nested})
nous tirons la formule
\begin{multline}
\label{Hm/ttl/nested}
\gtsHm=
\gtsKc\!
{\iint\limits_\gtsCSurface}\!\sqrt{g}\,{\gtsd\Omega}\:%
\left[
\gtsJr\ %
g^{ij}h_{\alpha\beta}\,
T_{i\hphantom{\alpha}}^{\hphantom{i}\alpha}
T_{j\hphantom{\beta}}^{\hphantom{j}\beta}
+
{\textstyle{\frac{1}{2}}}
\left[\gtsMC-\gtsSMC\right]^2
\right]\\
+
8\pi\gtsJ\left|\gtsPontrjagin\right|
+
2\pi\gtsKg\left(\gtsgenus-1\right),
\end{multline}
où nous avons introduit
la \wiit{constante relative de couplage} $\gtsJr\equiv\gtsJ/\gtsKc$.
L'écriture (\ref{Hm/ttl/nested}) du Hamiltonien total $\gtsHm$
place pertinemment sur un même pied
les \witextit{interactions élastique} et \witextit{magnétique} :
\wih{interaction ! elastique@élastique}\wih{interaction ! magnétique}
la métrique $g_{ij}$ et le paramètre d'ordre $\gtsOP$
interagissent.
Par \wibf[configuration ! géométrique]{configuration géométrique}
nous désignerons un couple $\left(g_{ij},\gtsOP\right)$
qui minimise le \wiit[Hamiltonien ! total]{Hamiltonien total} $\gtsHm$,
c'est-à-dire un couple $\left(g_{ij},\gtsOP\right)$ solution
des \wibf[equations@équations ! d'Euler-Lagrange]{équations d'Euler-Lagrange}
\swih[équations d'Euler-Lagrange]{Euler-Lagrange}
déduites du Hamiltonien $\gtsHm$.
Ainsi la double \wiit{levée topologique} (\ref{diag/Mbr/Hm})
se combine-t-elle en une levée double :
\begin{equation}
\label{diag/Mbr/final}
\begin{CD}
\gtsSurface
  @>\gtsHm>\text{configuration géométrique}>
\left(g_{ij},\gtsOP\right)
\in
T_{\gtsgenus}\times\pi^2_{\gtsgenus\gtsPontrjagin}.
\end{CD}
\end{equation}
\textlatin{A priori}
les \wiit[configuration ! géométrique]{configurations géométriques}
$\left(g_{ij},\gtsOP\right)$ ne saturent pas
le \wiit[Hamiltonien ! total]{Hamiltonien total} $\gtsHm$ :
les configurations géométriques
dont l'énergie totale $\gtsEg$ est supérieure
à l'\wibf[energie@énergie ! mini@\textlatin{minimum minimorum} topologique]%
  {énergie \textlatin{minimum minimorum} topologique}
\begin{equation}
\label{GF/Em}
\gtsTE
\equiv
8\pi\gtsJ\left|\gtsPontrjagin\right|
+
2\pi\gtsKg\left(\gtsgenus-1\right)
\end{equation}
sont dites \wiit[configuration ! frustrée]{frustrées}.
Cette \wibf[frustration ! géométrique]{frustration géométrique}
apparaît clairement comme
la compétition entre
les deux \wiit[classification topologique]{ordres topologiques},
et l'excès d'énergie qui la caractérise comme une énergie
d'\wibf[interaction ! topologique]{interaction topologique}.

Pour résumer,
le \wiit[Hamiltonien ! total]{Hamiltonien total} $\gtsHm$ combine
la double \wibf[classification topologique]{classification topologique}
de la \witextit{membrane magnétique souple compactifiée} $\gtsCSurface$ :
\wih{membrane ! souple compactifiée}\wih{membrane ! magnétique}
\begin{itemize}
  \item l'énergie des configurations géométriques
    appartenant à la même \wiit{classe d'homotopie}
    est minorée par
    l'\wiit[energie@énergie ! mini@\textlatin{minimum minimorum} topologique]%
      {énergie \textlatin{minimum minimorum} topologique} $\underline{\gtsEg}$
    nécessaire pour retourner le paramètre d'ordre $\gtsOP$
    et pour former les trous topologiques;
  \item une \wibf[frustration ! géométrique]{frustration géométrique}
    émerge \textlatin{a priori}
    de la compétition entre les deux ordres topologiques,
    les configurations géométriques
    (qui vérifient les
    \wiit[equations@équations ! d'Euler-Lagrange]{équations d'Euler-Lagrange}
    déduites de $\gtsHm$)
    ne minimisant pas nécessairement \textlatin{a priori}
    l'\wiit[energie@énergie ! magnétique]{énergie magnétique}
    et/ou l'\wiit[energie@énergie ! elastique@élastique]{énergie élastique}.
\end{itemize}
Il est évident que des illustrations simples
doivent permettre de clarifier
le mécanisme de cette frustration géométrique.

\section{Membranes de topologie sphérique}
\begin{introduction}
Dans cette partie, nous envisageons
des membranes magnétiques souples compactifiées
de \wi[genre topologique ! sphérique]{genre topologique sphérique}
et de forme spontanée simple :
des \wi[brisure de symétrie]{brisures de symétrie}
successives révèlent une physique de plus en plus exotique.

Je met en évidence
le mécanisme de la \wiit[frustration ! géométrique]{frustration géométrique}
en \wibf[configuration ! magnétique gelée]{gelant} une configuration magnétique
de la \wiit[membrane ! souple rigidifiée]{membrane souple rigidifiée}.
\end{introduction}

\subsection{La sphère : une forte dégénérescence}
\begin{emphasise}
Pour une \wiit[membrane ! souple rigidifiée]{membrane souple rigidifiée}
($\gtsKc=\infty$),
le \wiit[Hamiltonien ! total]{Hamiltonien total} $\gtsHm$ se réduit
au \wiit[Hamiltonien ! magnétique]{Hamiltonien magnétique} ${\gtsHm}_{mag}$ :
les \wiit[configuration ! géométrique]{configurations géométriques}
se confondent alors avec
les \wi[configuration ! topologique auto-duale]%
  {configurations topologiques auto-duales}.
\end{emphasise}

Une représentation naturelle de la sphère rigide,
en \wiit[coordonnées ! sphériques]{coordonnées sphériques}
$\left(\rho,\theta,\varphi\right)$,
est
\begin{equation}
\label{S/coords/sph}
\rho=r,
\end{equation}
avec $r$ le rayon de la sphère rigide.
La métrique de la membrane magnétique s'écrit alors
\begin{equation}
\label{S/g}
g=r^2
\left[
\gtsd\theta\!\otimes\!\gtsd\theta
+\sin^2 \theta\ \gtsd\varphi\!\otimes\!\gtsd\varphi
\right];
\end{equation}
ainsi $g^{\theta\varphi}=g^{\varphi\theta}=0$ et nous avons
\begin{equation}
\label{S/g/nested}
g^{\theta\theta}\!\sqrt{g}=\sin\theta,
\qquad
g^{\varphi\varphi}\!\sqrt{g}=\frac{1}{\sin\theta}.
\end{equation}
Aussi l'expression (\ref{Hm/mg/nested})
du \wiit[Hamiltonien ! magnétique]{Hamiltonien magnétique}
s'écrit
\begin{equation}
\label{S/Hm/mg/crude}
\gtsHm_{mag}\!=\!\gtsJ\!%
\int\limits_{0}^{+\pi}\!\!\gtsd\theta\!\!
\int\limits_{-\pi}^{+\pi}\!\!\gtsd\varphi\!
\left[\!\vphantom{\left[\frac{\sin^2\gtsMTheta}{\sin\theta}\right]^2}
\left[\sin\theta\;\gtsMTheta_{\theta}^2%
  +\frac{\sin^2\gtsMTheta}{\sin\theta}\;\gtsMPhi_{\varphi}^2\right]
\!+\!
\left[\frac{\gtsMTheta_{\varphi}^2}{\sin\theta}%
  +\sin\theta\:\sin^2\gtsMTheta\;\gtsMPhi_{\theta}^2\right]
\!\right],
\end{equation}
où les indices représentent les dérivés partielles.
Notons que la densité Hamiltonienne magnétique
de la \wiit[sphère ! rigide magnétique]{sphère rigide magnétique}
est indépendante du rayon $r$ de la sphère.
De la fonctionnelle (\ref{S/Hm/mg/crude})
nous tirons aisément
les \wiit[equations@équations ! auto-duales]{équations auto-duales}
de la \wiit[sphère ! rigide magnétique]{sphère rigide magnétique} :
\wih{description ! géométrique}
\begin{subequations}
\label{S/Hm/mg/SD/geo}
\begin{align}
\label{S/Hm/mg/SD/geo/theta}
\sin\theta\;\gtsMTheta_{\theta}&=%
  \hphantom{-}\gtsSign\;\sin\gtsMTheta\;\gtsMPhi_{\varphi},\\
\label{S/Hm/mg/SD/geo/phi}
\gtsMTheta_{\varphi}&=%
  -\gtsSign\;\sin\theta\:\sin\gtsMTheta\;\gtsMPhi_{\theta}.
\end{align}
\end{subequations}
Leur \wiit[description ! algébrique]{versions algébriques}
prennent la forme
\begin{subequations}
\label{S/Hm/mg/SD/alg}
\begin{align}
\label{S/Hm/mg/SD/alg/re}
\gtsMWx_{\gtsrswx}&=+\gtsSign\;\gtsMWy_{\gtsrswy},\\
\label{S/Hm/mg/SD/alg/im}
\gtsMWx_{\gtsrswy}&=-\gtsSign\;\gtsMWy_{\gtsrswx}.
\end{align}
\end{subequations}
\begin{subequations}
\label{S/RS/ster}
Les nouvelles coordonnées
$\left(\gtsrswx,\gtsrswy\right)$ correspondent
aux \wiit[coordonnées ! stéréographiques]{coordonnées stéréographiques}
par rapport au pôle sud
de la \wi[sphère ! rigide]{sphère rigide} projetée
sur la sphère unité~$S^2$;
nous lisons
\begin{align}
\label{S/RS/ster/u}
\gtsrswx&=\tan{\scriptstyle{\frac{1}{2}}}\theta\;\cos\varphi,\\
\label{S/RS/ster/v}
\gtsrswy&=\tan{\scriptstyle{\frac{1}{2}}}\theta\;\sin\varphi.
\end{align}
\end{subequations}
Reconnaissant
le \wiit[Cauchy-Riemann~(critère de)]{critère de Cauchy-Riemann},
nous affirmons que $\gtsMW$ est
une \wiit[fonction ! méromorphe]{fonction méromorphe}
\cite{Whittaker}
de la variable complexe
$\gtsrsw\equiv\gtsrswx+\gtsSign\ci\gtsrswy$,
le signe $\gtsSign$ correspondant au signe
de la \wiit[indice de Pontrjagin]{charge topologique} $\gtsPontrjagin$.
Par conséquent, il existe une infinité
de \wiit[configuration ! topologique auto-duale]%
  {configuration topologique auto-duale}
sur la \wi[sphère ! magnétique rigide]{sphère magnétique rigide};
la configuration topologique auto-duale
$\gtsMW\left(\gtsrsw\right)=\gtsrsw$
est communément
appelée la \wibf[configuration ! \guillemets{hérisson}]%
  {\guillemets{\emph{configuration hérisson}}}
\cite{Fradkin,Thouless}.

\begin{emphasise}
Par la suite,
la recherche de
\wibf[configuration ! symétrique]{configurations symétriques}
permet de maintenir les calculs abordables.
Dans le cas présent,
il s'agit de choisir arbitrairement
un des axes de rotation de la sphère spontanée
puis de rechercher
les \wiit[configuration ! symétrique]{configurations symétriques}
suivant cet axe.
\end{emphasise}

Compte tenu des conventions déjà utilisées
et en introduisant la nouvelle coordonnée
$\zeta\equiv{\pi/2}-\theta$
afin de souligner la symétrie par rapport au plan équatorial,
les \wiit[configuration ! symétrique]{configurations symétriques}
sont ici telles que
\begin{equation}
\label{S/sym/OP/dif}
\gtsMTheta_{\varphi}=\gtsMPhi_{\zeta}=0
\end{equation}
avec comme \wiit[conditions ! aux pôles]{conditions aux pôles}
\begin{equation}
\label{S/sym/OP/pole}
\gtsMTheta\left(\zeta=-{\textstyle\frac{\pi}{2}}\right)=0
\quad\text{et}\quad
\gtsMTheta\left(\zeta=+{\textstyle\frac{\pi}{2}}\right)=0\left[\pi\right].
\end{equation}
Le \wiit[Hamiltonien ! magnétique]{Hamiltonien magnétique}
(\ref{S/Hm/mg/crude})
s'écrit alors
\begin{equation}
\label{S/sym/Hm/mg/crude}
\gtsHm_{mag}=\gtsJ\!%
\int\limits_{-\frac{\pi}{2}}^{+\frac{\pi}{2}}\!\!\gtsd\zeta\!\!
\int\limits_{-\pi}^{+\pi}\!\!\gtsd\varphi\!
\left[\cos\zeta\;\gtsMTheta_{\zeta}^2%
  +\frac{\sin^2\gtsMTheta}{\cos\zeta}\;\gtsMPhi_{\varphi}^2
\right].
\end{equation}
Les \wiit[equations@équations ! d'Euler-Lagrange]{équations d'Euler-Lagrange}
des configurations symétriques
qui extrémisent le Hamiltonien magnétique ($\delta{\gtsHm_{mag}}=0$)
sur la \wiit[sphère ! rigide magnétique]{sphère rigide magnétique}
se déduisent de (\ref{S/sym/Hm/mg/crude}) sans difficulté :
\begin{subequations}
\label{S/sym/Hm/mg/EL}
\begin{align}
\label{S/sym/Hm/mg/EL/MPhi}
\gtsMPhi_{\varphi\varphi}&=0,\\
\label{S/sym/Hm/mg/EL/MTheta}
\cos\zeta\;\gtsd_{\zeta}\!\left[\cos\zeta\:\gtsMTheta_{\zeta}\right]
&=\sin\gtsMTheta\,\cos\gtsMTheta\;\gtsMPhi_{\varphi}^{2}.
\end{align}
\end{subequations}
L'intégration de l'équation (\ref{S/sym/Hm/mg/EL/MPhi})
est immédiate :
\begin{subequations}
\label{S/sym/Hm/mg/EL/int1}
\begin{equation}
\label{S/sym/Hm/mg/EL/int1/MPhi}
\gtsMPhi_{\varphi}=\gtsq_{\varphi}
\qquad
\gtsq_{\varphi}\in\mathbb{Z};
\end{equation}
par conséquent (\ref{S/sym/Hm/mg/EL/MTheta}) devient
\begin{equation}
\label{S/sym/Hm/mg/EL/MTheta/prime}
\cos\zeta\;\gtsd_{\zeta}\!\left[\cos\zeta\:\gtsMTheta_{\zeta}\right]
=\gtsq_{\varphi}^{2}\:\sin\gtsMTheta\,\cos\gtsMTheta.
\end{equation}
\end{subequations}
Cette nouvelle équation (\ref{S/sym/Hm/mg/EL/MTheta/prime})
s'intégre sans effort en multipliant ses deux membres
par $\gtsMTheta_{\zeta}$;
nous lisons
\begin{subequations}
\label{S/sym/Hm/mg/EL/int1/MTheta}
\begin{equation}
\label{S/sym/Hm/mg/EL/int1/MTheta/crude}
\cos^{2}\zeta\;\gtsMTheta_{\zeta}^{2}
=\gtsq_{\varphi}^{2}\:\sin^{2}\gtsMTheta+\gtsm
\qquad
\gtsm\in\left[0,+\infty\right[.
\end{equation}
Or les \wiit[conditions ! aux pôles]{conditions aux pôles}
(\ref{S/sym/OP/pole}) imposent
\begin{equation}
\label{S/sym/Hm/mg/EL/m}
\gtsm=0,
\end{equation}
aussi nous aurions dû lire
\begin{equation}
\label{S/sym/Hm/mg/EL/int1/MTheta/nested}
\cos\zeta\;\gtsMTheta_{\zeta}
=\gtsSign\;\gtsq_{\varphi}\:\sin\gtsMTheta.
\end{equation}
\end{subequations}
Conséquence immédiate :
les \wiit[configuration ! topologique auto-duale]{configurations auto-duales}
recouvrent toutes les
\wiit[configuration ! magnétique symétrique]%
  {configurations magnétiques symétriques}
puisque (\ref{S/sym/Hm/mg/EL/int1/MPhi})
et (\ref{S/sym/Hm/mg/EL/int1/MTheta/nested})
vérifient sans équivoque
l'équation auto-duale (\ref{S/Hm/mg/SD/geo/theta}).
Des calculs analytiques fastidieux
montrent que l'\wiit[description ! algébrique]{écriture algébrique}
des configurations magnétiques auto-duales
est de la forme
\begin{equation}
\label{S/sym/Hm/mg/EL/alg}
\gtsMW\left(\gtsrsw\right)=\gtsrsw^{\gtsq_{\varphi}}.
\end{equation}
Pour $\gtsSign\gtsq_{\varphi}=1$,
nous retrouvons
la \wiit[configuration ! \guillemets{hérisson}]%
  {\guillemets{configuration hérisson}}
\begin{equation}
\label{S/sym/Hm/mg/EL/int2/MTheta/hedgehog}
\gtsMTheta={\textstyle\frac{\pi}{2}}-\zeta=\theta.
\end{equation}
En outre,
des calculs ennuyeux de géométrie différentielle montrent que
les \wiit[deformation@déformation ! symétrique]{déformations symétriques}
de la sphère souple sont d'ordre trois au moins :
les \wiit[configuration ! magnétique symétrique]%
  {configurations magnétiques symétriques}
ne déformerons donc pas
une \wibf[membrane ! souple rigidifiée]{sphère souple suffisamment rigide}
$\left(\gtsJr\approx{1}\right)$.
Autrement dit,
les configurations géométriques symétriques
de la sphère souple magnétique
ne sont pas \wiit[configuration ! frustrée]{frustrées}.

\clearpage
\subsection{Le cylindre infini : des solitons assouvis}
\wih{soliton ! assouvi}
Considérons maintenant
une \wiit[membrane ! souple compactifiée]%
  {membrane magnétique souple compactifiée}
$\gtsCSurface$ toujours de
\wiit[genre topologique ! sphérique]{genre topologique sphérique}
mais dont la forme spontanée
n'admet qu'un seul axe de révolution :
\wih{brisure de symétrie}
une section de cylindre droit.

En \wiit[coordonnées ! cylindriques]{coordonnées cylindriques}
$\left(\rho,\varphi,\gtsz\right)$,
une telle \wiit[membrane ! souple rigidifiée]{membrane rigidifiée}
est parfaitement décrite par
\begin{subequations}
\label{C/coords/cyl}
\begin{gather}
\label{C/coords/cyl/radius}
\rho=r,\\
\label{C/coords/cyl/z}
-\Delta\gtsz\leqslant\gtsz\leqslant+\Delta\gtsz,
\end{gather}
\end{subequations}
où les constantes réelles positives $r$ et $\Delta\gtsz$
sont respectivement le rayon du cylindre
et la demi-longueur de la section.
La métrique de la membrane magnétique est alors de la forme
\begin{equation}
\label{C/g}
g=r^2
\gtsd\varphi\!\otimes\!\gtsd\varphi
  +\gtsd\gtsz\!\otimes\!\gtsd\gtsz;
\end{equation}
aussi $g^{\varphi\gtsz}=g^{\gtsz\varphi}=0$ et nous avons
\begin{equation}
\label{C/g/nested}
g^{\varphi\varphi}\!\sqrt{g}=\frac{1}{r},
\qquad
g^{\gtsz\gtsz}\!\sqrt{g}={r}.
\end{equation}
D'emblée nous étudions
les \wiit[configuration ! magnétique symétrique]%
  {configurations symétriques}
telles que
\begin{equation}
\label{C/sym/OP/dif}
\gtsMTheta_{\varphi}=\gtsMPhi_{\gtsz}=0;
\end{equation}
dans ces conditions
le \wiit[Hamiltonien ! magnétique]{Hamiltonien magnétique}
(\ref{Hm/mg/nested}) s'écrit
\begin{equation}
\label{C/sym/Hm/mg/crude}
\gtsHm_{mag}=\gtsJ\!%
\int\limits_{-\Delta\gtsz}^{+\Delta\gtsz}\!\!\gtsd\gtsz\!\!
\int\limits_{-\pi}^{+\pi}\!\!\gtsd\varphi\!
\left[r\;\gtsMTheta_{\gtsz}^2%
  +\frac{\sin^2\gtsMTheta}{r}\;\gtsMPhi_{\varphi}^2
\right].
\end{equation}
Une expression sans dimension de (\ref{C/sym/Hm/mg/crude})
s'obtient en  effectuant
un changement d'échelle suivant l'axe des $\gtsz$ :
\begin{equation}
\label{C/sym/Hm/mg/scaled}
\gtsHm_{mag}=\gtsJ\!%
\int\limits_{-\Delta\zeta}^{+\Delta\zeta}\!\!\gtsd\zeta\!\!
\int\limits_{-\pi}^{+\pi}\!\!\gtsd\varphi\!
\left[\gtsMTheta_{\zeta}^2%
  +\sin^2\gtsMTheta\;\gtsMPhi_{\varphi}^2
\right],
\end{equation}
avec
\begin{equation}
\label{C/sym/scaling}
\zeta\equiv\gtsz/r
\quad\text{et}\quad
\Delta\zeta\equiv\Delta\gtsz/r.
\end{equation}
Enfin,
la membrane magnétique est \wiit[compactification]{compactifiée}
en imposant une densité Hamiltonienne nulle aux bords :
\wih{conditions ! aux bords}
\begin{subequations}
\label{C/sym/OP/border}
\begin{gather}
\label{C/sym/OP/border/MThetaP}
\gtsMTheta_{\zeta}\left(\zeta=\pm\Delta\zeta\right)=0,\\
\label{C/sym/OP/border/MTheta}
\gtsMTheta\left(\zeta=-\Delta\zeta\right)=0
\quad\text{et}\quad
\gtsMTheta\left(\zeta=+\Delta\zeta\right)=0\left[\pi\right].
\end{gather}
\end{subequations}
Les \wiit[equations@équations ! auto-duales symétriques]%
  {équations auto-duales symétriques}
d'une section du
\wi[cylindre ! magnétique rigide]{cylindre magnétique rigide}
se déduisent facilement
de la fonctionnelle (\ref{C/sym/Hm/mg/scaled}) :
\wih{description ! géométrique}
\begin{equation}
\label{C/sym/Hm/mg/SD/crude}
\gtsMTheta_{\zeta}=\gtsSign\;\sin\gtsMTheta\;\gtsMPhi_{\varphi}.
\end{equation}
En outre
les relations (\ref{C/sym/OP/dif}) imposent
le dédoublement
\begin{subequations}
\label{C/sym/Hm/mg/SD/geo}
\begin{align}
\label{C/sym/Hm/mg/SD/geo/MPhi}
\gtsMPhi_{\varphi}&=\hphantom{\gtsSign}\;\gtsq_{\varphi},\\
\label{C/sym/Hm/mg/SD/geo/MTheta}
\gtsMTheta_{\zeta}&=\gtsSign\;\gtsq_{\varphi}\;\sin\gtsMTheta,
\end{align}
\end{subequations}
avec $\gtsq_{\varphi}\in\mathbb{Z}$.
Enfin
la solution de l'équation (\ref{C/sym/Hm/mg/SD/geo/MTheta})
est le \wiit{soliton}
\begin{subequations}
\label{C/sym/Hm/mg/soliton}
\begin{equation}
\label{C/sym/Hm/mg/soliton/geo}
\gtsMTheta\left(\zeta\right)=%
2\arctan\left[{\exp\left(\gtsSign\gtsq_{\varphi}\zeta\right)}\right].
\end{equation}
Par conséquent
les \wiit[conditions ! aux bords]{conditions aux bords}
(\ref{C/sym/OP/border}) ne sont satisfaites que si
\begin{equation}
\label{C/sym/Hm/mg/soliton/border}
\Delta\zeta=+\infty,
\end{equation}
\end{subequations}
les solitons (\ref{C/sym/Hm/mg/soliton/geo}) recouvrant
alors $\left|\gtsq_{\varphi}\right|$ fois la sphère~unité~$S^2$.
En d'autres termes,
seul le cylindre de longueur infinie admet
des \wiit[configuration ! magnétique auto-duale symétrique]%
  {configurations magnétiques auto-duales symétriques}.
Avant d'approfondir le cas des cylindres de longueur finie,
achevons celui du cylindre infini.

Puisque,
compte tenu du changement d'échelle (\ref{C/sym/scaling}) et
de l'expression
des \wiit[soliton ! magnétique]{solitons magnétiques} (\ref{C/sym/Hm/mg/soliton}),
le rayon $r$ et la coordonnée $\gtsz$
apparaissent respectivement comme
le \wibf[paramètre ! de forme pertinent]{paramètre de forme pertinent}
et la \wibf{coordonnée curviligne pertinente},
le rayon $r$ de la membrane souple rigidifiée
doit être libéré suivant l'axe des $\gtsz$ :
\begin{equation}
\label{C/sym/el}
r\left(\gtsz\right)=%
r_{0}\:\left[1+\gtsTL\left(\gtsz\right)\right],
\end{equation}
où $r_{0}$ représente le \wiit[rayon ! spontané]{rayon spontané}
et la fonction $\gtsTL$
la \wiit[deformation@déformation ! locale]{déformation continue locale}.
\wih{deformation@déformation ! continue}
L'orthogonalité de la métrique rigidifiée est conservée
tandis que les formules (\ref{C/g/nested}) deviennent
\begin{equation}
\label{C/g/el/nested}
g^{\varphi\varphi}\!\sqrt{g}=
  \frac{\sqrt{1+r_{0}^2\gtsTL_{\gtsz}^2}}{r},
\qquad
g^{\gtsz\gtsz}\!\sqrt{g}=
  \frac{r}{\sqrt{1+r_{0}^2\gtsTL_{\gtsz}^2}}.
\end{equation}
Considérant
des \wiit[deformation@déformation ! de faible amplitude]%
  {déformations continues de faible amplitude},
nous développons la densité
du \wiit[Hamiltonien ! elastique@élastique]{Hamiltonien élastique}
(\ref{Hm/el/heuristic/nested})
jusqu'à l'ordre deux
en $\gtsTL$, $\gtsTL_{\gtsz}$ et $\gtsTL_{\gtsz\gtsz}$ :
\begin{equation}
\label{C/sym/Hm/el/nested}
\gtsHm_{el}=
{\textstyle{\frac{1}{8}}}\pi\gtsKc\!
  \int\limits_{-\Delta\zeta}^{+\Delta\zeta}\!\!\!\gtsd\zeta\,\gtsTL^2,
\end{equation}
avec des notations naturelles
et en omettant le terme topologique.
\pagebreak

\begin{emphasise}
Si une des configurations magnétiques
de la membrane souple $\gtsCSurface$ rigidifiée
est \wiit[configuration ! magnétique gelée]{gelée},
le \wiit[Hamiltonien ! magnétique]{Hamiltonien magnétique}
$\gtsHm_{mag}$ de la membrane souple $\gtsCSurface$
devient alors une fonctionnelle
de la seule \wibf[deformation@déformation ! locale]{déformation} $\gtsTL$.
Dans ce contexte
les déformations $\gtsTL$ qui minimisent
le Hamiltonien total $\gtsHm$
permettent de se faire une idée
des \wibf[deformation@déformation ! effective]%
  {déformations effectivement induites}
par la distribution de spins et subies par
la \wiit[membrane ! souple compactifiée]%
  {membrane souple compactifiée} $\gtsCSurface$.
\end{emphasise}

Pour la métrique souple (\ref{C/g/el/nested}),
un développement limité de la densité
du \wiit[Hamiltonien ! magnétique]{Hamiltonien magnétique}
(\ref{Hm/mg/nested})
jusqu'à l'ordre deux en $\gtsTL$ et $\gtsTL_{\gtsz}$
donne :
\begin{multline}
\label{C/sym/Hm/mg/frozen/nested}
\gtsHm_{mag}=
\gtsJ\!
\int\limits_{-\Delta\zeta}^{+\Delta\zeta}\!\!\gtsd\zeta\!\!
\int\limits_{-\pi}^{+\pi}\!\!\gtsd\varphi\,
\left[\!\vphantom{\left[\gtsMTheta_{\zeta}^2\right]^2}
  \left[\gtsMTheta_{\zeta}^2 +\sin^2 \gtsMTheta\ \gtsMPhi_{\varphi}^2
    \right]%
\right.\\
\left.\vphantom{\left[\gtsMTheta_{\zeta}^2\right]^2}
  -\left(-\gtsTL+{\textstyle\frac{1}{2}}\gtsTL_{\zeta}^2\right)
    \left[\gtsMTheta_{\zeta}^2 -\sin^2 \gtsMTheta\ \gtsMPhi_{\varphi}^2
      \right]
  +{\gtsTL^2\:\sin^2 \gtsMTheta\ \gtsMPhi_{\varphi}^2}
\right],
\end{multline}
pour une \wiit[configuration ! magnétique gelée]%
  {configuration magnétique symétrique gelée}
\wih{configuration ! magnétique symétrique}
quelconque.
Dès lors les déformations $\gtsTL$ induites
par les solitons magnétiques (\ref{C/sym/Hm/mg/soliton})
\wiit[configuration ! magnétique gelée]{gelés}
et subies par le cylindre souple magnétique infini
doivent extrémiser le \wiit[Hamiltonien ! total]{Hamiltonien total}
\begin{equation}
\label{C/sym/Hm/infty/frozen/nested}
\gtsHm=
2\pi\gtsJ\gtsq_{\varphi}^2\!
\int\limits_{-\infty}^{+\infty}\!\!\gtsd\zeta\,
\left[
\left[2+\gtsTL^2\right]
\sin^2 \gtsMTheta
+\frac{1}{16\gtsJr\gtsq_{\varphi}^2}\;\gtsTL^2
\right],
\end{equation}
somme du Hamiltonien élastique (\ref{C/sym/Hm/el/nested})
et du Hamiltonien magnétique (\ref{C/sym/Hm/mg/frozen/nested})
pour ces configurations magnétiques.
Or il apparaît clairement que seule
la \wiit[deformation@déformation ! triviale]{déformation triviale}
$\gtsTL=0$
minimise cette fonctionnelle (\ref{C/sym/Hm/infty/frozen/nested}) :
les \wiit[soliton ! magnétique]{solitons magnétiques}
(\ref{C/sym/Hm/mg/soliton})
ne déforment pas le
\wi[cylindre ! souple magnétique infini]{cylindre souple magnétique infini}.

\clearpage
\subsection{Sections de cylindre : déformation solitonique}
\label{chp/HSES/HSECS}
\wih{deformation@déformation ! solitonique}
Dans la section précédente,
nous montrons qu'il n'existe pas
de configuration magnétique symétrique auto-duale
sur les sections finies de cylindre magnétique rigide :
ce n'est pas pour autant qu'il n'existe pas
de \wiit[configuration ! topologique symétrique]%
  {configuration topologique symétrique}.

\enlargethispage*{4\baselineskip}
De telles configurations doivent satisfaire
les \wiit[conditions ! aux bords]{conditions aux bords} (\ref{C/sym/OP/border})
et minimiser
le \wiit[Hamiltonien ! magnétique]{Hamiltonien magnétique}
(\ref{C/sym/Hm/mg/scaled}),
c'est-à-dire vérifier
les \wiit[equations@équations ! d'Euler-Lagrange]%
  {équations d'Euler-Lagrange}
\wih{description ! géométrique}
\begin{subequations}
\label{C/sym/Hm/mg/EL}
\begin{align}
\label{C/sym/Hm/mg/EL/MPhi}
\gtsMPhi_{\varphi\varphi}&=0,\\
\label{C/sym/Hm/mg/EL/MTheta}
\gtsMTheta_{\zeta\zeta}
&=\sin\gtsMTheta\,\cos\gtsMTheta\;\gtsMPhi_{\varphi}^{2}.
\end{align}
\end{subequations}
L'équation (\ref{C/sym/Hm/mg/EL/MPhi}) s'intégrant sans effort,
nous écrivons
\begin{subequations}
\label{C/sym/Hm/mg/EL/nested}
\begin{align}
\label{C/sym/Hm/mg/EL/nested/MPhi}
\gtsMPhi_{\varphi\hphantom{\zeta}}&=\gtsq_{\varphi},\\
\label{C/sym/Hm/mg/EL/nested/MTheta}
\gtsMTheta_{\zeta\zeta}&=\gtsq_{\varphi}^2\;\sin\gtsMTheta\,\cos\gtsMTheta,
\end{align}
\end{subequations}
avec $\gtsq_{\varphi}\in\mathbb{Z}$
le nombre de recouvrement de la sphère~unité~$S^2$
autour de l'axe de révolution.
Notons au passage que ce système dédoublé (\ref{C/sym/Hm/mg/EL/nested})
reproduit le dédoublement auto-dual (\ref{C/sym/Hm/mg/SD/geo}).
De surcroît
l'équation (\ref{C/sym/Hm/mg/EL/nested/MTheta}),
au changement d'échelle $\varrho\equiv\gtsq_{\varphi}\zeta$ près,
n'est rien d'autre que
l'\wibf[sinus-Gordon ! equation@équation simple de]%
  {équation simple de sinus-Gordon}~(\textsc{sg})\,%
\footnote{\label{fnt/SG}%
Le lecteur est invité à consulter l'\appendixname~\ref{app/SG}
pour se familiariser (rapidement)
avec l'\witextit{équation de sinus-Gordon}
et les notations utilisées.\\
Néanmoins, pour le lecteur averti :
$\gtsSG\left(\cdot\mid\gtsm\right)$
correspond à la solution croissante de
l'\witextit{équation simple de sinus-Gordon} (\textsc{sg})
localisée à l'origine et
caractérisée par le \emph{paramètre} $\gtsm$,
la \witextit{quasi quart-période de sinus-Gordon} $\gtsK$
vérifiant $\gtsK\left(\gtsm\right)=\EllipticK\left(1+\gtsm\right)$
avec $\EllipticK$ l'intégrale elliptique de première espèce.
};
la solution de l'équation (\ref{C/sym/Hm/mg/EL/nested/MTheta})
est donc le \wibf[soliton ! périodique]{soliton périodique}
\begin{subequations}
\label{C/sym/Hm/mg/EL/soliton}
\begin{equation}
\label{C/sym/Hm/mg/EL/soliton/geo}
\gtsMTheta\left(\zeta\right)=%
\gtsSign\gtsSG\left(\gtsq_{\varphi}\zeta\mid\gtsm\right)%
-{\textstyle{\frac{\pi}{2}}}\:\delta_{\text{paire},\gtsq_{\zeta}},
\end{equation}
avec $\gtsq_{\zeta}\in\mathbb{Z}$
le nombre de recouvrement de la sphère~unité~$S^2$
le long de l'axe de révolution.
Les \wiit[conditions ! aux bords]{conditions aux bords}
(\ref{C/sym/OP/border/MTheta}) ne sont satisfaites que si
le \wibf[sinus-Gordon ! paramètre]{paramètre} $\gtsm$ vérifie
\begin{equation}
\label{C/sym/Hm/mg/EL/soliton/border}
\gtsm=\gtsK^{-1}\left(
    \left|\frac{\gtsq_{\varphi}}{\gtsq_{\zeta}}\right|\Delta\zeta%
  \right),
\end{equation}
les \wiit[soliton ! périodique]{solitons périodiques}
  (\ref{C/sym/Hm/mg/EL/soliton/geo})
ayant alors la \wiit[indice de Pontrjagin]{charge topologique}
\begin{equation}
\label{C/sym/Hm/mg/EL/soliton/Q}
\gtsQ=\gtsq_{\varphi}\gtsq_{\zeta}.
\end{equation}
\end{subequations}
Observons pour finir que les résultats du
\wiit[cylindre ! rigide infini]{cylindre rigide infini}
se retrouvent aisément
en faisant tendre $\Delta\zeta$ vers l'infini,
tandis que le rayon $r$ et la coordonnée $\gtsz$
\wih{paramètre ! de forme pertinent}
\wih{coordonnée curviligne pertinente}
gardent leur pertinence.
\pagebreak

\begin{figure}[p]
\parbox{.35\linewidth}{%
  \begin{center}
    \includegraphics[height=\textheight]{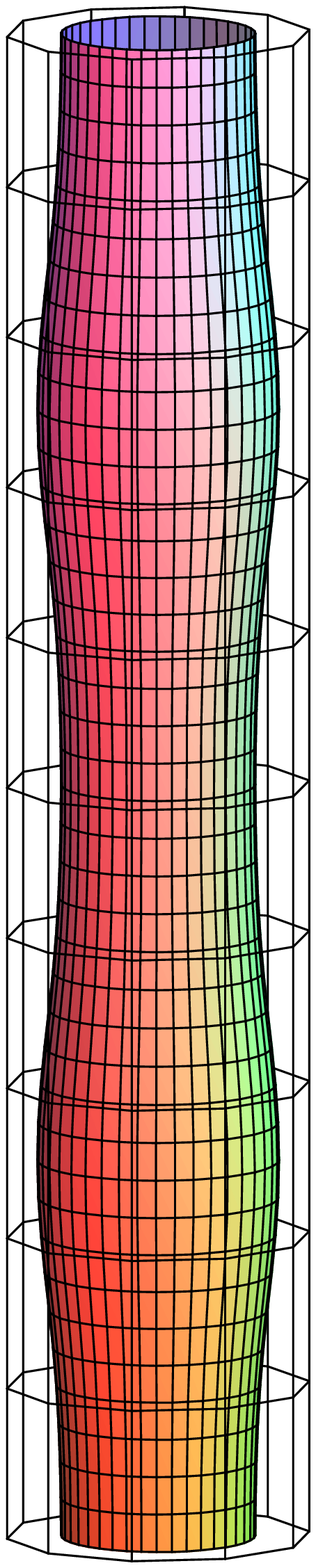}
  \end{center}
  }
\parbox[b]{.65\linewidth}{%
  \caption[Section de cylindre magnétique souple : déformation]{%
    \wiit[deformation@déformation ! symétrique]{Déformation symétrique}
    d'une section de cylindre magnétique souple
    en présence d'un $2\pi$-soliton magnétique
    (\ref{C/sym/Hm/mg/EL/soliton}) :
    l'unité de longueur est le rayon $r$ de la section rigidifiée
    (représentée par la grille extérieure),
    la demi-longueur relative $\Delta\zeta\equiv\Delta\gtsz/r$
    est égale à $5$,
    la \wiit{constante relative de couplage}
    $\gtsJr\equiv\gtsJ/\gtsKc$ à $1/2$,
    enfin la \wiit[deformation@déformation ! locale]{déformation locale} $\gtsTL$
    est magnifiée par un facteur $10$.\\
    Le \wiit[soliton ! frustré]{soliton frustré} se libère
    en rétrécissant globalement la membrane magnétique;
    les deux renflements correspondent aux deux vagues du soliton :
    son énergie magnétique $\gtsEg_{mag}$ s'y concentrant
    indépendamment de la géométrie de la membrane,
    il refoule l'énergie magnétique accumulée en ses branches.
    }
  }
  \label{fig/ECS}
\end{figure}
\clearpage

Afin d'étudier la \wiit[frustration ! géométrique]{frustration géométrique}
des solitons magnétiques (\ref{C/sym/Hm/mg/EL/soliton}),
introduisons $\gtsEr_{mag}$ le rapport
de leur \wiit[energie@énergie ! magnétique]{énergie magnétique} $\gtsEg_{mag}$
sur leur \wiit[energie@énergie ! topologique]{énergie topologique}
$\gtsTE_{\gtsPontrjagin}$ (\ref{BPB/Em/mg}) :
\begin{equation}
\label{C/sym/Hm/mg/EL/soliton/Er/def}
\gtsEr_{mag}\equiv\frac{\gtsEg_{mag}}{\gtsTE_{\gtsPontrjagin}}.
\end{equation}
Le calcul explicite de $\gtsEr_{mag}$ donne
\begin{subequations}
\label{C/sym/Hm/mg/EL/soliton/Er}
\begin{align}
\gtsEr_{mag}
\label{C/sym/Hm/mg/EL/soliton/Er/nested}
&=\EllipticE\left(1+\gtsm\right)+
  {\textstyle\frac{1}{2}}\gtsm\:
  \EllipticK\left(1+\gtsm\right),\\
\label{C/sym/Hm/mg/EL/soliton/Er/conv}
&=\sqrt{1+\gtsm}\left[
  \EllipticE\left(\frac{1}{1\!+\!\gtsm}\right)\!-\,
  {\textstyle\frac{1}{2}}\frac{\gtsm}{1\!+\!\gtsm}\:
  \EllipticK\left(\frac{1}{1\!+\!\gtsm}\right)
\right],
\end{align}
\end{subequations}
où les fonctions $\EllipticK$ et $\EllipticE$ désignent
respectivement l'intégrale elliptique de première
et de seconde espèce \cite{Whittaker,Lawden}.
L'énergie relative $\gtsEr_{mag}$
des solitons magnétiques (\ref{C/sym/Hm/mg/EL/soliton})
ne dépend ainsi que du \wiit[sinus-Gordon ! paramètre]{paramètre} $\gtsm$.
Qui plus est,
$\gtsEr_{mag}$ croît strictement de $1$ vers $\infty$
lorsque $\gtsm$ varie de $0$ vers $\infty$.
Le paramètre $\gtsm$ se révèle donc comme
le \wibf[paramètre ! pertinent]{paramètre pertinent}
de la \wiit[frustration ! géométrique]{frustration géométrique} :
plus $\gtsm$ est grand
plus les solitons magnétiques (\ref{C/sym/Hm/mg/EL/soliton})
sont \wiit[soliton ! frustré]{frustrés},
le cas $\gtsm=0$ correspondant à un système assouvi.
Rétrospectivement
la relation (\ref{C/sym/Hm/mg/EL/soliton/border})
permet de mieux saisir le mécanisme de frustration
des sections finies de cylindre magnétique rigide :
les solitons magnétiques (\ref{C/sym/Hm/mg/EL/soliton})
tendent à se retourner autour de l'axe de révolution
($\left|\gtsq_{\varphi}\right|$~grand)
et à s'étendre le long de l'axe de révolution
($\left|\gtsq_{\zeta}\right|$~petit et $\Delta\zeta$~grand).
Conséquence immédiate :
les solitons magnétiques (\ref{C/sym/Hm/mg/EL/soliton})
sont frustrés parce que \wibf[soliton ! confiné]%
  {confinés}~($\Delta\zeta<\infty$).

Maintenant les \wiit[deformation@déformation ! locale]{déformations} $\gtsTL$
subies par les sections de cylindre souples magnétiques
et induites par les solitons magnétiques frustrés
(\ref{C/sym/Hm/mg/EL/soliton})
\wiit[configuration ! magnétique gelée]{gelés}
doivent extrémiser
le \wi[Hamiltonien ! total]{Hamiltonien total}
\begin{multline}
\label{C/sym/Hm/gen/frozen/nested}
\gtsHm=
2\pi\gtsJ\gtsq_{\varphi}^2\!
\int\limits_{-\Delta\zeta}^{+\Delta\zeta}\!\!\gtsd\zeta\,
\left[\vphantom{\left(\frac{1}{16\gtsJr\gtsq_{\varphi}^2}\right)^2}
\left(2+\gtsm\right)
-\gtsm\left[-\gtsTL+{\textstyle\frac{1}{2}}\gtsTL_{\zeta}^2\right]
\right.\\
\left.\vphantom{\left(\frac{1}{16\gtsJr\gtsq_{\varphi}^2}\right)^2}
-\left(1+\gtsm\right)\left[2+\gtsTL^2\right]
\JacobiSN^2\left(\gtsq_{\varphi}\zeta\mid{1+\gtsm}\right)
+
\left(1+\frac{1}{16\gtsJr\gtsq_{\varphi}^2}\right)\!\gtsTL^2
\right],
\end{multline}
qui doit être lu comme la version frustrée
du Hamiltonien total (\ref{C/sym/Hm/infty/frozen/nested}).
Les \wiit[equations@équations ! d'Euler-Lagrange]{équations d'Euler-Lagrange}
de cet Hamiltonien (\ref{C/sym/Hm/gen/frozen/nested})
prennent la forme
\begin{subequations}
\label{C/sym/Hm/gen/frozen/EL}
\begin{equation}
\label{C/sym/Hm/gen/frozen/EL/eqn}
\gtsTL_{\varrho\varrho}+
\left[\left(1+\gtsLm\right)\!\gtsLA
-\gtsLm\gtsLB\;\JacobiSN^2\left(\varrho\mid\gtsLm\right)
\right]\gtsTL
=\sqrt{\gtsLm}\ \gtsLj,
\end{equation}
où nous avons posé
\begin{align}
\gtsLm&=1+\gtsm,\\
\gtsLA&=\frac{2}{\gtsm\gtsq_{\varphi}^2}
\left[1+\frac{1}{16\gtsJr\gtsq_{\varphi}^2}
\right],\\
\gtsLB&=\frac{2}{\gtsm\gtsq_{\varphi}^2},\\
\gtsLj&=\frac{-1}{\gtsq_{\varphi}^2\sqrt{1+\gtsm}}.
\end{align}
\end{subequations}
Ici nous avons également introduit
la nouvelle variable $\varrho\equiv\gtsq_{\varphi}\zeta$.
L'équation différentielle linéaire inhomogène du second ordre
(\ref{C/sym/Hm/gen/frozen/EL})
n'est rien d'autre
l'\wibf[equation@équation ! inhomogène de Lamé]%
  {équation inhomogène de Lamé}~(\textsc{il})
avec un terme inhomogène constant\,%
\footnote{\label{fnt/IL}%
Le lecteur est encouragé à lire l'\appendixname~\ref{app/IL}
qui aborde
l'\witextit{équation inhomogène de Lamé}~(\textsc{il})
en se bornant au contexte rencontré.\\
Toutefois, pour le lecteur averti :
$\gtsIL\left(\cdot\mid\Jacobim;\gtsLA,\gtsLB,\gtsLj\right)$
désignera la fonction (bornée) associée à la solution minimale
de l'équation aux différences linéaire du second ordre déduite
de l'\wiit[equation@équation ! inhomogène de Lamé]%
  {équation inhomogène de Lamé}~(\textsc{il})
suivant la méthode communément utilisée
pour construire les fonctions dites de Lamé
(solutions de l'\wiit[equation@équation ! homogène de Lamé]%
  {équation homogène de Lamé}).
};
une solution raisonnable
de l'équation (\ref{C/sym/Hm/gen/frozen/EL})
est donc la \wibf[deformation@déformation ! solitonique]{déformation solitonique}
\begin{equation}
\label{C/sym/Lame}
\gtsTL\left(\zeta\right)=
\gtsIL\left(
    \gtsq_{\varphi}\ \zeta\mid 1+\gtsm;\gtsLA,\gtsLB,\gtsLj
  \right).
\end{equation}
La \wibf[frustration ! géométrique]{frustration géométrique}
apparaît ostensiblement
dans l'équation d'Euler-Lagrange (\ref{C/sym/Hm/gen/frozen/EL}) :
$\gtsTL=0$ n'est pas solution.
Concentrons-nous à présent sur le mécanisme de déformation.
D'après les relations
(\ref{C/sym/scaling}), (\ref{C/sym/Hm/mg/EL/soliton/border})
et (\ref{C/sym/Hm/mg/EL/soliton/Er})
nous savons que
l'\wiit[energie@énergie ! magnétique]{énergie magnétique}
$\gtsEg_{mag}$ des solitons (\ref{C/sym/Hm/mg/EL/soliton})
décroît en fonction du rayon $r$ :
les solitons tendent à rétrécir radialement
la section de cylindre souple magnétique.
D'un autre côté
le Hamiltonien élastique (\ref{C/sym/Hm/el/nested})
essaie de maintenir sa \wi[forme ! spontanée]{forme spontanée}.
Par conséquent
la compétition entre
l'\witextit{énergie magnétique} et l'\witextit{énergie élastique}
\wih{energie@énergie ! magnétique}\wih{energie@énergie ! elastique@élastique}
se traduit par une réduction du rayon $r$.
Or l'\wiit[energie@énergie ! soliton@du soliton]{énergie du soliton}
\swih[energie@énergie ! soliton@du soliton]{soliton}
est essentiellement localisée sur sa \wiit[soliton ! vague du]{vague}
\swih[soliton]{vague}
($\gtsMTheta$ proche de ${\textstyle{\frac{\pi}{2}}}$)
selon (\ref{C/sym/Hm/mg/scaled}) :
la déformation est ainsi moins importante sur la vague du soliton
puisque l'énergie élastique est isotrope.
Pour interpréter les conséquences géométriques
de cette \wibf{libération géométrique} définissons
la \wibf[dilatation relative ! du cylindre]%
  {dilatation relative du cylindre}
$\gtsrd_{C}$ par
\begin{equation}
\label{C/rd/def}
\gtsrd_{C}\equiv\frac{r}{r_{0}}=1+\gtsTL.
\end{equation}
\begin{figure}[th]
  \begin{center}
  \includegraphics[width=.98\linewidth]{sgtsCrdDeltazeta.mps}
  \caption[Sections de cylindre magnétiques souples : dilatation]{%
    La \wiit[dilatation relative ! du cylindre]{dilatation relative} $\gtsrd_{C}$
    de sections de cylindre magnétiques souples en présence
    d'un $\pi$-soliton magnétique (\ref{C/sym/Hm/mg/EL/soliton}) :
    la dilatation relative $\gtsrd_{C}$ définie par (\ref{C/rd/def})
    est représentée en fonction de $\zeta/\Delta\zeta=\gtsz/\Delta\gtsz$
    pour différentes demi-longueurs relatives $\Delta\zeta=\Delta\gtsz/r$,
    la \wiit{constante relative de couplage}
    étant fixée $\left(\gtsJr\equiv\gtsJ/\gtsKc=1/16\right)$.\\
    $\bullet$ Le cas dégénéré $\left(\Delta\zeta=\infty\right)$ correspond
    au cylindre magnétique souple infini :
    le soliton magnétique n'étant pas \wiit[soliton ! confiné]{confiné}
    peut s'étendre sans restriction aucune
    tout le long de l'axe de révolution
    et son énergie magnétique $\gtsEg_{mag}$ atteindre
    son \wiit[energie@énergie ! mini@\textlatin{minimum minimorum} topologique]%
      {énergie \textlatin{minimum minimorum} topologique}
    $\gtsTE_{\gtsPontrjagin}$.\\
    $\bullet$ Les autres cas $\left(\Delta\zeta<\infty\right)$ illustrent
    le mécanisme de \wiit{libération géométrique} :
    le soliton magnétique \wiit[soliton ! confiné]{confiné}
    ne pouvant pas s'épanouir suivant l'axe de rotation
    refoule son excès d'énergie magnétique
    $\Delta\gtsEg_{mag}=\gtsEg_{mag}-\gtsTE_{\gtsPontrjagin}$
    en rétrécissant radialement la membrane magnétique souple;
    l'énergie du soliton étant localisée en sa vague
    et l'énergie élastique isotrope,
    les déformations sont plus prononcées en ses branches.
    }
  \label{fig/C/Dz}
  \end{center}
\end{figure}

\section{Membranes de topologie torique}
\wih{genre topologique ! torique}
\subsection{Le tore : une dégénérescence solitonique}
\label{chp/HSES/HSRT}
La représentation la plus naturelle du tore
est donnée en \wiit[coordonnées ! cylindriques]{coordonnées cylindriques}
$\left(\rho,\xi,\gtsz\right)$ :
\begin{equation}
\label{T/coords/n}
\rho=R+r\cos\varphi,
\qquad
\gtsz=r\sin\varphi,
\end{equation}
où le \wiit[rayon ! revolution@de révolution]{rayon de révolution} $R$
et le \wiit[rayon ! axiale]{rayon axiale} $r$ vérifient $0<r<R$,
l'angle $\varphi$ variant de $-\pi$ à $\pi$.
Toutefois,
il s'avère plus commode d'utiliser
la représentation suivante
\cite{Zhongcan}
\begin{equation}
\label{T/coords/pp}
\rho=\frac{a\sinh b}{\cosh b -\cos \eta},
\qquad
\gtsz=\frac{a\sin \eta}{\cosh b -\cos \eta},
\end{equation}
où les nouveaux paramètres constants $a$ et $b$
sont des réels positifs,
tandis que l'\wiit[angle ! poloïdale]{angle poloïdale} $\eta$
varie de $-\pi$ à $\pi$.
Les relations
\begin{equation}
\label{T/coords/pp/par}
a=\sqrt{\!\left(R+r\right)\!\left(R-r\right)}
\quad\text{et}\quad
\cosh b=\frac{R}{r}
\end{equation}
fournissent une interprétation géométrique
immédiate des nouveaux paramètres :
appelons $a$ le \wibf[rayon ! géométrique]{rayon géométrique}
et $b$ l'\wibf[angle ! d'excentricité]{angle d'excentricité}.
Réciproquement,
les paramètres naturels $R$ et $r$ sont donnés par
\begin{equation}
\label{T/coords/n/par}
R=\frac{a}{\tanh b}
\quad\text{et}\quad
r=\frac{a}{\sinh b}.
\end{equation}
Enfin,
la transformation entraîne la relation
\begin{subequations}
\label{T/coords/n/pp}
\begin{align}
\tan{\scriptstyle{\frac{1}{2}}}\eta
&=\tanh{\scriptstyle{\frac{1}{2}}}b\;%
  \tan{\scriptstyle{\frac{1}{2}}}\varphi,\\
&=\sqrt{\frac{R-r}{R+r}}\tan{\scriptstyle{\frac{1}{2}}}\varphi.
\end{align}
\end{subequations}
La métrique de la membrane magnétique
en \wiit[coordonnées ! péri-polaires]{coordonnées péri-polaires}
$\left(\xi,\eta\right)$ s'écrit alors
\begin{equation}
\label{T/g}
g=\frac{a^2}{\left(\cosh b -\cos \eta\right)^2}
\left[
  \sinh^2 b\ \gtsd\xi\!\otimes\!\gtsd\xi+\gtsd\eta\!\otimes\!\gtsd\eta
\right],
\end{equation}
ainsi $g^{\xi\eta}=g^{\eta\xi}=0$ et nous avons
\begin{equation}
\label{T/g/nested}
g^{\xi\xi}\!\sqrt{g}=\frac{1}{\sinh b},
\qquad g^{\eta\eta}\!\sqrt{g}={\sinh b}.
\end{equation}
Pour le tore rigide l'expression (\ref{Hm/mg/nested})
du \wiit[Hamiltonien ! magnétique]{Hamiltonien magnétique}
s'écrit donc littéralement
\begin{equation}
\label{T/Hm/mg/crude}
\gtsHm_{mag}\!=\!\gtsJ\!%
\int\limits_{-\pi}^{+\pi}\!\!\gtsd\xi\!\!
\int\limits_{-\pi}^{+\pi}\!\!\gtsd\eta\!
\left[\!\vphantom{\left[\frac{\sin^2\gtsMTheta}{\sin\theta}\right]^2}
\left[\frac{\gtsMTheta_{\xi}^2}{\sinh b}%
  +{\sinh b}\:\sin^2\gtsMTheta\;\gtsMPhi_{\eta}^2\right]
\!+\!
\left[{\sinh b}\;\gtsMTheta_{\eta}^2%
  +\frac{\sin^2\gtsMTheta}{\sinh b}\;\gtsMPhi_{\xi}^2\right]
\!\right].
\end{equation}
Un changement d'échelle approprié suivant l'axe des $\gtsz$ donne
\begin{equation}
\label{T/Hm/mg/scaled}
\gtsHm_{mag}\!=\!\gtsJ\!%
\int\limits_{-\Delta\zeta}^{+\Delta\zeta}\!\!\gtsd\zeta\!\!
\int\limits_{-\pi}^{+\pi}\!\!\gtsd\eta\!
\left[\!\vphantom{\left[\gtsMPhi_{\xi}^2\right]^2}
\left[\gtsMTheta_{\zeta}^2+\sin^2\gtsMTheta\;\gtsMPhi_{\eta}^2\right]
\!+\!
\left[\gtsMTheta_{\eta}^2+\sin^2\gtsMTheta\;\gtsMPhi_{\zeta}^2\right]
\!\right],
\end{equation}
avec
\begin{equation}
\label{T/Hm/mg/scaling}
\zeta\equiv{\sinh b}\;\xi
\quad\text{et}\quad
\Delta\zeta\equiv{\sinh b}\;\pi.
\end{equation}
De cet Hamiltonien (\ref{T/Hm/mg/scaled}) nous déduisons
les \wiit[equations@équations ! auto-duales]{équations auto-duales}
du tore rigide magnétique :
\wih{description ! géométrique}
\begin{subequations}
\label{T/Hm/mg/SD/geo}
\begin{align}
\label{T/Hm/mg/SD/geo/theta}
\gtsMTheta_{\zeta}&=%
  \hphantom{-}\gtsSign\;\sin\gtsMTheta\;\gtsMPhi_{\eta},\\
\label{T/Hm/mg/SD/geo/phi}
\gtsMTheta_{\eta}&=%
  -\gtsSign\;\sin\gtsMTheta\;\gtsMPhi_{\zeta}.
\end{align}
\end{subequations}
Sans effort nous écrivons
leur \wiit[description ! algébrique]{versions algébriques}
\begin{subequations}
\label{T/Hm/mg/SD/alg}
\begin{align}
\label{T/Hm/mg/SD/alg/re}
\gtsMWx_{\zeta}&=+\gtsSign\;\gtsMWy_{\eta},\\
\label{T/Hm/mg/SD/alg/im}
\gtsMWx_{\eta}&=-\gtsSign\;\gtsMWy_{\zeta}.
\end{align}
\end{subequations}
En vertu du
\wiit[Cauchy-Riemann~(critère de)]{critère de Cauchy-Riemann},
$\gtsMW$ est une \wiit[fonction ! méromorphe]{fonction méromorphe}
\cite{Whittaker}
de la variable complexe
$\omega\equiv\zeta+\gtsSign\ci\eta$,
le signe $\gtsSign$ étant le signe
de la charge topologique~$\gtsPontrjagin$.
Introduisons le réseau
$\gtsLattice(2\Delta\zeta,\ci 2\pi)\equiv
\{2m\Delta\zeta+\ci 2n\pi\mid m,n\in\mathbb{Z}\}$
et notons que
le tore est homéomorphe à
${\mathbb{C}}/\gtsLattice(2\Delta\zeta,\ci 2\pi)$ :
$\gtsMW$ est donc une \wiit[fonction ! elliptique]{fonction elliptique}
\cite{Whittaker,Lawden}.
Par conséquent la \wi[indice de Pontrjagin]{charge topologique}
$\gtsPontrjagin$ de la configuration $\gtsMW$ correspond
à l'\witextit{ordre} de la fonction elliptique $\gtsMW$
\cite{Whittaker}, au signe près.
Or l'ordre d'une fonction elliptique
est toujours supérieur ou égal à deux;
nous avons donc
\begin{equation}
\label{T/Q}
\left|\gtsPontrjagin\right|\geqslant2
\end{equation}
en accord
avec le \wiit[Eells-Wood~(théorème de)]{théorème de Eells-Wood}
\cite{ROHM}.
En outre,
la représentation des fonctions elliptiques à l'aide des
\wiit[sigmafonction@$\sigma$-fonctions de Weierstrass]%
  {$\sigma$-fonctions de Weierstrass}
\cite{Whittaker,Lawden}
autorise une écriture formelle
des configurations topologiques auto-duales :
\wih{configuration ! topologique auto-duale}
\begin{subequations}
\label{T/Hm/mg/SD/formal}
\begin{equation}
\label{T/Hm/mg/SD/formal/gen}
\gtsMW\left(\omega\right)=
\tan{\scriptstyle{\frac{1}{2}}}\gtsMTheta_{0}\;
  \gtse^{\ci\gtsMPhi_{0}}\:
\prod_{n=1}^{\left|\gtsPontrjagin\right|}\!
  \frac{\WeierstrassSigma\left(\omega-z_{n}\right)}
    {\WeierstrassSigma\left(\omega-p_{n}\right)},
\end{equation}
où les zéros $z_{n}$ et les pôles $p_{n}$
doivent vérifier la règle de sélection
\begin{equation}
\label{T/Hm/mg/SD/formal/cond}
\sum_{n=1}^{\vert\gtsQ\vert}z_{n}=
\sum_{n=1}^{\vert\gtsQ\vert}p_{n}.
\end{equation}
Dans la foulé,
écrivons leur densité
\wiit[Hamiltonien ! magnétique]{Hamiltonienne magnétique} $\gtsHmd_{mag}$
à l'aide des \wiit[zetafonction@$\zeta$-fonctions de Weierstrass]%
  {$\zeta$-fonctions de Weierstrass} :
\begin{align}
\gtsHmd_{mag}
&=\gtsJ\!
\left[
  \frac{\left|\gtsMW_{\omega}\right|}%
    {1+{\left|\gtsMW\right|}^2}
\right]^2\nonumber\\
\label{T/Hmd/mg/SD/formal/gen}
&=\gtsJ\!
\left[
  \frac{\left|\gtsMW\right|}{1+{\left|\gtsMW\right|}^2}
\right]^2
\left|
  \sum_{n=1}^{\left|\gtsPontrjagin\right|}%
    \WeierstrassZeta(\omega-z_{n})-%
  \sum_{n=1}^{\left|\gtsPontrjagin\right|}%
    \WeierstrassZeta(\omega-p_{n})
\right|^2.
\end{align}
\end{subequations}

\subsection{Sections de tore : déformation solitonique}
\wih{deformation@déformation ! solitonique}
Restreignons nous maintenant
à une membrane magnétique souple compactifiée $\gtsCSurface$
dont la forme spontanée est une section de tore.

\enlargethispage*{2\baselineskip}
En \wiit[coordonnées ! péri-polaires]{coordonnées péri-polaires}
$\left(\xi,\eta\right)$ précédemment introduites,
une telle membrane rigidifiée
\wih{membrane ! souple rigidifiée}
est décrite en restreignant l'angle de rotation :
\begin{subequations}
\label{TS/coords/pp}
\begin{equation}
\label{TS/coords/pp/xi}
-\Delta\xi\leqslant\xi\leqslant+\Delta\xi,
\end{equation}
où le demi-angle $\Delta\xi$ doit vérifier
\begin{equation}
\label{TS/coords/pp/Delta}
0<\Delta\xi<\pi.
\end{equation}
\end{subequations}
Encore une fois nous nous bornerons
à étudier les \wiit[configuration ! magnétique symétrique]%
  {configurations symétriques} :
\begin{equation}
\label{TS/sym/OP/dif}
\gtsMTheta_{\eta}=\gtsMPhi_{\xi}=0;
\end{equation}
le \wiit[Hamiltonien ! magnétique]{Hamiltonien magnétique}
(\ref{Hm/mg/nested}) prend alors la forme attendue
\begin{equation}
\label{TS/sym/Hm/mg/scaled}
\gtsHm_{mag}\!=\!\gtsJ\!%
\int\limits_{-\Delta\zeta}^{+\Delta\zeta}\!\!\gtsd\zeta\!\!
\int\limits_{-\pi}^{+\pi}\!\!\gtsd\eta\!
\left[\gtsMTheta_{\zeta}^2+\sin^2\gtsMTheta\;\gtsMPhi_{\eta}^2\right],
\end{equation}
avec
\begin{equation}
\label{TS/sym/Hm/mg/scaling}
\zeta\equiv{\sinh b}\;\xi
\quad\text{et}\quad
\Delta\zeta\equiv{\sinh b}\;\Delta\xi.
\end{equation}
\clearpage\noindent
Enfin, la section de tore est \wiit[compactification]{compactifiée}
en imposant une densité Hamiltonienne nulle aux bords :
\wih{conditions ! aux bords}
\begin{subequations}
\label{TS/sym/OP/border}
\begin{gather}
\label{TS/sym/OP/border/MThetaP}
\gtsMTheta_{\zeta}\left(\zeta=\pm\Delta\zeta\right)=0,\\
\label{TS/sym/OP/border/MTheta}
\gtsMTheta\left(\zeta=-\Delta\zeta\right)=0
\quad\text{et}\quad
\gtsMTheta\left(\zeta=+\Delta\zeta\right)=0\left[\pi\right].
\end{gather}
\end{subequations}

Il saute aux yeux que le \wi{magnétisme}
des sections de tore magnétiques souples rigidifiées
est similaire à celui
des sections de cylindre magnétiques souples rigidifiées
étudié dans la {\subsectionname}~\ref{chp/HSES/HSECS} :
seule la forme spontanée des membranes souples diffère.
Autrement dit,
les géométries ne sont pas similaires
mais
le choix des métriques (\ref{C/g}) et (\ref{T/g})
rend le magnétisme similaire.
Par contre
la différence entre les formes spontanée est intrinsèque :
les \wiit[courbure ! principale]{courbures principales} sont
globales (\textlatin{i.e.} constantes)
pour les sections de cylindre,
locales pour les sections de tore.
Il en découle
des \wiit[paramètre ! de forme pertinent]{paramètres de forme pertinents}
de nature différente.
Pour les sections de cylindre,
le paramètre de forme pertinent
est le rayon $r$ : une distance.
Pour les sections de tore,
le paramètre de forme pertinent
est l'\wiit[angle ! d'excentricité]{angle d'excentricité} $b$ :
un rapport de distance.
Nous nous attendons donc à trouver
des lois de déformation similaires
mais aux conséquences géométriques dissemblables :
nous allons montrer qu'elles sont semblables.

D'après (\ref{TS/sym/Hm/mg/scaling}) et par analogie,
l'angle d'excentricité $b$
(le \wiit[paramètre ! de forme pertinent]{paramètre de forme pertinent})
doit être libéré en fonction
de l'angle $\xi$
(la \wiit{coordonnée curviligne pertinente}) :
\begin{equation}
\label{TS/sym/el}
b\left(\xi\right)=b_{0}+\gtsTL\left(\xi\right),
\end{equation}
avec $b_{0}$ l'\wiit[angle ! d'excentricité spontané]{angle d'excentricité spontané}
et $\gtsTL$ la \wiit[deformation@déformation ! locale]{déformation locale}.
La métrique reste orthogonale
et les relations (\ref{T/g/nested}) se lisent
\begin{equation}
\label{T/g/el/nested}
g^{\xi\xi}\!\sqrt{g}=\frac{1}{\sqrt{\,\sinh^2 b+\gtsTL_{\xi}^2}},
\qquad g^{\eta\eta}\!\sqrt{g}={\sqrt{\,\sinh^2 b+\gtsTL_{\xi}^2}}.
\end{equation}
De long calculs fastidieux conduisent,
en utilisant des outils classiques
de géométrie différentielle \cite{Struik},
à l'expression suivante de
la \wiit[courbure ! moyenne locale]{courbure moyenne locale} :
\begin{multline}
\label{T/mc}
2a\:\gtsMC\left(\xi,\eta\right)=\!
\frac{\sinh b}{\sqrt{\,\sinh^2 b+\gtsTL_{\xi}^2}}
\left[\sinh b+\frac{\cosh b\cos\eta-1}{\sinh b}\right]\\
+\frac{\cosh b-\cos\eta}%
{{\left[\sinh^2 b+\gtsTL_{\xi}^2\right]}^{\frac{3}{2}}}
\left[\sinh b\gtsTL_{\xi\xi}-\cosh b\gtsTL_{\xi}^2\right].
\end{multline}
Clairement,
pour des \wiit[deformation@déformation ! suffisamment lisse]%
  {déformations suffisamment lisses},
le second terme s'évanouit
et le facteur d'échelle du premier terme tend vers un :
l'expression qui s'en dégage correspond alors
à la \wiit[courbure ! moyenne rigidifiée]{courbure moyenne rigidifiée}.
En conséquence de quoi,
nous essayons comme
\wiit[courbure ! moyenne spontanée]{courbure moyenne spontanée}
non nulle $\gtsSMC$ la fonction
\begin{multline}
\label{T/Hm/el/smc}
2a\:\gtsSMC\left(\xi,\eta\right)=
\sinh b_{0}+\frac{\cosh b_{0}\cos\eta-1}{\sinh b_{0}}\\
+\frac{\cosh b-\cos\eta}%
  {{\left[\sinh^2 b+\gtsTL_{\xi}^2\right]}^{\frac{3}{2}}}
\left[\widetilde{\gtsel}_{2}\sinh b\gtsTL_{\xi\xi}
  -\widetilde{\gtsel}_{1}\cosh b\gtsTL_{\xi}^2\right],
\end{multline}
avec $\widetilde{\gtsel}_{1}$ et $\widetilde{\gtsel}_{2}$
des \wiit[constante phénoménologique]{constantes phénoménologiques}.
Pour parachever notre modélisation, développons
la densité du Hamiltonien élastique (\ref{Hm/el/heuristic/nested})
jusqu'à l'ordre deux
en $\gtsTL$, $\gtsTL_{\xi}$ et $\gtsTL_{\xi\xi}$ :
\begin{subequations}
\label{T/Hm/el/2nd}
\begin{equation}
\label{T/Hm/el/2nd/nested}
\gtsHm_{el}=
{\textstyle{\frac{1}{2}}}\pi\gtsKc\!
\int\limits_{-\Delta\zeta}^{+\Delta\zeta}\!\!\!\gtsd\zeta\,
\left[
  \gtsCst_{2}\gtsTL^2
  +2\gtsCst_{1}\gtsel_{2}\gtsTL_{\zeta}^2
  +\gtsel_{2}^2\gtsTL_{\zeta\zeta}^2
\right],
\end{equation}
avec
\begin{align}
\gtsel_{2}&=1-\widetilde{\gtsel}_{2},\\
\gtsCst_{1}&=\frac{1+\sinh b_{0}\cosh b_{0}}{\sinh^2 b_{0}},\\
\gtsCst_{2}&=\frac{1+\sinh b_{0}\cosh b_{0}\left(3+\sinh^2 b_{0}\right)}%
  {\sinh^4 b_{0}}.
\end{align}
\end{subequations}
Précisons le changement d'échelle effectué :
\begin{equation}
\label{TS/scaling}
\zeta\equiv{\sinh b_{0}}\;\xi
\quad\text{et}\quad
\Delta\zeta\equiv{\sinh b_{0}}\;\Delta\xi.
\end{equation}
Les \wiit[deformation@déformation ! spontanée]{déformations spontanées} $\gtsTL$
qui extrémisent cet Hamiltonien élastique (\ref{T/Hm/el/2nd})
satisfont aux \wiit[equations@équations ! d'Euler-Lagrange]%
  {équations d'Euler-Lagrange} suivantes
\cite{GiaquintaI}
\begin{equation}
\label{T/Hm/el/2nd/EL}
\gtsCst_{2}\gtsTL-2\gtsCst_{1}\gtsel_{2}\gtsTL_{\zeta\zeta}%
  +\gtsel_{2}^2\gtsTL_{\zeta\zeta\zeta\zeta}=0.
\end{equation}
Pour interdire toutes déformations spontanées
nous imposerons
\begin{equation}
\label{T/Hm/el/2nd/EL/smoothness}
\gtsel_{2}=0.
\end{equation}
Finalement,
pour des \witextit{déformations continues de faible amplitude},
\wih{deformation@déformation ! de faible amplitude}
\wih{deformation@déformation ! continue}
le \wiit[Hamiltonien ! elastique@élastique]{Hamiltonien élastique}
(\ref{Hm/el/heuristic/nested}) s'écrit
\begin{subequations}
\label{TS/Hm/el}
\begin{align}
\gtsHm_{el}
\label{TS/Hm/el/nested}
&={\textstyle{\frac{1}{2}}}\pi\gtsKc\gtsCst_{2}\!
\int\limits_{-\Delta\zeta}^{+\Delta\zeta}\!\!\!\gtsd\zeta\,\gtsTL^2,\\
\label{TS/Hm/el/crude}
&={\textstyle{\frac{1}{2}}}\pi\gtsKc\gtsCst_{2}\!
\int\limits_{-\Delta\zeta}^{+\Delta\zeta}\!\!\!\gtsd\zeta\,
{\left(b-b_{0}\right)}^2.
\end{align}
\end{subequations}

Les déformations $\gtsTL$ subies par
les sections de tore souples magnétiques
et induites par les solitons magnétiques frustrés gelés
extrémisent ainsi le \wiit[Hamiltonien ! total]{Hamiltonien total}
\begin{multline}
\label{TS/Hm/frozen/nested}
\gtsHm=
2\pi\gtsJ\gtsq_{\eta}^2\!
\int\limits_{-\Delta\zeta}^{+\Delta\zeta}\!\!\gtsd\zeta\,
\left[\vphantom{\left(\frac{1}{16\gtsJr\gtsq_{\eta}^2}\right)^2}
\left(2+\gtsm\right)
-\gtsm\left[\coth b_{0}\,\gtsTL+
  {\textstyle\frac{1}{2}}\gtsTL_{\zeta}^2\right]
\right.\\
-\left(1+\gtsm\right)\left[2+\coth^2 b_{0}\,\gtsTL^2\right]
\JacobiSN^2\left(\gtsq_{\eta}\zeta\mid{1+\gtsm}\right)\\
+\left.\vphantom{\left(\frac{1}{16\gtsJr\gtsq_{\eta}^2}\right)^2}
\left(1+{\textstyle\frac{1}{2}}\gtsm+\frac{1+\gtsm}{\sinh^2 b_{0}}%
  +\frac{\gtsCst_{2}}{4\gtsJr\gtsq_{\eta}^2}\right)\!\gtsTL^2
\right].
\end{multline}
Les \wiit[equations@équations ! d'Euler-Lagrange]{équations d'Euler-Lagrange}
prennent alors la forme escomptée
\begin{subequations}
\label{TS/Hm/frozen/EL}
\begin{equation}
\label{TS/Hm/frozen/EL/eqn}
\gtsTL_{\varrho\varrho}+
\left[\left(1+\gtsLm\right)\!\gtsLA
-\gtsLm\gtsLB\;\JacobiSN^2\left(\varrho\mid\gtsLm\right)
\right]\gtsTL
=\sqrt{\gtsLm}\ \gtsLj,
\end{equation}
où
\begin{align}
\gtsLm&=1+\gtsm,\\
\gtsLA&=\frac{1}{\gtsm\gtsq_{\eta}^2}
\left[1+\frac{2}{2+\gtsm}
  \left(\frac{1+\gtsm}{\sinh^2 b_{0}}
    +\frac{\gtsCst_{2}}{\gtsJr\gtsq_{\eta}^2}
\right)\right],\\
\gtsLB&=2\frac{\coth^2 b_{0}}{\gtsm\gtsq_{\eta}^2},\\
\gtsLj&=\frac{\coth b_{0}}{\gtsq_{\eta}^2\sqrt{1+\gtsm}}.
\end{align}
\end{subequations}
La nouvelle variable vérifiant ici $\varrho\equiv\gtsq_{\eta}\zeta$.
Comme pressenti
les lois de déformation des sections de tore
sont similaires à celles des sections de cylindre.
\pagebreak

\begin{figure}[!ht]
  \begin{center}
  \includegraphics[width=.98\linewidth]{sgtsCTSrdKappa.mps}
  \caption[Sections de tore magnétiques souples : dilatation]{%
    La \wiit[dilatation relative ! du tore]{dilatation relative} $\gtsrd_{T}$
    d'une section magnétique souple du \wiit{tore de Clifford}
    (\figurename~\ref{fig/torus/Clifford})
    en présence d'un $\pi$-soliton magnétique :
    la dilatation relative $\gtsrd_{T}$ définie par (\ref{T/rd/def})
    est représentée en fonction de l'angle de rotation $\xi$
    pour différentes valeurs
    de la \wiit{constante relative de couplage} $\gtsJr\equiv\gtsJ/\gtsKc$,
    le demi-angle de rotation étant fixée
    $\left(\Delta\xi=\pi/2\right)$.\\
    $\bullet$ Le cas dégénéré $\left(\gtsJr=0\right)$
    correspond à la section souple rigidifiée
    $\left(\gtsKc\gg\gtsJ\right)$.\\
    $\bullet$ Les autres cas $\left(\gtsJr\approx{1}\right)$ illustrent
    le mécanisme de \wiit{libération géométrique} :
    le soliton magnétique fortement confiné $\left(\sinh b_{0}=1\right)$
    \wih{soliton ! confiné}
    refoule son excès d'énergie magnétique
    $\Delta\gtsEg_{mag}=\gtsEg_{mag}-\gtsTE_{\gtsPontrjagin}$
    en rétrécissant globalement la membrane magnétique souple;
    l'énergie du soliton étant localisée en sa vague
    et l'énergie élastique isotrope,
    la contraction est moins importante en sa vague
    et la membrane souple l'épouse d'autant plus
    que la \wiit{constante relative de couplage} $\gtsJr$ est grande.
    }
  \label{fig/T/rd}
  \end{center}
\end{figure}
\clearpage

\textlatin{A fortiori}
les changements d'échelle (\ref{C/sym/scaling})
et (\ref{T/Hm/mg/scaling})
dictent la \wiit[similitude géométrique]{relation de similitude} entre
les \wiit[paramètre ! de forme pertinent]{paramètres de forme pertinents}
des sections de cylindre (le rayon $r$)
et des sections de tore (l'angle d'excentricité $b$);
en utilisant une notation évidente,
nous écrivons
\begin{equation}
\label{C-T/sim}
r\leftrightarrow\frac{1}{\sinh b}.
\end{equation}
En vertu de cette \wibf{similitude géométrique} (\ref{C-T/sim}),
le mécanisme de déformation des sections de tore
se déduit littéralement de celui des section de cylindre :
la compétition entre l'énergie magnétique
et l'énergie élastique entraîne un accroissement
de l'angle d'excentricité $b$,
ce processus s'amplifiant aux branches solitoniques.
Observant que le produit
du \wiit[rayon ! extérieur]{rayon extérieur} $\overline{R}\equiv{R+r}$
et du \wiit[rayon ! intérieur]{rayon intérieur} $\underline{R}\equiv{R-r}$
reste constant
($\overline{R}\,\underline{R}=a^2$ selon (\ref{T/coords/pp/par}))
définissons la
\wibf[dilatation relative ! du tore]{dilatation relative du tore}
$\gtsrd_{T}$ par
\begin{equation}
\label{T/rd/def}
\gtsrd_{T}\equiv%
  \frac{\overline{R}}{{\overline{R}}_{0}}=%
  \left(\frac{\underline{R}}{{\underline{R}}_{0}}\right)^{-1},
\end{equation}
où ${\overline{R}}_{0}$ et ${\underline{R}}_{0}$
sont associées à la \wiit[forme ! spontanée]{forme spontanée}.
Les relations de transformation (\ref{T/coords/n/par})
permettent d'établir une relation simple
entre la \wiit[dilatation relative ! du tore]%
  {dilatation relative du tore}
$\gtsrd_{T}$ et l'\wiit[angle ! d'excentricité]{angle d'excentricité} $b$ :
\begin{equation}
\label{T/rd/nested}
\gtsrd_{T}=
\frac{\tanh{\scriptstyle{\frac{1}{2}}}b_{0}}
  {\tanh{\scriptstyle{\frac{1}{2}}}b}.
\end{equation}

\subsection{Déformations symétriques du tore de Clifford}
\begin{figure}[bh]
  \begin{center}
  \includegraphics[width=\linewidth]{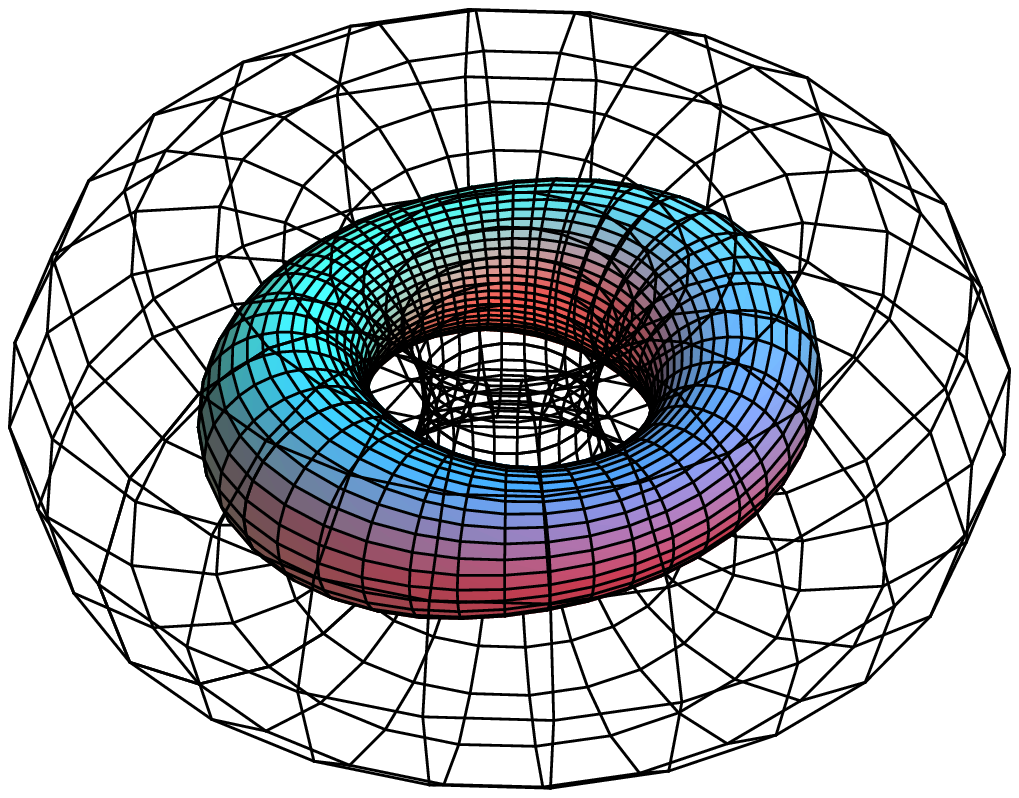}
  \caption[Tore de Clifford magnétique souple : déformation symétrique]{%
    \wiit[deformation@déformation ! symétrique]{Déformation symétrique}
    du tore de Clifford magnétique souple
    en présence d'un $2\pi$-soliton magnétique symétrique :
    l'angle d'excentricité $b_{0}$ du tore spontané
    (représentée par la grille extérieure)
    vérifie $\sinh b_{0}=1$ par définition
    (\figurename~\ref{fig/torus/Clifford}),
    la \wiit{constante relative de couplage}
    $\gtsJr\equiv\gtsJ/\gtsKc$ est égale à $2$,
    enfin la déformation locale $\gtsTL$
    est magnifiée par un facteur $10$.\\
    Le \wiit[soliton ! frustré]{soliton frustré}
    se libère\wih{libération géométrique}
    de sa symétrie imposée
    en contractant globalement la membrane magnétique;
    les deux renflements diamétralement opposés
    correspondent aux deux \wiit[soliton ! vague du]{vagues} du soliton :
    son énergie magnétique $\gtsEg_{mag}$ s'y concentrant
    indépendamment de la géométrie de la membrane,
    il refoule l'énergie magnétique accumulée en ses branches.
    }
  \label{fig/ECT}
  \end{center}
\end{figure}

Le tore de Clifford magnétique souple serait assouvi
puisqu'il saturerait
son \wiit[energie@énergie ! elastique@élastique]{énergie élastique}
selon la \wiit[conjecture de Clifford]{conjecture de Clifford}
(\figurename~\ref{fig/torus/Clifford})
tout en saturant
son \wiit[energie@énergie ! magnétique]{énergie magnétique}
du fait de sa dégénérescence solitonique
(\subsectionname~\ref{chp/HSES/HSRT}).
Par déformations symétriques du tore de Clifford
nous entendrons les déformations
subies par le tore de Clifford magnétique souple
et induites par un soliton magnétique symétrique :
en fait il s'agit du cas dégénéré $\Delta\xi=\pi$
d'une section du tore de Clifford
vu dans la {\subsectionname} précédente.
La \figurename~\ref{fig/ECT} illustre cette approche.

\ProvidesFile{gtsConclusion.tex}[1999/05/20 v0.5b GTS project: Conclusion]
\specialchapter{Conclusion}
\begin{introduction}
La première partie esquisse la construction
de nouveaux degrés de liberté antiferromagnétiques
pour un réseau régulier de basse dimension.
Des polytopes réguliers du réseau
ils héritent de la hiérarchie des algèbres~à~divisions :
le nouveau paramètre d'ordre local
suit la hiérarchie algébrique satisfaite
par le modèle $\sigma$~non-linéaire.
Le réseau de l'état antiferromagnétique
(le \guillemets{sur-réseau})
apparaît alors comme un réseau de n{\oe}uds
(les \guillemets{sur-n{\oe}uds})
avec liens
(le champs potentiel vecteur
associé au champs de jauge locale).
Une étude approfondie
du nouveau paramètre d'ordre local
pris à la limite continue
semble un préliminaire nécessaire
au passage à une théorie de champs continue.
Notons enfin que
l'argument naïf qui consiste en dimension deux
(algèbre des complexes~$\mathbb{C}$)
à envoyer le plan-temps~compactifié~$S^3$
sur la sphère~unité~$S^2$
pour dégager l'invariant topologique de Hopf
n'est plus pertinent en dimension trois
(algèbre des quaternions~$\mathbb{H}$)
puisque la hiérarchie algébrique impose
alors d'envoyer
la sphère~unité~$S^7$ sur la sphère~unité~$S^4$
\cite{Ebbinghaus,Nakahara}:
une étude du réseau cubique antiferromagnétique
devrait permettre d'affiner la prescription esquissée,
d'autant plus que réseau cubique antiferromagnétique
peut-être soit assouvi soit frustré
\cite{UPTAFA}.
En dimension quatre
(algèbre des octonions~$\mathbb{O}$),
l'invariant topologique de Hopf
décompte les recouvrements de la sphère~unité~$S^8$
lorsque la sphère~unité~$S^{15}$ est parcourue :
une étude du réseau $4$-cubique et $4$-triangulaire
\cite{Coxeter}
devrait permettre de celer la prescription
et une meilleur compréhension
du mécanisme algébrique sous- jasent
puisque l'algèbre des octonions~$\mathbb{O}$
n'est ni associative ni commutative
mais seulement alternative
---
l'alternativité étant commune
aux quatre algèbres réelles à divisions~$\mathbb{A}_{\gtsDimension}$.
\bigskip

La seconde partie s'évertue à décrire
la membrane magnétique
comme un système physique
possédant une double obstruction topologique.
La frustration géométrique s'interprète alors
comme la compétition entre les deux ordres topologiques.
En outre cette approche suggère la construction
d'un Hamiltonien élastique qui admettrait
une décomposition \guillemets{à la Bogomol'nyi},
c'est-à-dire qui lèverait
la classification topologique des surfaces
comme le modèle $\sigma$~non-linéaire
(le Hamiltonien magnétique)
lève la classification des configurations magnétiques.
Le Hamiltonien heuristique de Helfrich-Willmore
qui mimique une telle décomposition a permis
de mieux saisir le mécanisme de libération géométrique.
Les déformations d'une membrane souple ont été appréhendées en gelant
une configuration magnétique de ladite membrane souple rigidifiée :
un rétrécissement global
avec des renflements localisés aux retournements des spins
a été mis en évidence pour les membranes magnétiques frustrées.
Ce résultat demande à être confirmé
en élaborant un Hamiltonien élastique
indépendant de la topologie de la membrane souple
(c'est-à-dire \guillemets{à la Bogomol'nyi})
et en abordant la compétition
entre les deux ordres topologiques conjointement.
Notons que l'existence d'un tel Hamiltonien élastique soulèverait
l'existence d'Hamiltoniens \guillemets{à la Bogomol'nyi}.
\bigskip

Le processus microscopique
qui se dégage de la prescription suggérée
dans la première partie
et la compétition topologique introduite
dans la deuxième partie
élargissent le champs d'investigation initial
en fournissant une origine microscopique
à la transmutation de Fermi-Bose
et un moyen de traquer la physique exotique
étudiée lors de cette thèse.
Des simulations numériques du modèle de Heisenberg
sur des réseaux bidimensionnels
souples spontanément courbes
devrait ainsi exhiber
des distributions topologiques de spins
dont l'ordonnancement local permettrait
de préciser le processus microscopique.
De même l'étude des transitions de phase
entre les différents états antiferromagnétiques topologiques
devrait également s'avérer utile.
\end{introduction}

\appendix
\ProvidesFile{gtsAppendixA.tex}[1999/06/17 v0.8b GTS project: Appendix A]
\chapter{L'équation de sinus-Gordon}
\label{app/SG}
\begin{introduction}
Cet {\appendixname} présente des solutions périodiques
de l'\wi[sinus-Gordon ! equation@équation double de]{équation double de sinus-Gordon}%
(\textsc{dsg}) obtenues élégamment à partir de solutions périodiques
de l'\wi[sinus-Gordon ! equation@équation simple de]{équation simple de sinus-Gordon}
(\textsc{sg}).
\end{introduction}

\section{L'équation simple de sinus-Gordon}
\label{app/SG/SSG}
\enlargethispage*{2\baselineskip}
Considérons
l'\wibf[sinus-Gordon ! equation@équation simple de]{équation de sinus-Gordon}
(\textsc{sg})
\swih[sinus-Gordon]{equation@équation ! sinus@de sinus-Gordon}
\begin{equation}
\label{SG/MTheta}
\gtsMTheta_{\varrho\varrho}=\sin\gtsMTheta\cos\gtsMTheta.
\end{equation}
\begin{subequations}
\label{SG/res}
En intégrant une première fois
l'équation différentielle du second degrés (\ref{SG/MTheta}),
nous lisons
\begin{equation}
\label{SG/res/int1}
\gtsMTheta_{\varrho}^2=\sin^2 \gtsMTheta +\gtsm
\qquad
\gtsm\in\left[0,+\infty\right[,
\end{equation}
où le \wibf[sinus-Gordon ! paramètre]{paramètre} $\gtsm$
\swih[sinus-Gordon]{paramètre}
est une constante d'intégration.
Il est alors aisé d'écrire l'équation satisfaite par
la fonction réciproque $\varrho=\varrho\left(\gtsMTheta\right)$ :
\begin{equation}
\label{SG/res/inv}
\frac{\gtsd\varrho}{\gtsd\gtsMTheta}=
\frac{\gtsSign}{\sqrt{\gtsm +\sin^2 \gtsMTheta}}
\qquad
\gtsSign=\pm{1}.
\end{equation}
Notons $\gtsSG\left(\cdot\mid\gtsm\right)$
la fonction croissante solution de (\ref{SG/MTheta})
caractérisée par
le \wiit[sinus-Gordon ! paramètre]{paramètre} $\gtsm$
et fixée à l'origine,
c'est-à-dire vérifiant $\gtsSG\left(0\mid\gtsm\right)=\frac{\pi}{2}$.
La fonction réciproque de $\gtsSG\left(\cdot\mid\gtsm\right)$
est donnée par
l'\wiit[intégrale canonique elliptique]{intégrale elliptique}
\begin{equation}
\label{SG/res/def}
\gtsSG^{-1}\left(\theta\mid\gtsm\right)=
\int\limits_{\frac{\pi}{2}}^{\theta}\!\!
  \frac{\gtsd\varphi}{\sqrt{\gtsm +\sin^2\varphi}},
\end{equation}
solution immédiate de (\ref{SG/res/inv}).
Ainsi les solutions \textsc{sg} envisagées s'écriront-elles,
avec des notations naturelles,
\begin{equation}
\label{SG/res/gsol}
\theta\left(\varrho\right)=
  \gtsSign\gtsSG\left(\varrho\mid\gtsm\right)+\gtsConst.
\end{equation}
\end{subequations}
\noindent
En observant que
\begin{equation*}
\gtsSG^{-1}\left(\theta+\pi\mid\gtsm\right)+
    2\gtsSG^{-1}\left(0\mid\gtsm\right)=
  \gtsSG^{-1}\left(\theta\mid\gtsm\right),
\end{equation*}
nous définissons
la \wibf[sinus-Gordon ! quasi quart-période]{quasi quart-période}
\swih[sinus-Gordon]{quasi quart-période}
$\gtsK$ par
l'\wiit[intégrale canonique elliptique]{intégrale elliptique complète}
\begin{equation}
\label{SG/qp/def}
\gtsK\left(\gtsm\right)=
  \int\limits_{0}^{\frac{\pi}{2}}\!\!
    \frac{\gtsd\varphi}{\sqrt{\gtsm +\sin^2\varphi}},
\end{equation}
qui décroît strictement de $\infty$ vers $0$ en fonction
du \wiit[sinus-Gordon ! paramètre]{paramètre} $\gtsm$.
\begin{subequations}
\label{SG/rel}
La fonction $\gtsSG\left(\cdot\mid\gtsm\right)$
vérifie donc la relation
\begin{equation}
\label{SG/rel/qp}
\gtsSG\left(\varrho+2\gtsq\gtsK\mid\gtsm\right)=
  \gtsSG\left(\varrho\mid\gtsm\right)+\gtsq\pi
\qquad
\gtsq\in\mathbb{Z};
\end{equation}
de plus,
\begin{equation}
\label{SG/rel/id}
\gtsSG\left(-\gtsK\right)=0,
\quad\gtsSG\left(0\right)={\textstyle\frac{\pi}{2}},
\quad\gtsSG\left(\gtsK\right)=\pi;
\end{equation}
enfin, par construction,
\begin{equation}
\label{SG/rel/sym}
{\textstyle\frac{\pi}{2}}-%
  \gtsSG\left(-\varrho\mid\gtsm\right)=
\gtsSG\left(\varrho\mid\gtsm\right)-%
  {\textstyle\frac{\pi}{2}}.
\end{equation}
\end{subequations}
Clairement le \wiit[sinus-Gordon ! paramètre]{paramètre} $\gtsm$
contrôle
la \wibf[sinus-Gordon ! quasi période]{quasi périodicité}
\swih[sinus-Gordon]{quasi période}
des solutions.
Lorsque le \wiit[sinus-Gordon ! paramètre]{paramètre} $\gtsm$ tend vers $0$,
cette \wiit[sinus-Gordon ! quasi période]{quasi périodicité}
tend vers l'infini, et la solution $\gtsSG\left(\cdot\mid\gtsm\right)$
devient un \wibf{soliton} localisé à l'origine.

\begin{emphasise}
\wih{trigonométrie jacobienne}
L'\wibf[sinus-Gordon ! equation@équation simple de]{équation de sinus-Gordon}
a été résolue sans effort aucun,
mais les relations ainsi obtenues sont peu commodes à utiliser
en \wiit[calcul symbolique/numérique]{calcul symbolique ou numérique}.
Des expressions explicites peuvent être écrites en termes
d'\wibf[intégrale canonique elliptique]{intégrales canoniques elliptiques}
et des \wibf[fonctions ! elliptiques de Jacobi]{fonctions elliptiques de Jacobi}
\swih[fonctions]{Jacobi}
\cite{Whittaker,Lawden}.
\end{emphasise}

En effet,
en effectuant le changement de variable judicieux \cite{Sutcliffe}
\linebreak[4]
$\sin \gtsMTheta=\JacobiDN\left(\gtsu\mid 1+\gtsm\right)$,
l'équation différentielle originelle (\ref{SG/MTheta})
prend la forme simple $\gtsu_{\varrho}^2=1$.
Immédiatement, nous lisons
\begin{equation}
\label{SG/Jacobi/sin}
\sin\gtsSG\left(\varrho\mid\gtsm\right)=
  \JacobiDN\left(\varrho\mid 1+\gtsm\right).
\end{equation}
En substituant cette relation dans (\ref{SG/res/int1}),
nous obtenons
\begin{equation}
\label{SG/Jacobi/int1}
\gtsSG_{\varrho}^2\left(\varrho\mid\gtsm\right)=
\left(1+\gtsm\right)\;\JacobiCN^2\left(\varrho\mid 1+\gtsm\right).
\end{equation}
\begin{subequations}
\label{SG/Jacobi/cos}
La trigonométrie classique et jacobienne permettent
de calculer la variante cosinus de (\ref{SG/Jacobi/sin}) :
\begin{align}
\label{SG/Jacobi/cos/crude}
\cos\gtsSG\left(\varrho\mid\gtsm\right)
&=-\sqrt{1\!+\!\gtsm}\:
  \JacobiSN\!\left(\varrho\mid1\!+\!\gtsm\right),\\
\label{SG/Jacobi/cos/nested}
&=-\JacobiSN\left(\varrho\sqrt{1\!+\!\gtsm}%
  \mid1/{1\!+\!\gtsm}\right).
\end{align}
\end{subequations}
Pour finir,
écrivons la variante en tangente de l'angle moitié :
\begin{equation}
\label{SG/Jacobi/tan}
\tan{{\scriptstyle{\frac{1}{2}}}
  \gtsSG\left(\varrho\mid\gtsm\right)}=
\JacobiND\left(\varrho\mid 1+\gtsm\right)
  +\sqrt{1+\gtsm}\;\JacobiSD\left(\varrho\mid 1+\gtsm\right),
\end{equation}
qui prend la forme dégénérée,
quand $\gtsm$ tend vers $0$,
$\tan{{\scriptstyle{\frac{1}{2}}}
\gtsSG\left(\varrho\mid0\right)}=\gtse^{\varrho}$.

Concernant
la \wiit[sinus-Gordon ! quasi quart-période]{quasi quart-période}
$\gtsK\left(\gtsm\right)$,
le rapprochement
des relations (\ref{SG/rel/qp}) et (\ref{SG/Jacobi/sin})
permet de l'identifier avec la quart-période $\EllipticK\left(1+\gtsm\right)$
des \wiit[fonctions ! elliptiques de Jacobi]{fonctions elliptiques de Jacobi}
$\JacobiPQ\left(\cdot\mid 1+\gtsm\right)$.
Précisément, nous avons
\begin{subequations}
\label{SG/qp/ei}
\begin{align}
\label{SG/qp/ei/crude}
\gtsK\left(\gtsm\right)
&=\EllipticK\left(1+\gtsm\right),\\
\label{SG/qp/ei/nested}
&=\frac{1}{\sqrt{1+\gtsm}}\:\EllipticK\left(\frac{1}{1+\gtsm}\right).
\end{align}
\end{subequations}
En \wiit[calcul symbolique/numérique]{calcul symbolique},
$\EllipticK$ correspond
à l'\wiit[intégrale canonique elliptique]{intégrale elliptique de première espèce}.

\section{L'équation double de sinus-Gordon}
\label{app/SG/DSG}
L'\wibf[sinus-Gordon ! equation@équation double de]{équation double de sinus-Gordon}
(\textsc{dsg}) s'obtient en ajoutant un terme de couplage
à l'\wiit[sinus-Gordon ! equation@équation simple de]{équation simple de sinus-Gordon} :
\begin{equation}
\label{DSG/MTheta}
\gtsMTheta_{\varrho\varrho}=
  \sin \gtsMTheta \cos \gtsMTheta
  +\gtsgm\sin \gtsMTheta.
\end{equation}
Le changement de variable
\begin{equation}
\label{DSG/MTheta/cv}
\gtsMTg\equiv\tan{\scriptstyle{\frac{1}{2}}}\gtsMTheta
\qquad
\gtsMTg\in\left[0,+\infty\right[
\end{equation}
transforme
l'équation différentielle non linéaire (\ref{DSG/MTheta})
sous la forme plus commode
\begin{equation}
\label{DSG/tg}
\left(1+\gtsMTg^2\right)\,\gtsMTg_{\varrho\varrho}\!=\!
2\gtsMTg\left[
  \gtsMTg_{\varrho}^2+
    {\scriptstyle{\frac{1}{2}}}\!\left(1+\gtsgm\right)-
    {\scriptstyle{\frac{1}{2}}}\!\left(1-\gtsgm\right)\gtsMTg^2
\right].
\end{equation}
Cette équation différentielle admet
deux solutions triviales de type \wiit{soliton} :
le double soliton
$\gtsMTg=\cosh{\frac{1}{\sqrt{2}}\varrho}$ ($\gtsgm=\frac{1}{2}$)
et le soliton
$\gtsMTg=\gtse^{\varrho}$ ($\gtsgm=0$).

Clairement, la transformation
$(\varrho,\gtsMTg)\to(\lambda\varrho,\mu\gtsMTg)$
et l' absence de couplage $\gtsgm$ ne modifient pas
la nature de la nouvelle équation \textsc{dsg} (\ref{DSG/tg}).
Cette remarque suggère fortement d'injecter
dans (\ref{DSG/tg}) la fonction essai
\begin{equation}
\label{DSG/trial}
\gtsMTg=
\tan{{\scriptstyle{\frac{1}{2}}}\gtsSA}\;
\tan{{\scriptstyle{\frac{1}{2}}}
  \gtsSG\left({\varrho/\gtsSL}\mid\gtsm\right)}.
\end{equation}
De cette manipulation fastidieuse se dégagent
un système d'équations dont s'extraient les expressions
de $\gtsSL$ et de $\tan{{\scriptstyle{\frac{1}{2}}}\gtsSA}$ :
\begin{subequations}
\label{DSG/Jacobi/rel}
\begin{align}
\label{DSG/Jacobi/rel/length}
\gtsSL\!&=\!
\sqrt{\frac{2+\gtsm-%
    \sqrt{\gtsm^2\!+\!4\left(1\!+\!\gtsm\right)\gtsgm^2}}
  {2\left(1-\gtsgm^2\right)}},\\
\label{DSG/Jacobi/rel/angle}
\tan{{\scriptstyle{\frac{1}{2}}}\gtsSA}\!&=\!\!
\sqrt{\frac{1+\gtsgm}{1-\gtsgm}\;
\frac{2\gtsgm+\gtsm-%
    \sqrt{\gtsm^2\!+\!4\left(1\!+\!\gtsm\right)\gtsgm^2}}
  {2\gtsgm-\gtsm+%
    \sqrt{\gtsm^2\!+\!4\left(1\!+\!\gtsm\right)\gtsgm^2}}}.
\end{align}
\end{subequations}
Lorsque le couplage $\gtsgm$ disparaît,
la longueur $\gtsSL$ et
$\tan{{\scriptstyle{\frac{1}{2}}}\gtsSA}$
tendent vers l'unité :
sans surprise la fonction essai (\ref{DSG/trial})
approche la fonction \textsc{sg} associée.
Ainsi ces solutions périodiques se révèlent-elles
être une généralisation
des solutions $\gtsSG\left(\cdot\mid\gtsm\right)$.
Notons $\gtsSG\left(\cdot\mid\gtsm,\gtsgm\right)$
les solutions (\ref{DSG/trial});
la formule de l'angle moitié correspondante s'écrit
\begin{equation}
\label{DSG/Jacobi/tan}
\tan{{\scriptstyle{\frac{1}{2}}}%
  \gtsSG\left(\varrho\mid\gtsm,\gtsgm\right)}=
\tan{{\scriptstyle{\frac{1}{2}}}\gtsSA}\;
\left[
  \JacobiND\left(\varrho/\gtsSL\mid 1+\gtsm\right)
  +\sqrt{1+\gtsm}\;
  \JacobiSD\left(\varrho/\gtsSL\mid 1+\gtsm\right)
\right],
\end{equation}
où l'\wiit[angle ! à l'origine]{angle à l'origine} $\gtsSA$
et la \wiit[soliton ! longueur solitonique]{longueur solitonique} $\gtsSL$
\swih[soliton]{longueur solitonique}
vérifient (\ref{DSG/Jacobi/rel}).
En conséquence de quoi,
la \wibf[sinus-Gordon ! quasi quart-période généralisée]%
  {quasi quart-période généralisée}
$\gtsK\left(\gtsm,\gtsgm\right)$ satisfait
\begin{equation}
\label{DSG/qp}
\gtsK\left(\gtsm,\gtsgm\right)=
  \gtsSL\:\gtsK\left(\gtsm\right).
\end{equation}
\begin{subequations}
\label{DSG/rel}
La relation (\ref{SG/rel/qp}) se généralise alors comme suit :
\begin{equation}
\label{DSG/rel/qp}
\gtsSG\left(\varrho+2\gtsq\gtsK\mid\gtsm,\gtsgm\right)=
\gtsSG\left(\varrho\mid\gtsm,%
  {\left(-1\right)}^{\gtsq}\gtsgm\right)+\gtsq\pi
\qquad
\gtsq\in\mathbb{Z};
\end{equation}
de plus, les identités (\ref{SG/rel/id}) deviennent
\begin{equation}
\label{DSG/rel/id}
\gtsSG\left(-\gtsK\right)=0,
\quad\gtsSG\left(0\right)=\gtsSA,
\quad\gtsSG\left(\gtsK\right)=\pi;
\end{equation}
finalement, la relation de parité (\ref{SG/rel/sym}) prend la forme
\begin{equation}
\label{DSG/rel/sym}
{\textstyle\frac{\pi}{2}}-%
  \gtsSG\left(-\varrho\mid\gtsm,\gtsgm\right)=
\gtsSG\left(\varrho\mid\gtsm,-\gtsgm\right)-%
  {\textstyle\frac{\pi}{2}}.
\end{equation}
\end{subequations}

\section{\'Evaluation numérique du paramètre}
\label{app/SG/SGM}
\wih{soliton ! paramètre}
En physique les \witextbf{conditions aux limites}
\wih{conditions ! aux pôles}\wih{conditions ! aux bords}
sont généralement fixées.
Ce qui revient ici à se donner
la \wiit[sinus-Gordon ! quasi quart-période]{quasi quart-période}
$\gtsK$.
Or il n'existe pas dans la littérature
de relation ou d'algorithme numérique
permettant d'évaluer la fonction réciproque $\EllipticK^{-1}$ :
une \wibf[calcul symbolique/numérique]{méthode numérique}~\textlatin{ad hoc}
doit être suggérée.

Pour commencer,
notons que
l'\wiit[intégrale canonique elliptique]{intégrale elliptique de première espèce}
$\EllipticK\left(\Jacobim\right)$ peut s'exprimer sous la forme
d'une \wiit[fonction ! hypergéométrique]{fonction hypergéométrique}
\cite{Abramowitz,Whittaker} :
\begin{equation*}
\EllipticK\left(\Jacobim\right)=\frac{\pi}{2}
  \hypergeom\left(\frac{1}{2},\frac{1}{2};1;\Jacobim\right).
\end{equation*}
\pagebreak

\noindent
Cette remarque et la relation (\ref{SG/qp/ei/nested})
incitent à écrire $\gtsK$ comme une série entière
du \wiit{module réduit} $\gtsrm=1/\sqrt{1+\gtsm}$.
Ces considérations conduisent à l'expression
\begin{equation}
\label{SG/qp/psd/nested}
{\textstyle\frac{2}{\pi}}\gtsK=
  \sum_{n=0}^{\infty}
    {\left[
      \frac{{\left(\frac{1}{2}\right)}_{n}}{{\left(1\right)}_{n}}
    \right]}^2\:\gtsrm^{2n+1},
\end{equation}
où les \wiit[Pochhammer~(symboles de)]{symboles de Pochhammer}
\cite{Abramowitz} sont utilisés.
Cette série entière est alors
\wiit[inversion formelle]{formellement inversée}
pour obtenir la série entière représentant
la fonction $\gtsrm\left(\gtsK\right)$.
Un scripte (codé en \MapleV)
permettant de calculer
les premiers coefficients de cette série
s'écrirait comme suit :
\begin{center}
\includescript{sgtsSGM.mbj}
\end{center}
De cette façon, la relation (\ref{SG/qp/psd/nested}) permet
l'\wiit[calcul symbolique/numérique]{évaluation numérique}
de la fonction réciproque $\gtsK^{-1}$
pour des valeurs numériques suffisamment petites.

Pour des valeurs trop grandes,
les inégalités suivantes \cite{GDAnderson}
\begin{equation}
\ln\left[\frac{2^2}{\sqrt{1-\Jacobim}}\right]
\leqslant\EllipticK\left(\Jacobim\right)\leqslant
\ln\left[\frac{e^2}{\sqrt{1-\Jacobim}}\right]
\qquad
\left[0\!<\!\Jacobim\!<\!1\right]
\end{equation}
s'avèrent utiles (et nécessaires) pour fournir
à notre procédure \MapleProc{fsolve} favorite
un intervalle initial suffisamment fin
pour résoudre (avec succès)
\wih{calcul symbolique/numérique}
l'équation $\gtsK=\gtsrm\:\EllipticK\left(\gtsrm^2\right)$
numériquement.

\clearpage
\begin{figure}[!ht]
  \begin{center}
  \includegraphics[width=.98\linewidth]{sgtsSGM.mps}
  \caption[Fonction réciproque %
      de la quasi quart-période de sinus-Gordon]{%
    La \wiit[sinus-Gordon ! quasi quart-période]{fonction réciproque %
      de la quasi quart-période de sinus-Gordon} $\gtsK$.
    }
  \label{fig/SG/SGM}
  \end{center}
\end{figure}
\hfill

\ProvidesFile{gtsAppendixB.tex}[1999/06/17 v0.8b GTS project: Appendix B]
\chapter{L'équation inhomogène de Lamé}
\label{app/IL}

\begin{introduction}
Cet {\appendixname} se contente de présenter
une solution particulière
de l'\wi[equation@équation ! inhomogène de Lamé]{équation inhomogène de Lamé}%
(\textsc{il}) avec un terme inhomogène constant.
La construction de cette solution particulière
s'inspire de la construction
des fonctions dites de Lamé
\wih{fonction ! Lame@de Lamé}
\swih[fonction]{Lamé}
\swih[equation@équation]{Lamé}
\cite{ErdelyiLm,Arscott}.
\end{introduction}

\section{L'équation aux différences}
\label{app/ILE/de}
Soit
l'\wibf[equation@équation ! inhomogène de Lamé]%
  {équation inhomogène de Lamé} (\textsc{il})
\begin{equation}
\label{ILE/TL}
\gtsTL_{\varrho\varrho}+
\left[\left(1+\gtsLm\right)\!\gtsLA
-\gtsLm\gtsLB\;\JacobiSN^2\left(\varrho\mid\gtsLm\right)
\right]\gtsTL
=\sqrt{\gtsLm}\ \gtsLj,
\end{equation}
où les amplitudes $\gtsLA$ et $\gtsLB$ sont des paramètres constants.
\textlatin{A priori},
le terme inhomogène $\gtsLj$ est un réel non-nul
et le \wibf[equation@équation ! inhomogène de Lamé ! paramètre]{paramètre}
\swih[equation@équation inhomogène de Lamé]{paramètre}
$\gtsLm$ un réel positif.

Comme la \wiit{transformation réelle de Jacobi}
\swih[transformation réelle de Jacobi]{Jacobi}
$\left(\varrho,\Jacobim\right)%
\!\rightarrow\!\left(\varrho\sqrt{\Jacobim},\Jacobim^{-1}\right)$
laisse
l'équation différentielle linéaire du second ordre (\ref{ILE/TL})
inchangée,
nous pouvons supposé $\gtsLm$ compris entre $0$ et $1$.
En outre,
puisque la fonction
$\JacobiSN^2\left(\varrho\mid\Jacobim\right)$ est paire,
l'équation différentielle (\ref{ILE/TL})
reste inchangée sous la transformation de parité
 $\varrho\!\rightarrow\!-\varrho$,
donc une solution particulière pertinente
sera une fonction paire en $\varrho$.
De manière analogue,
la parité en $\varrho-\EllipticK\left(\Jacobim\right)$
de $\JacobiSN^2\left(\varrho\mid\Jacobim\right)$
invite à chercher une solution particulière
également paire en $\varrho-\EllipticK\left(\Jacobim\right)$.
Ces remarques préliminaires faites,
et tenant compte de la nature même de (\ref{ILE/TL}),
cherchons une solution particulière sous la forme
\begin{equation}
\label{ILE/fml}
\sum_{n=0}^{\infty}
\gtsILa_{n}\ \JacobiSN^{2n}\left(\varrho\mid\Jacobim\right).
\end{equation}
\pagebreak

\noindent
La substitution de la fonction formelle (\ref{ILE/fml})
dans (\ref{ILE/TL}) impose
à la suite des coefficients réels $\gtsILa_{n}$ la relation
de \wibf[recurrence@récurrence à trois termes]{récurrence à trois termes}
\cite{ErdelyiLm,Arscott,Gautschi,Lorentzen,WHPress}
suivante :
\begin{subequations}
\label{ILE/fml/ttrr}
\begin{multline}
\label{ILE/fml/ttrr/rec}
\gtsILa_{n+1}
-\left(1+\Jacobim\right)\;
\frac{4n^2-\gtsLA}{\left(2n+2\right)\!\left(2n+1\right)}
\;\gtsILa_{n}\\
+\Jacobim\;
\frac{\left(2n-2\right)\!\left(2n-1\right)-\gtsLB}
  {\left(2n+2\right)\!\left(2n+1\right)}
\;\gtsILa_{n-1}=0
\end{multline}
pour $n=1,2,\ldots$;
la relation d'amorçage étant
\begin{equation}
\label{ILE/fml/ttrr/init}
2\gtsILa_{1}+\left(1+\Jacobim\right)\!\gtsLA\;\gtsILa_{0}=
\sqrt{\Jacobim}\ \gtsLj.
\end{equation}
\end{subequations}
L'équation caractéristique de (\ref{ILE/fml/ttrr})
se lit immédiatement :
\begin{equation}
\label{ILE/fml/ttrr/ce}
\gtsILa^2-\left(1+\Jacobim\right)\;\gtsILa+\Jacobim=0;
\end{equation}
elle admet deux racines évidentes, $\Jacobim$ et $1$.
Donc d'après le \wiit[Perron~(théorème de)]{théorème de Perron}
\cite{Gautschi,WHPress},
pour $0\leqslant \Jacobim<1$ (\textlatin{i.e.} $\Jacobim\neq1$),
l'équation aux différences linéaire du second ordre (\ref{ILE/fml/ttrr})
admet deux solutions linéairement indépendantes, à savoir
la \wibf[solution ! minimale]{solution minimale} et
la \wibf[solution ! dominante]{solution dominante},
dont les rapports $\gtsILa_{n+1}/\gtsILa_{n}$
tendent, respectivement, vers $\Jacobim$ et $1$.

\enlargethispage*{1\baselineskip}
\section{La solution minimale}
\label{app/ILE/ms}
\'Ecrivons la relation de récurrence (\ref{ILE/fml/ttrr/rec})
sous la forme
\begin{equation}
\label{ILE/fml/ttrr/rec/frac}
\frac{\gtsILa_{n}}{\gtsILa_{n-1}}=
\frac{{\scriptstyle -\Jacobim}
    {\frac{\left(2n-2\right)\left(2n-1\right)-\gtsLB}
      {\left(2n+2\right)\left(2n+1\right)}}}
  {{\scriptstyle -\left(1+\Jacobim\right)}
    {\frac{4n^2-\gtsLA}{\left(2n+2\right)\left(2n+1\right)}
    +{\gtsILa_{n+1}}/{\gtsILa_{n}}}}.
\end{equation}
Si la nouvelle relation de récurrence (\ref{ILE/fml/ttrr/rec/frac})
est réitéré indéfiniment à partir d'un certain rang $n\geqslant1$,
une suite de fractions est alors engendrée \cite{Lorentzen}.
Cette suite,
selon le \wiit[Pincherle~(théorème de)]{théorème de Pincherle}
\cite{Gautschi,Lorentzen,WHPress},
converge vers le rapport $\gtsILa_{n}/\gtsILa_{n-1}$
où les coefficients $\gtsILa_{n}$ et $\gtsILa_{n-1}$
sont effectivement ceux de
la \wiit[solution ! minimale]{solution minimale}.
\begin{subequations}
\label{ILE/sol/coef}
Notons $\gtsILr_{n}$ la \wiit{fraction continue} précédente,
les coefficients $\gtsILa_{n}$ sont alors donnés par
\begin{equation}
\label{ILE/sol/coef/gn}
\gtsILa_{n}=\gtsILa_{0}\;\prod_{p=1}^{n}\!\gtsILr_{p}
\qquad \left[n=1,2,\ldots\right],
\end{equation}
avec, selon (\ref{ILE/fml/ttrr/init}),
\begin{equation}
\label{ILE/sol/coef/init}
\gtsILa_{0}=
\frac{\sqrt{\Jacobim}\ \gtsLj}
  {2\gtsILr_{1}+\left(1+\Jacobim\right)\!\gtsLA};
\end{equation}
\pagebreak

\noindent
de plus, avec des notations usuelles \cite{Lorentzen},
le rapport $\gtsILr_{n}$ s'écrit
\begin{equation}
\label{ILE/sol/ratio}
\gtsILr_{n}=
\confrac_{p=n}^{\infty}\!
\left(
\frac{{\scriptstyle -\Jacobim}
    {\frac{\left(2p-2\right)\left(2p-1\right)-\gtsLB}
      {\left(2p+2\right)\left(2p+1\right)}}}
  {{\scriptstyle -\left(1+\Jacobim\right)}
    {\frac{4p^2-\gtsLA}{\left(2p+2\right)\left(2p+1\right)}}}
\right)
\qquad
\left[n=1,2,\ldots\right].
\end{equation}
\end{subequations}

Nous adopterons la notation
$\gtsIL\left(\cdot\mid\Jacobim;\gtsLA,\gtsLB,\gtsLj\right)$
pour désigner la solution particulière de (\ref{ILE/TL})
associée à la solution minimale
de l'équation aux différences (\ref{ILE/fml/ttrr}).
La relation (\ref{ILE/sol/coef/gn}) permet d'écrire
cette solution particulière
\wiit[forme ! Horner@de Horner~(sous la)]{sous la forme de Horner}
\swih[forme]{Horner}
\begin{equation*}
\gtsIL\left(\varrho\mid\Jacobim;\gtsLA,\gtsLB,\gtsLj\right)=
\gtsILa_{0}\;
\biggl[1+\gtsILr_{1}\;\JacobiSN^2\left(\varrho\mid\Jacobim\right)
  \Bigl[1+\gtsILr_{2}\;\JacobiSN^2\left(\varrho\mid\Jacobim\right)
    [\cdots]
  \Bigr]
\biggr],
\end{equation*}
qui doit être penser comme une \guillemets{forme de Horner continue}.
Nous écrirons donc, avec des notations évidentes,
\begin{equation}
\label{ILE/Hrnr}
\gtsIL\left(\varrho\mid\Jacobim;\gtsLA,\gtsLB,\gtsLj\right)=
\gtsILa_{0}\;
\conHorner_{n=1}^{\infty}\!
\left[
1+\gtsILr_{n}\;\JacobiSN^2\left(\varrho\mid\Jacobim\right)
\right].
\end{equation}
L'expression (\ref{ILE/Hrnr}) n'est vraie que pour $0<\Jacobim<1$.
En fait,
l'invariance de
l'\wiit[equation@équation ! inhomogène de Lamé]%
  {équation inhomogène de Lamé} (\ref{ILE/TL})
par la \wiit{transformation réelle de Jacobi} :
\begin{equation}
\gtsIL\left(\varrho\mid\Jacobim;\gtsLA,\gtsLB,\gtsLj\right)=
\gtsIL\left(
  \varrho\sqrt{\Jacobim}\mid\Jacobim^{-1};\gtsLA,\gtsLB,\gtsLj
\right).
\end{equation}

\section{\'Evaluation numérique}
\label{app/ILE/cas}
La construction de
$\gtsIL\left(\varrho\mid\Jacobim;\gtsLA,\gtsLB,\gtsLj\right)$
se prête naturellement à l'élaboration
d'un \wibf[calcul symbolique/numérique]{algorithme numérique}.
Cependant,
l'évaluation de la \wiit{fraction continue} (\ref{ILE/sol/ratio}),
autrement dit des rapports $\gtsILr_{n}$ dans (\ref{ILE/Hrnr}),
apparaît clairement comme la principale difficulté :
l'\wiit[Lentz~(algorithme modifié de)]%
  {algorithme de Lentz dans sa version modifiée}
\cite{WHPress,Lentz,ThompsonBarnett,WBJones}
doit être suggéré.

\backmatter
\bibliographystyle{gtsAlpha}
\bibliography{gts}
\printauthorindex
\printwordindex
\listoffigures
\ProvidesFile{gtsPublications.tex}[1999/05/24 v0.9b GTS project: Publications]
\specialchapter{Publications}
\label{chp/publications}
\section*{Publications}
\begin{publication}
\bibitem{HSETS-p}
  \foreignlanguage{english}{%
    \emph{Heisenberg spins on an elastic torus section}},\\
  Phys. Lett. A \textbf{248} (1998), no.~5-6, 439--444,\\
  avec \surname{Rossen} \name{Dandoloff}.
\bibitem{HSCS-p}
  \foreignlanguage{english}{%
    \emph{Heisenberg spins on a cylinder section}},\\
  \emph{à paraître} \textit{in} Internat. J. Modern Phys. B (1999),\\
  avec \surname{Rossen} \name{Dandoloff}
  et \surname{Avadh} \name{Saxena}.
\end{publication}

\section*{En Préparation}
\begin{publication}
\bibitem{ZZZ-p}
  \foreignlanguage{english}{%
    \emph{Antiferromagnetic Heisenberg spins
      on a triangular lattice:\\
      the continuum limit}},\\
  avec \surname{Rossen} \name{Dandoloff}.
\bibitem{HST-p}
  \foreignlanguage{english}{%
    \emph{Heisenberg spins on a torus}},
  avec \surname{Rossen} \name{Dandoloff}.
\end{publication}

\ProvidesFile{gtsPostamble.tex}[1999/06/21 v0.9b GTS project: Postamble]
\specialchapter{Informations}
\label{chp/informations}
\begin{itemize}
  \item[$\bullet$] \url{perso.wanadoo.fr/jgmbenoit/}
  \item[$\bullet$] \email{jerome.benoit@ptm.u-cergy.fr}
\end{itemize}

\lastmatter
\end{document}